\documentclass[prd,
			altaffilletter,
			twocolumn,
			superscriptaddress,
			floatfix,
			nofootinbib]{revtex4} 
\usepackage[dvipsnames]{xcolor}
\usepackage{amsfonts,
			amsmath,
			graphicx, 
			bm,
			tikz-feynman,
			float,
			mathtools,
			bbold,
			url,
			todonotes,
			ifpdf,
			multirow,
			footmisc,
			pgfplots,
			braket,
			slashed}

\usepackage[colorlinks=true,
			linkcolor=red,
			urlcolor=red,
			citecolor=red]{hyperref}
			
\usetikzlibrary{positioning,decorations.markings}
\tikzset{main node/.style={circle,fill=yellow!40,draw,minimum size=0.65cm,inner sep=0pt}, }
\tikzset{hang node/.style={circle,fill=red!40,draw,minimum size=0.5cm,inner sep=0pt}, }

\usepackage[yyyymmdd,hhmmss]{datetime}

\def\today{\number\day\space\ifcase\month\or
January\or February\or March\or April\or May\or June\or
July\or August\or September\or October\or November\or December\fi
\space\number\year}
\newcount\mins \newcount\hours
\def\now{\hours=\time \mins=\time
	\divide\hours by60 \multiply\hours by60 \advance\mins by-\hours
	\divide\hours by60 
	\number\hours:\ifnum\mins<10 0\fi\number\mins }

\newcommand{\cambridge}{Department of Applied Mathematics and Theoretical Physics, University of Cambridge, Cambridge, CB3 0WA, UK}
\newcommand{\glasgow}{SUPA, School of Physics and Astronomy, University of Glasgow, Glasgow, G12 8QQ, UK}

\graphicspath{{./figures/}}
\begin{document}
%\expandafter\show\the\font
%\stop
%\title{Form factors for the processes $B_c^+ \to D^0 \ell^+ \nu_{\ell}$ and $B_c^+ \to D_s^+ \ell^+ \ell^- (\nu \overline{\nu})$ from lattice QCD\\\rough{version: \today, \currenttime}}
\title{Form factors for the processes $B_c^+ \to D^0 \ell^+ \nu_{\ell}$ and $B_c^+ \to D_s^+ \ell^+ \ell^- (\nu \overline{\nu})$ from lattice QCD}

\author{Laurence~J. \surname{Cooper}} 
\email[]{Laurence.Cooper@glasgow.ac.uk}
\affiliation{\glasgow}
\affiliation{\cambridge}

\author{Christine~T.~H.~\surname{Davies}} 
\email[]{Christine.Davies@glasgow.ac.uk}
\affiliation{\glasgow}

%\author{Judd \surname{Harrison}} 
%%\email[]{Judd.Harrison@glasgow.ac.uk}
%\affiliation{\glasgow}

%\author{Javad \surname{Komijani}} 
%\affiliation{\tehran}

\author{Matthew \surname{Wingate}}
\email[]{M.Wingate@damtp.cam.ac.uk}
\affiliation{\cambridge}

\collaboration{HPQCD Collaboration}
\email[URL: ]{http://www.physics.gla.ac.uk/HPQCD}
\begin{abstract}
	\noindent We present results of the first lattice QCD calculations of the  weak matrix elements for the decays $B_c^+ \to D^0 \ell^+ \nu_{\ell}$, $B_c^+ \to D_s^+ \ell^+ \ell^-$ and $B_c^+ \to D_s^+ \nu \overline{\nu}$. Form factors across the entire physical $q^2$ range are then extracted and extrapolated to the continuum limit with physical quark masses.
	%We combine our form factors for $B_c \to D$ with CKM matrix elements to predict the semileptonic decay rate 
	%%update
	%$\Gamma(B_c^+ \to D^0 \overline{\ell} \nu_{\ell}) = $ \rough{XYZ} $ \times 10^9 \,\text{s}^{-1}$.
	%\rough{The lattice QCD uncertainty is comparable to the CKM uncertainty here.}
	Results are derived from correlation functions computed on MILC collaboration gauge configurations with three different lattice spacings and including 2+1+1 flavours of sea quarks in the Highly Improved Staggered Quark (HISQ) formalism. HISQ is also used for all of the valence quarks.
	The uncertainty on the decay widths from our form factors for $B_c^+ \to D^0 \ell^+ \nu_{\ell}$ is similar in size to that from the present value for $V_{ub}$.
	We obtain the ratio $\Gamma (B_{c}^{+} \rightarrow D^0 \mu^{+} \nu_{\mu}) /\left|\eta_{\mathrm{EW}} V_{u b}\right|^{2}=4.43(63) \times 10^{12} \mathrm{~s}^{-1}$. 
	Combining our form factors with those found previously by HPQCD for $B_{c}^{+} \rightarrow J / \psi \mu^{+} \nu_{\mu}$, we find $\left|V_{cb}/V_{ub} \right|^2 \Gamma( B_c^+ \to D^0 \mu^+ \nu_\mu )/\Gamma(B_{c}^{+} \rightarrow J / \psi \mu^{+} \nu_{\mu}) = 0.257(36)_{B_c \to D}(18)_{B_c \to J/\psi}$.
	We calculate the differential decay widths of $B_c^+ \to D_s^+ \ell^+ \ell^-$ across the full $q^2$ range, and give integrated results in $q^2$ bins that avoid possible effects from charmonium and $u \overline{u}$ resonances.
	For example, we find that the ratio of differential branching fractions integrated over the range $q^2 = 1 \; \mathrm{GeV}^2 - 6 \; \mathrm{GeV}^2$ for $B_c^+ \to D_s^+ \mu^+ \mu^-$ and $B_{c}^{+} \rightarrow J / \psi \mu^{+} \nu_{\mu}$ is $6.31{\tiny }(90)_{B_c \to D_s}(65)_{B_c \to J/\psi} \times 10^{-6}$.
	We also give results for the branching fraction of $B_c^+ \to D_s^+ \nu \overline{\nu}$.
	Prospects for reducing our errors in the future are discussed.
\end{abstract}

\maketitle

%===========================================================================

\section{Introduction}

In this paper, we use lattice QCD methods to calculate the form factors that capture the non-perturbative physics of the pseudoscalar $B_c^+$ meson decaying weakly into either $D^0 \ell^+ \nu_{\ell}$, $D_s^+ \ell^+ \ell^-$ or $D_s^+ \nu \overline{\nu}$.
This is the first time that these calculations have been performed.
%Lattice QCD provides a framework for studying non-perturbative phenomenology of hadrons.
To ascertain the successes and shortcomings of the Standard Model's description of the physics observed in experiment, it is essential to produce predictions from the Standard Model at high precision that fully incorporate the non-perturbative strong interaction phenomenology of hadrons.
Lattice QCD provides a route towards achieving this for the weak matrix elements studied here.

We present the first lattice QCD calculation of the form factors $f_{0}$ and $f_+$ for the vector current matrix elements for $B_c^+ \to D^0 \ell^+ \nu_{\ell}$ throughout the entire range of physical momentum transfer squared, $q^2$.
An accurate prediction from the Standard Model of the normalisation and shape of the form factors for $B_c^+ \to D^0 \ell^+ \nu_{\ell}$ will complement observations of this process from experiment and ultimately lead to a new exclusive determination of the CKM matrix element $|V_{ub}|$ in the future.
LHCb expect~\cite{LHCbImplications} that Upgrade II will make it possible to have a measurement of $B_{c}^{+} \rightarrow D^{0} \mu^{+} \nu_{\mu}$ with sufficient accuracy to offer a competitive determination of $V_{ub}$.
Further scrutiny of $V_{ub}$ is needed to address the long-standing unresolved tension between inclusive and exclusive determinations (for example, see world averages of $V_{ub}$ from both inclusive and exclusive determinations in~\cite{HFLAV:2019otj}).
Exclusive determinations of $V_{ub}$ using form factors from lattice QCD have so far been focussed on the semileptonic decays $B \to \pi$, $B_s \to K$ and $\Lambda_b \to p$, so determining $V_{ub}$ via semileptonic $B_c \to D$ will offer another data point.
We also consider the branching fraction ratio of $B_c \to D$ and, using form factors from~\cite{Harrison:2020gvo}, the process $B_c \to J/\psi$.
This allows the combination $V_{ub}/ V_{cb}$ to be examined given experimental information on this ratio.
%using HPQCD's heavy-HISQ method \cite{McNeile:2010ji,McNeile:2011ng}.

Alongside our calculation of the form factors for $B_c^+ \to D^0 \ell^+ \nu_{\ell}$, we also carry out a lattice QCD calculation of the form factors $f_{0}$, $f_+$ and $f_T$ for the vector and tensor current matrix elements of the rare processes $B_c^+ \to D_s^+ \ell^+ \ell^-$ and $B_c^+ \to D_s^+ \nu \overline{\nu}$.
These semileptonic decays are examples of flavour-changing, neutral current (FCNC) processes and they are of interest in their own right.
Such processes are not allowed at tree-level in the Standard Model, thus contributions from physics beyond the Standard Model may be more visible than with tree-level decays.
Therefore, FCNC transitions are an important avenue towards understanding the validity of the Standard Model. 

%These form factors allow us to predict branching ratios $\mathcal{B} (B_c^+ \to D^0 \ell^+ \nu_{\ell})$ and $\mathcal{B} (B_c^+ \to D_s^+ \ell^+ \ell^-)$.
%The former deecay occurs via a $b \to u$ transition allowed at tree-level in effective weak theory.
%The latter occurs via a $b \to s$ transition which is loop-suppressed.
%For $B_c^+ \to D^0 \ell^+ \nu_{\ell}$, comparisons with experiment could lead to a new determination of the CKM matrix element $V_{ub}$.
%With our Standard Model theory predictions of $B_c^+ \to D_s^+ \ell^+ \ell^-$, new physics contributing to this process can be investigated.

%Lattice QCD provides a framework for studying non-perturbative phenomenology of hadrons.
%To ascertain the successes and shortcomings of the Standard Model's description of the physics observed in experiment, it is essential to produce predictions from the Standard Model at high precision.
%In this paper, we use lattice QCD methods to calculate the form factors that capture the non-perturbative physics of the pseudoscalar $B_c^+$ meson decaying weakly into either $D^0 \ell^+ \nu_{\ell}$ or $D_s^+ \ell^+ \ell^-$ for the first time.
%Our form factors can be combined with observables from experiment to probe the Standard Model.
%\rough{INSERT - mention LHCb here?}

The form factors calculated here are part of an ongoing programme by HPQCD to study weak decays of mesons containing a bottom quark.
Our ultimate aim is to determine Standard Model contributions at high enough precision such that comparison with experiment reveals or constrains new physics scenarios.
We are now in an era in which fully relativistic lattice QCD calculations of decays of mesons containing bottom quarks are achievable.
We use the Highly Improved Staggered Quark formalism (HISQ)~\cite{Follana:2006rc} that is specifically designed to have small discretisation errors.
The large mass of the $b$ quark requires very fine lattices to control discretisation effects.
We simulate with bottom quarks at their physical mass on our finest lattice, and unphysically light bottom quarks on the coarser lattices.
Together this data informs the limit of vanishing lattice spacing and physical quark masses through HPQCD's \lq heavy-HISQ\rq~strategy. 
Recent calculations that have established the method for determining semileptonic form factors include~\cite{McLean:2019sds,McLean:2019qcx,Cooper:2020wnj,Harrison:2020gvo,Parrott:2020vbe,Harrison:2021tol}.
%\rough{\emph{[other citations required here?]}}

%After present the form factors $f_{0,+}$ for $B_c \to D_l$ and $f_{0,+,T}$ for $B_c \to D_s$ in the continuum and with physical masses.
%By using PDG~\cite{PDG} values for various CKM matrix elements, we plot differential branching fractions and find integrated quantities.

We also investigate strategies for improving on this first calculation of the form factors for $B_c \to D$ and $B_c \to D_s$.
These methods will inform the strategy for other future calculations of heavy-to-light quarks decays.
Form factors with smaller uncertainties will offer a more powerful examination of the precision flavour physics we envisage.
To minimise cost, we try these improvements in the $B_c \to D_s$ case only.

The sections in this paper are organised as follows:
\begin{itemize}
	\item Section~\ref{sec:calc_details} gives a comprehensive description of how the  form factors across the entire physical range of 4-momentum transfer are obtained from lattice correlation functions.
	%	Firstly, in Section~\ref{sec:matElts_to_ffs} we motivate the different lattice correlation functions we construct by describing our strategy to extract the form factors from a selection matrix elements.
	%	The suite of lattice correlation functions from which the matrix elements are obtained are discussed in Section~\ref{sec:latt_corr_funcs}, and the choice of parameters and $SU(3)$ gauge field configurations are the focus of Section~\ref{sec:ensembles_params}.
	%	Our methods for fitting the correlation functions are given in Section~\ref{sec:fit_correls}.
	%	Finally, Section~\ref{sec:fit_ffs} shows how we fit the lattice data for the form factors, derived from the matrix extracted in the previous section, to then obtain results in the continuum limit with physical masses.
	Results from fitting the correlation functions are attached to this paper.
	Appendix~\ref{sec:fitting_analysis} discusses intermediate results from the correlation function fitting and form factor fits.
	
	%	Effective simulation energies and amplitudes from the correlation functions are discussed in Section~\ref{sec:sim_energies_amps}.
	%	These are helpful in choosing suitable priors for groundstate quantities, though we ensure that prior widths are much broader than the results obtained by the fit.
	%	The different choices involved when fitting the correlation functions are interrogated in Section~\ref{sec:stab_corrfit} where we demonstrate that the  results for the matrix elements are stable and robustly determined.
	%	After reporting in Section~\ref{sec:report_ZV} values obtained for the multiplicative renormalisation $Z_V$ for the local non-conserved vector current, in Section~\ref{sec:ffs_fit_analysis} we probe the sensitivity of the form factor fits. We find that our results our stable.
	
	\item In Section~\ref{sec:results}, we present our form factors obtained from taking the physical-continuum limit of the lattice data.
	We plot and tabulate observables found from combining our form factors with CKM matrix elements and known Wilson coefficients.
	Details of the form factor fits are presented in Appendix~\ref{sec:ffs_fit_analysis}.
	Appendix~\ref{app:reconstruct_ff} gives the means for the reader to reconstruct our form factors.
	
	\item In Section~\ref{sec:prospects}, we investigate extensions to our calculations that aim to improve the precision of our determination of the physical-continuum form factors in a future update.
	These discussions will guide other calculations of heavy-to-light decay processes in the future.
	
	%	Firstly, in Section~\ref{sec:ex5}, we introduce a finer lattice, the \emph{exafine} lattice, on which $b$ quarks at their physical mass can be simulated. Collecting data on this lattice helps resolve the dependence on the heavy quark mass of the form factor fit function. We perform preliminary gathering of correlation function data and observe substantial error reduction on the introduction of exafine lattice data to our fits of the form factors.
	%	
	%	Secondly, in Section~\ref{sec:fplus_from_spatVec}, we address the error on $f_+$ near zero-recoil. We investigate extracting $f_+$ lattice data via matrix elements of the \emph{spatial} vector current instead of the temporal component of the vector current. Analysis of the appropriate correlation functions on our coarsest lattice indicates substantial improvement on the error of $f_+$ at high-$q^2$ should this approach be expanded to all lattices.
\end{itemize}

\section{Calculation Details} \label{sec:calc_details}

\subsection{Form factors}

%Comparing experimental observation of $B_c^+ \to D^0 \ell^+ \nu_{\ell}$ with expectations from theory provides a direct (exclusive) evaluation of the CKM matrix element $V_{ub}$.
%Exclusive determinations of $V_{ub}$ from lattice QCD have so far been focussed on the semileptonic decays $B \to \pi$, $B_s \to K$ and $\Lambda_b \to p$, so determining $V_{ub}$ via semileptonic $B_c \to D^0$ will offer another data point.
%There is long-standing tension between exclusive determinations (from specific decays, e.g. $B \to \pi$) of $V_{ub}$ and the inclusive determination (summing over all possible final hadron states)~\cite{Amhis:2019ckw}.
%The value for $V_{ub}$ derived from the form factors in this study can be compared to existing values to further explore any historical discrepancy.

%\rough{INSERT - comments on LHCb and $B_c \to D$ (?). Can't find a specific reference for this, so will probably have to instead make some vague statement along the lines of \lq these decays could soon be measured by LHCb\rq.}

%The semileptonic decay $B_c^+ \to D_s^+ \ell^+ \ell^-$ is an example of a flavour-changing-neutral-current (FCNC) process and is of interest in its own right.
%Such processes are disallowed at tree-level in the Standard Model, thus contributions from physics beyond the Standard Model may be more significant than with tree-level decays.
%Therefore, FCNC transitions are an important avenue towards understanding the validity of the Standard Model. 
%\rough{INSERT - anything more specific to $B_c \to D_s$?}
Our calculations use equal-mass $u$ and $d$ quarks.
The corresponding quark flavour is denoted as $l$.
In this paper, we use the shorthand $B_c \to D_l$ and $B_c \to D_s$ to label the two different decays considered here.
The subscript on the $D_l$ and $D_s$ mesons denotes the flavour of the daughter quark that arises from the decay of the parent $b$ quark.

The form factors $f_0$ and $f_+$ are defined through the vector current matrix element
\begin{align} \label{form factors}
	\langle  D_{l(s)} (\boldsymbol{p}_2) & | V^{\mu} | B_c (\boldsymbol{p}_1) \rangle =  f_0^{l(s)} (q^2) \Bigg[ \frac{M_{B_c}^2 - M_{D_{l(s)}}^2}{q^2}q^{\mu} \Bigg] \nonumber \\
	&+ f_+^{l(s)} (q^2) \Bigg[ p_2^{\mu} + p_1^{\mu} - \frac{M_{B_c}^2 - M_{D_{l(s)}}^2}{q^2}q^{\mu} \Bigg] 
\end{align}
where $q = p_1 - p_2$ is the 4-momentum transfer, and, since we study the transitions $B_c \to D_l$ and $B_c \to D_s$ in tandem throughout this article, we will use the notation $f_{0,+}^{l}$ and $f_{0,+,T}^{s}$ respectively to differentiate between their form factors.

The semileptonic weak decay $B_c^+ \to D^0 \ell^+ \nu_{\ell}$ is facilitated by a $b \to uW^-$ quark transition.
Ignoring isospin breaking effects and possible long-distance QED corrections, the differential decay rate is related to the form factors through
\begin{align} \label{eqn:BcD_decay_rate}
	&\frac{d \Gamma}{d q^2} = \eta_{\mathrm{EW}}^2  |V_{ub}|^2 \frac{G_F^2}{24 \pi^3} \Big( 1 - \frac{m_{\ell}^2}{q^2} \Big)^2 |\boldsymbol{q}| \times \nonumber \\
	&\hspace{2mm}\Bigg[ \Big( 1 + \frac{m_{\ell}^2}{2q^2}  \Big) |\boldsymbol{q}|^2 f_+^l (q^2)^2
	+ \frac{3m_{\ell}^2}{8 q^2} \frac{(M_{B_c}^2 - M_D^2)^2}{M_{B_c}^2} f_0^l (q^2)^2 \Bigg].
\end{align}
This is proportional to $\eta_{\mathrm{EW}}^2 |V_{ub}|^2$, where the factor $\eta_{\mathrm{EW}} = 1.0062(16)$ is the electroweak correction to $G_F$~\cite{Sirlin:1981ie} and we use the same value as in~\cite{Harrison:2020gvo} for $B_{c}^{+} \rightarrow J / \psi \ell^{+} \nu_{\ell}$.
The mass of the lepton in the final state is $m_{\ell}$.
The contribution of $f_0$ is suppressed by the lepton mass and so is only relevant for the decay mode $B_c^+ \to D^0 \tau^+ \nu_{\tau}$.
The physical range of momentum transfer
\begin{align}
	m_{\ell}^2 < q^2 < (M_{B_c} - M_D)^2 = 19.4 \hspace{1mm} \mathrm{GeV}^2
\end{align}
is large here because of the large mass of the $b$ quark.

The short-distance physics of the FCNC transition $B_c \to D_s$ is described by form factors $f_{0,+}$ of the vector current $\overline{s} \gamma^{\mu} b$ and the form factor $f_T$ of the tensor operator $T^{\mu \nu} = \overline{s} \sigma^{\mu \nu} b$ where $2\sigma^{\mu \nu} = [ \gamma^{\mu}, \gamma^{\nu}]$.
The form factor $f_T$ is defined through the matrix element of the tensor operator
\begin{align}
	\braket{D_s (\boldsymbol{p}_2) | T^{k0} | B_c (\boldsymbol{p}_1) } = \frac{2i M_{B_c} p_2^k}{M_{B_c} + M_{D_s}} f_T^s (m_b;q^2).
\end{align}
The tensor form factor $f_T^s$ is scheme and scale dependent.
We will quote results in the $\overline{\mathrm{MS}}$ scheme at scale $4.8 \; \mathrm{GeV}$.
%The tensor form factor $f_T^s$ is scheme and scale dependent, and will
Within the Standard Model, the tensor form factor $f_T$ is relevant for the rare decay $B_c^+ \to D_s^+ \ell^+ \ell^-$ that proceeds via $b \to s$, but not for $B_c^+ \to D_s^+ \nu \overline{\nu}$ or the tree-level decay $B_{c}^{+} \rightarrow D^0 \ell^{+} \nu_{\ell}$.
The daughter quark for $B_c \to D_s$ is heavier than in the case of $B_c \to D$.
The computational expense of computing lattice quark propagators increases as the quark mass decreases, so computing the form factors for $B_c \to D_s$ amounts to a less expensive computation than for $B_c \to D$.
Hence, we compute the tensor form factor $f_T$ only for the process $B_c \to D_s$.
In the future, we intend to also calculate the tensor form factor for $b \to d$ processes.

From matrix elements of the scalar density and vector current on four different lattices with a selection of heavy and light quark masses, we fit the corresponding form factor data to obtain the form factors in the continuum limit with physical quark masses.
By combining existing values of CKM matrix elements $V_{ts}$ and $V_{tb}$, along with values of Wilson coefficients, we predict the decay rate for $B_c^+ \to D_s^+ \ell^+ \ell^-$ within the scope of Standard Model phenomenology.
The expression for the decay rate follows similarly to Section VII in~\cite{Bouchard:2013eph} for $B \to K \ell^+ \ell^-$ where we take the $\overline{\mathrm{MS}}$ scale to be $m_b$ for the tensor form factor.
We also predict the decay rate for $B_c^+ \to D_s^+ \nu \overline{\nu}$ using an expression similar to that for $B \to K  \nu \overline{\nu}$ in~\cite{Buras:2014fpa,Altmannshofer:2009ma}.

%The decay rate $\Gamma (B_c \to D l \nu_l)$, with corresponding form factors $f_0$ and $f_+$, can be expressed as $\Gamma = \Gamma_+ + \Gamma_0$ where 
%\begin{align}
%\Gamma_+ &= \frac{G_F^2 V_{ub}^2}{8\pi^3} \Big( 1 - \frac{m_l^2}{q^2} \Big) \Delta f_+ (q^2) \Bigg[ \Delta^2 \Big( \frac{1}{3} + \frac{m_l^2}{2} \Big) \Bigg] \\
%\Gamma_0 &= \frac{G_F^2 V_{ub}^2}{8\pi^3} \Big( 1 - \frac{m_l^2}{q^2} \Big) \Delta f_0 (q^2) \Bigg[ m_l^2 \frac{(M_{Bc}^2 - M_D^2)^2}{8M_{B_c}^2} \Bigg]
%\end{align}
%are the contributions from $f_+$ and $f_0$ respectively, and
%\begin{align}
%\Delta (q^2) = |\mathbf{p}_2| = \sqrt{\Bigg(\frac{M_{B_c}^2 + M_{D}^2 - q^2}{2M_{B_c}}\Bigg)^2 - M_{D}^2}.
%\end{align}
%Hence, precise determination of both $f_0$ and $f_+$ is of interest.
%(In contrast, semileptonic weak decays of the constituent charm quark inside the $B_c$ present narrower kinematics that rule out the case of $l=\tau$~\cite{Cooper:2020wnj}.)

\subsection{Ensembles and Parameters} \label{sec:ensembles_params}

We use ensembles with $2+1+1$ flavours of HISQ sea quark generated by the MILC Collaboration \cite{Bazavov:2010ru,Bazavov:2012xda,Bazavov:2015yea}.
Table~\ref{LattDesc1} presents details of the ensembles.
The Symanzik-improved gluon action used is that from \cite{Hart:2008sq}, where the gluon action is improved perturbatively through $\mathcal{O}(\alpha_s a^2)$ 
including the effect of dynamical HISQ sea quarks.
The lattice spacing is identified by comparing the physical value for the Wilson flow parameter~\cite{Borsanyi:2012zs} $w_0 = 0.1715(9)$ fm \cite{Dowdall:2013rya} with lattice values for $w_0 / a$ from \cite{Chakraborty:2016mwy} and \cite{Chakraborty:2014aca}.
The following calculations feature strange quarks at their physical mass and equal-mass up and down quarks, with mass denoted by $m_l$.
We use lattices with $m_s/m_l = 5$ in the sea  and also the physical value $m_s/m_l = 27.4$~\cite{FermilabLattice:2014tsy}.
The corresponding pion masses are tabulated in Table~\ref{LattDesc1}~\cite{Bazavov:2017lyh}.
Values for $M_{\pi} L$ (where $L=aN_x$) are also given in Table~\ref{LattDesc1} as an indicator of sensitivity to finite-volume-effects.
%These lattices have with $M_{\pi}L$ around $4.5$ and $M_{\pi}L = 3.7$ respectively~\cite{Bazavov:2017lyh}, also tabulated in Table~\ref{LattDesc1}.
In the more precise calculation of~\cite{Bouchard:2013eph} for the form factors for $B \to K$, finite volume effects were found to be small compared to final uncertainties.
Hence, we expect finite volume effects to be very small compared to the uncertainties we achieve in this first calculation, so we ignore them.
%For sets 1 to 5 in Table \ref{LattDesc1}, strange propagators were re-used from \cite{Koponen:2017fvm}, a study of the pseudoscalar meson electromagnetic form factor.
%Light propagators were re-used from \cite{Koponen:2015sgx}, an extension of \cite{Koponen:2017fvm} to the pion.
%The valence quark masses used for the HISQ propagators on these gluon configurations are given in Table~\ref{tab:valencemass}.
The valence strange and charm quark masses used here, also tabulated in Table~\ref{LattDesc1}, were tuned in \cite{Chakraborty:2014aca, Koponen:2017fvm} slightly away from the sea quark masses to yield results that more closely correspond to physical values.
Corrections due to the tuning of valence strange quark and charm quark masses away from the masses of the  sea quarks should, at leading order, simply amount to a correction linear in the sea mass mistuning which we allow for in our fit of the form factors (described in Section~\ref{sec:fit_ffs}).
We take the mass of valence $l$ quarks to be equal to the mass of the sea $l$ quarks.
We ignore isospin-breaking and QED effects in this first calculation.
The propagators were calculated using the MILC code~\cite{MILCgithub}. 
\begin{table*}
	\caption{Parameters for the MILC ensembles of gluon field configurations. The lattice spacing $a$ is determined from the Wilson flow parameter $w_0$~\cite{Borsanyi:2012zs}. 
		%to 5 and \cite{PhysRevD.91.054508} on set 6. 
		The physical value $w_0 = 0.1715(9) \; \mathrm{fm}$ was fixed from $f_{\pi}$ in \cite{Dowdall:2013rya}. 
		%The very-coarse lattices, sets 1 and 2, have $a \approx 0.15$ [fm], and the coarse lattices, sets 3 and 4, have $a \approx 0.12$ [fm]. 
		Sets 1 and 2 have $a \approx 0.09 \; \mathrm{fm}$.
		Set 3 has $a \approx 0.059 \; \mathrm{fm}$ and set 4 has  $a \approx 0.044\; \mathrm{fm}$.
		%1, 3, 5 and 6 
		Sets 1, 3 and 4 have unphysically massive light quarks such that $m_l/m_s = 0.2$.
		We give $M_{\pi}L$ and $M_{\pi}$ values for each lattice in the fifth and sixth columns~\cite{Bazavov:2017lyh}.
		In the seventh column, we give $n_{\text{cfg}}$, the number of configurations used for each set.
		We also use four different positions for the source on each configuration to increase statistics.
	}
	\begin{center}
	\begin{tabular}{c c c c c c c c c c c c c c c} 
		\hline\hline
		set & handle & $w_0/a$ & $N_x^3 \times N_t$ & $M_{\pi} L$ & $M_{\pi}~\mathrm{MeV}$ & $n_\text{cfg}$ & $am_l^{\text{sea}}$ & $am_s^{\text{sea}}$ & $am_c^{\text{sea}}$ & $am_l^{\text{val}}$ & $am_s^{\text{val}}$ & $am_c^{\text{val}}$ & $T$ \\ [0.1ex] 
		\hline
		%3 & coarse & 1.3826(11) & $24^3 \times 64$ & $1053$ & $-0.235$ & $0.0102$ & $0.0509$ & $0.635$ & $0.0102$ & $0.0541$ & $0.645$\\
		1 & fine & 1.9006(20) & $32^3 \times 96$ & $4.5$ & $316$ & $500$ & $0.0074$ & $0.037$ & $0.440$ & $0.0074$ & $0.0376$ & $0.450$ & $14,17,20$\\
		2 & fine-physical & 1.9518(17) & $64^3 \times 96$ & $3.7$ & $129$ & $500$ & $0.00120$ & $0.0364$ & $0.432$ & $0.00120$ & $0.036$ & $0.433$ & $14,17,20$ \\
		3 & superfine & $2.896(6)$ & $48^3 \times 144$ & $4.5$ & $329$ & $250$ & $0.0048$ & $0.024$ & $0.286$ & $0.0048$ & $0.0245$ & $0.274$ & $22,25,28$\\
		4 & ultrafine & $3.892(12)$ & $64^3 \times 192$ & $4.3$ & $315$ & $250$ & $0.00316$ & $0.0158$ & $0.188$ & $0.00316$ & $0.0165$ & $0.194$ & $31,36,41$\\
		%6 & superfine & 2.896(6) & $48^3 \times 144$ & $250$ & ? & $0.0048$ & $0.024$ & $0.286$ & $0.0048$ & $0.0234$ & $0.274$\\
		\hline\hline
	\end{tabular}
	\end{center}
	\label{LattDesc1}
\end{table*}

The numerical challenge of generating the finest lattices that we use here means that the ensembles do not fully explore the space of all possible topological charges.
The effects of topology freezing on meson phenomenology calculated on these lattices were explored in~\cite{Bernard:2017npd}.
It was found that a topological adjustment of $1\%$ is required for the $D$ meson decay constant on the ultrafine lattice (set 4).
The adjustment for $D_s$ is negligible and this is also expected to be the case for the $B_c$ meson.
The size of the errors achieved in our calculations here are such that effects from topological freezing (which could be of similar size for form factors as those seen for decay constants) are negligible and so we ignore them.
More accurate form factor calculations in the future may need to incorporate adjustments due to non-equilibrated topological charge distributions on the ultrafine and finer lattices.

The heavy-HISQ method sees all flavours of quark implemented with the HISQ \cite{Follana:2006rc} formalism.
This is a fully relativistic approach which involves calculations for a set of quark masses on ensembles of lattices with a range of fine lattice spacings, enabling a fit from which the physical result at the $b$ quark mass in the continuum can be determined.
In our heavy-HISQ method, we utilise a valence HISQ quark with mass $m_h$ that takes values between $m_c$ and $m_b$.
We will describe this quark as \lq heavy\rq.
In the limit of physical quark masses, the heavy quark will coincide with the $b$ quark.
Regarding the mesons that this quark forms with a constituent charm, strange or light quark, we adopt nomenclature for these mesons that is similar to mesons with a constituent bottom quark.
For example, we label the low-lying heavy-charm pseudoscalar meson as $H_c$.
If we were to take $m_h = m_b$, then this meson would coincide with the $B_c$ pseudoscalar meson.

\begin{figure}[t]
	%\captionsetup{singlelinecheck = false, justification=justified}
	\includegraphics[width=0.5\textwidth]{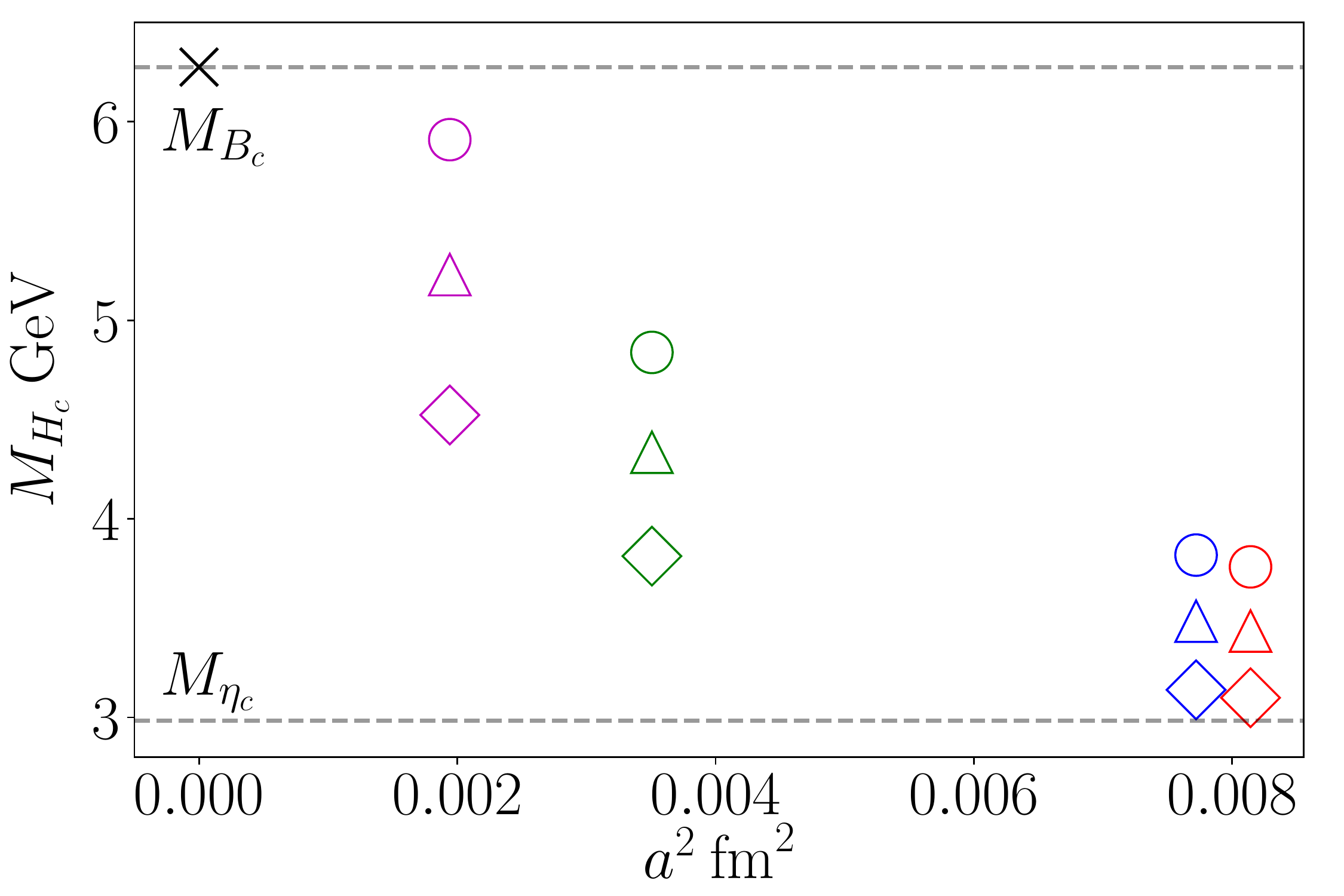}
	\caption{The mass $M_{H_c}$ of the heavy-charm pseudoscalar meson is plotted against the lattice spacing squared for each of the values $am_h = 0.5, 0.65, 0.8$ used in the heavy-HISQ calculation. Values for $M_{H_c}$ are obtained from fitting the correlation functions as described in Section~\ref{sec:fit_correls}. 
	%$M_{H_c}$ is a proxy for the bare lattice heavy quark mass $am_h$.
	The continuum-physical point is denoted by a cross at $a=0 \; \mathrm{fm}$ and $M_{H_c} = M_{B_c}$ from experiment~\cite{PDG}. Data from sets 1, 2, 3 and 4 are denoted by the colours red, blue, green and magenta respectively.
	Data for $am_h = 0.5, 0.65, 0.8$ can be identified by the diamond, triangle and circle markers respectively.
	These choices will be repeated in all subsequent plots.}
	\label{hhisqmassplot}
\end{figure}

This heavy-HISQ calculation uses bare heavy quark masses $am_h = 0.5, 0.65, 0.8$ on all four sets in Table~\ref{LattDesc1}.
The masses of the corresponding heavy-charm pseudoscalar mesons $H_c$ are plotted in Fig.~\ref{hhisqmassplot}.
The mass of the heaviest heavy-charm pseudoscalar meson is only 6\% %UPDATE21April
lighter than the physical $B_c$ meson.

Momentum is inserted only into the valence light(strange) quark of the $D_{l(s)}$ meson, thus the initial $H_c$ meson is always at rest on the lattice.
The momentum insertion is implemented through partially twisted boundary conditions~\cite{Sachrajda:2004mi, Guadagnoli:2005be} in the $(1 \: 1 \: 1)$ direction.
The twists used on each set are given in Table~\ref{LattDescHHISQmom}.
\begin{table}%[ht]
	\centering
	\caption{Twists used for heavy-HISQ calculation on each of the four sets given in Table \ref{LattDesc1}.
		%The heaviest mass used on the superfine lattice yields $m_{\eta_h}$ of roughly 75\% of the physical $m_{\eta_b}$ value.
		The twists are in the $(1\hspace{1mm}1\hspace{1mm}1)$ direction and defined in Eq.~\eqref{eqn:defn_theta}. The corresponding values of $q^2$ as a proportion of $q^2_{\mathrm{}max}$ are shown in Fig.~\ref{hhisq_prop_qSquaredMax}.}
	\begin{tabular}{ c  l  l} 
		\hline\hline
		% & \multicolumn{1}{c |}{$B_c \to D_l$} & \multicolumn{1}{c}{$B_c \to D_s$} \\
		set & twists $\theta$ for $B_c \to D_s$ &  twists $\theta$ for $B_c \to D_l$ \\ [0.1ex] 
		\hline
		1 &$0$, $0.4281$, $1.282$, $2.141$, $2.570$ & $0$, $0.4281$, $1.282$, $2.141$, $2.570$ \\
		2 & $0$, $0.8563$, $2.998$, $5.140$ & $0$, $3.000$, $5.311$ \\
		3 & $0$, $1.261$, $2.108$, $3.624$, $4.146$ & $0$, $1.261$, $2.108$, $2.666$ \\
		4 & $0$, $0.706$, $1.529$, $2.235$, $4.705$ & $0$, $0.706$, $1.529$, $2.235$, $4.705$ \\
		\hline\hline
	\end{tabular}
	\label{LattDescHHISQmom}
\end{table}
The twist angle $\theta$ is related to the three-momentum transfer $\boldsymbol{q}  = \boldsymbol{p}_1 - \boldsymbol{p}_2$ by
\begin{align}
	|\boldsymbol{q}| = \frac{\pi \theta \sqrt{3}}{a N_x}. \label{eqn:defn_theta}
\end{align}
For example, zero-twist ($\theta = 0$) corresponds to zero-recoil where $q^2$ takes its maximum physical value which we denote as $q^2_{\mathrm{max}}$.
In previous studies, such as Fig. 3 in~\cite{McLean:2019qcx}, it has been observed that the continuum dispersion relation is closely followed for mesons with staggered quarks, particularly on the finer lattices.
The twists we use allow a considerable proportion of the physical $q^2$ range to be probed.
Most of the twists in Table~\ref{LattDescHHISQmom} originate from a variety of past calculations in which the corresponding propagators were saved for future use.
%These twists were chosen to broadly target interesting kinematics of the particular process being considered at the time, however they have proved useful for many other projects since.

\begin{figure}
	\begin{center}
		\includegraphics[width=0.5\textwidth]{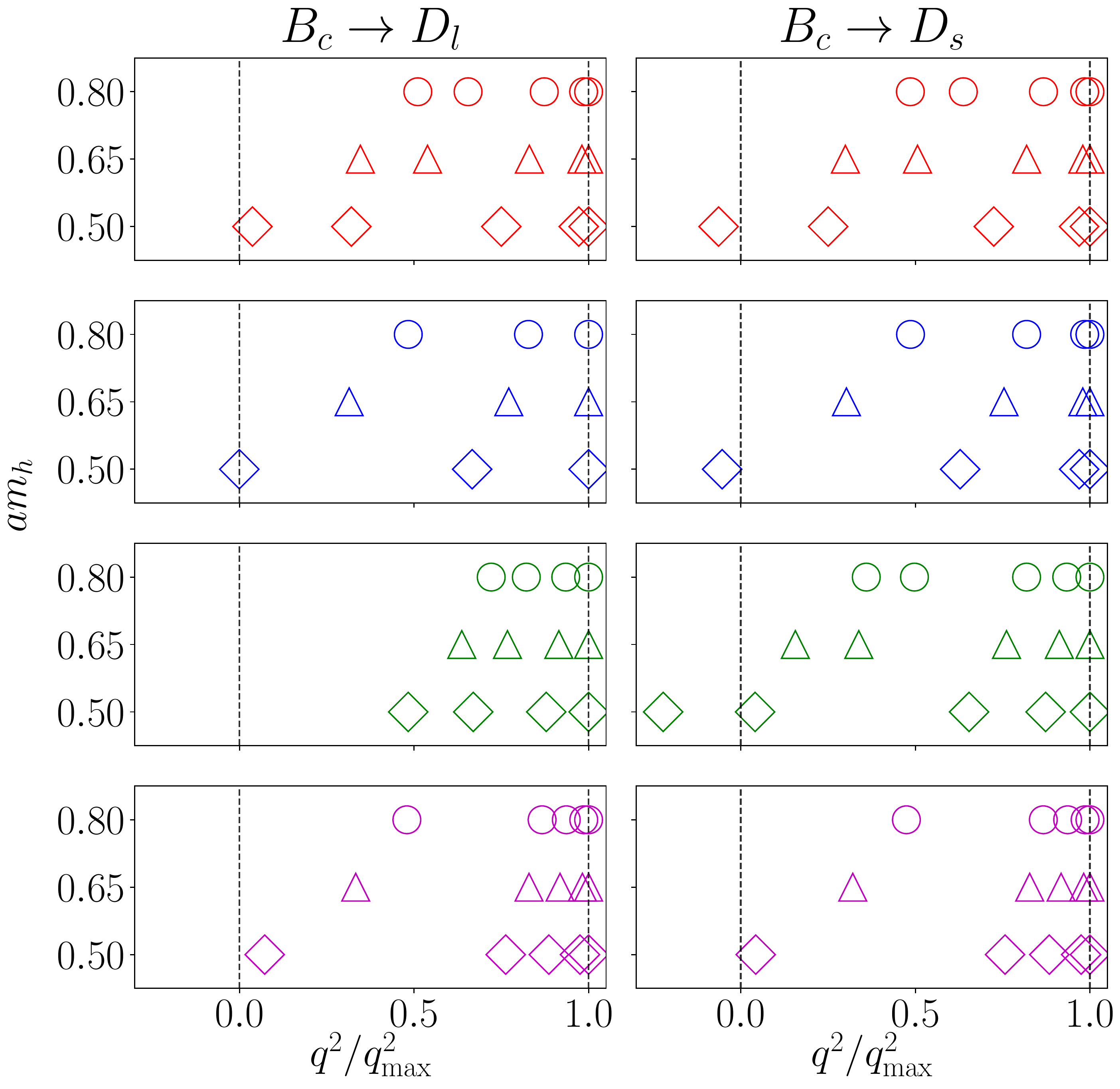}
		\caption{$q^2$ values on each set as a proportion of the maximum value $q^2_{\mathrm{max}} = (M_{H_c} - M_{D_{l(s)}})^2$. From top to bottom, data from sets 1, 2, 3 and 4 are displayed (see Table~\ref{LattDesc1}).
			For different $am_h$ on a given set, the same twists were used.
			%The corresponding twists are given in Table \ref{LattDescHHISQmom}. 
			As described in the caption for Fig.~\ref{hhisqmassplot}, data from sets 1, 2, 3 and 4 and heavy quark masses $am_h$ are denoted by different colours and marker styles. Values used here for the masses of the initial and final mesons are found from fits of correlation functions (to be discussed in Section~\ref{sec:fit_correls}).}
		\label{hhisq_prop_qSquaredMax}
	\end{center}
\end{figure}

Fig.~\ref{hhisq_prop_qSquaredMax} shows the $q^2$ realised by the twists in Table~\ref{LattDescHHISQmom}.
The values of $q^2 / q^2_{\text{max}}$ are given for each twist and heavy quark mass for both $H_c \to D_l$ and $H_c \to D_s$.
Twists that give negative $q^2$ are unphysical but will nevertheless aid the fits of the form factors across the physical range.
For all of the sets except one, all of the $q^2$ range is covered for the lightest heavy quark mass value $am_h = 0.5$ (recall that Fig.~\ref{hhisqmassplot} shows the corresponding mass of the heavy-charm pseudoscalar mesons).
For the finest lattice, set 4 in Table \ref{LattDesc1}, Fig.~\ref{hhisq_prop_qSquaredMax} shows for the largest heavy quark mass, close to $m_b$.

\subsection{Extracting form factors from matrix elements} \label{sec:matElts_to_ffs}

The conserved HISQ vector current is given explicitly in Appendix A of~\cite{Hatton:2019gha}.
It takes the form of a complicated linear combination of multi-link point-split operators.
Whilst the conserved current has the advantage that it does not require a multiplicative renormalisation factor, its form is unwieldy for lattice computations.
Hence, we elect to use simple local currents that are not conserved and determine the corresponding renormalisations.

Our calculation uses HISQ quarks exclusively.
In particular, since we use HISQ for both the parent heavy quark and the daughter light or strange quark, we can use the Partially Conserved Vector Current Ward identity to relate matrix elements of the renormalised local vector current $Z_V V^{\mu}_{\mathrm{local}}$ with matrix elements of the local scalar density through
\begin{align}
q_{\mu} \langle D_{l(s)} | V^{\mu}_{\text{local}} | H_c \rangle Z_V = (m_h - m_{l(s)})  \langle D_{l(s)} | S_{\text{local}} | H_c \rangle. \label{PCVClatt}
\end{align}
This holds since the mass and scalar density multiplicative renormalisation factors $Z_m$ and $Z_S$ satisfy $Z_{m} Z_{S}=1$.
Using Eq.~\eqref{PCVClatt} to determine $Z_V$ is a fully non-perturbative strategy.
Up to discretisation effects, the renormalisation factor is independent of $q^2$, and so it is sufficient to deduce its value at zero-recoil ($\boldsymbol{q} = \boldsymbol{0}$ and maximum $q^2$).
Using different staggered \lq tastes\rq\;of meson in Eq.~\eqref{PCVClatt} will contribute a discretisation error that is  accounted for when fitting the lattice form factor data.
At zero-recoil, Eq.~\eqref{PCVClatt} only features matrix elements of the scalar density and the \emph{temporal} component of the vector current, and so we do not compute matrix elements of the \emph{spatial} components of the vector current (though they will be considered in Section~\ref{sec:fplus_from_spatVec} as part of our investigation towards future improvements).

Combining Eqs.~\eqref{PCVClatt} and~\eqref{form factors} yields
\begin{align}
f_0^{l(s)} \big(q^2 \big) = \langle D_{l(s)} | S_{\text{local}} | H_c \rangle \frac{m_h - m_{l(s)}}{M_{H_c}^2 - M_{D_{l(s)}}^2}. \label{f0scalar}
\end{align}
We use Eq.~\eqref{f0scalar} to extract $f_0$ from the given combination of quark masses, meson masses and the matrix element of the scalar density.
%allowing extraction of $f_0$ solely from the scalar density matrix element.

Eq.~\eqref{form factors} for $\mu = 0$ can be trivially rearranged to yield
\begin{align} \label{fplusextract}
f_+^{l(s)} (q^2) = \frac{ Z_V \langle D_{l(s)} | V^{0}_{\text{local}} | H_c \rangle - q^0 f_0^{l(s)} (q^2)\frac{M_{H_c}^2 - M_{D_{l(s)}}^2}{q^2}}{ p_2^0 + p_1^0 - q^0 \frac{M_{H_c}^2 - M_{D_{l(s)}}^2}{q^2}}.
\end{align}
At zero-recoil, the denominator vanishes so $f_+$ cannot be extracted here.
%In practice, using Eq.~\eqref{fplusextract} near zero-recoil is problematic since both the numerator and denominator grow from 0 as $q^2$ is decreased from the maximum value at zero-recoil.
In practice, using Eq.~\eqref{fplusextract} near zero-recoil is problematic since both the numerator and denominator approach $0$ as $q^2$ increases towards its maximum value at zero-recoil.
%Hence, data for $f_+$ near zero-recoil is removed from the fits of the form factors.
This is discussed further in Appendix~\ref{sec:ffs_fit_analysis}.
(In Section~\ref{sec:fplus_from_spatVec}, we consider an alternative extraction of $f_+$ by using Eq.~\eqref{form factors} with $\mu \neq 0$.)

Finally, the tensor form factor is obtained through
\begin{align}
f_T^s \big(q^2 \big) = \frac{ Z_T \braket{D_s | T^{1,0}_{\mathrm{local}} | H_c} (M_{H_c} + M_{D_s}) }{2i M_{H_c} p_2^1}, \label{fTextract}
\end{align}
where $T^{1,0}_{\mathrm{local}}$ is the local tensor operator and $Z_T$ is its multiplicative renormalisation factor that takes the lattice tensor current to the $\overline{\mathrm{MS}}$ scheme.
We use values of the associated multiplicative renormalisation factor $Z_T$ obtained using the RI-SMOM intermediate scheme.
We give these values in Table~\ref{tab:Z_T}.
Values in the RI-SMOM scheme at scale $3 \; \mathrm{GeV}$ are converted to scale $4.8 \; \mathrm{GeV}$ in the $\overline{\mathrm{MS}}$ scheme.
Nonperturbative (condensate) artefacts in $Z_T$ in the RI-SMOM scheme were removed using analysis of the $J/\psi$ tensor decay constant~\cite{Hatton:2020vzp} .
%The corresponding values for $Z_T$ for our calculation of $B_c \to D_s$ will differ from those given in~\cite{Hatton:2020vzp} only by discretisation effects which are accounted for in the fits of lattice form factor data.
%
\begin{table}
	\centering
	\caption{Values used for the multiplicative renormalisation factor $Z_T$ of the tensor operator obtained from Tables VIII and IX in~\cite{Hatton:2020vzp} at scale $m_b$ in the $\overline{\mathrm{MS}}$ scheme. The set handles correspond to those given in Table~\ref{LattDesc1}. The top row gives the mean values of $Z_T$ and the rows beneath give the covariance matrix scaled by a factor of $10^5$.}
	\begin{tabular}{c c c}
		\hline\hline
		sets $1$ and $2$ & set $3$ & set $4$ \\ [0.1ex]
		\hline
		$0.9980$ & $1.0298$ & $1.0456$ \\
		\hline
		$0.6250$ & $0.6242$ & $0.6059$ \\
		& $0.6250$ & $0.6057$ \\
		& & $0.6250$ \\
		\hline\hline
	\end{tabular}
	\label{tab:Z_T}
\end{table}

\subsection{Euclidean Correlation Functions on the Lattice}	\label{sec:latt_corr_funcs}

We obtain the matrix elements discussed in Section~\ref{sec:matElts_to_ffs} from correlation functions on the lattice with ensembles and parameters specified in Section~\ref{sec:ensembles_params}.
We now describe the construction of these correlations functions.

To ensure that non-vanishing correlation functions are obtained when exclusively using staggered propagators in a heavy-HISQ calculation, operators at the source, sink and current insertion must be carefully selected so that the overall correlator is a taste singlet.
% and thus taste is conserved.
As we detail in Section~\ref{sec:fit_correls}, matrix elements of the scalar density, vector current and tensor operator are extracted from 3-point correlation functions whose constructions we now describe.

\begin{figure}
	\begin{center}
		\includegraphics[width=0.5\textwidth]{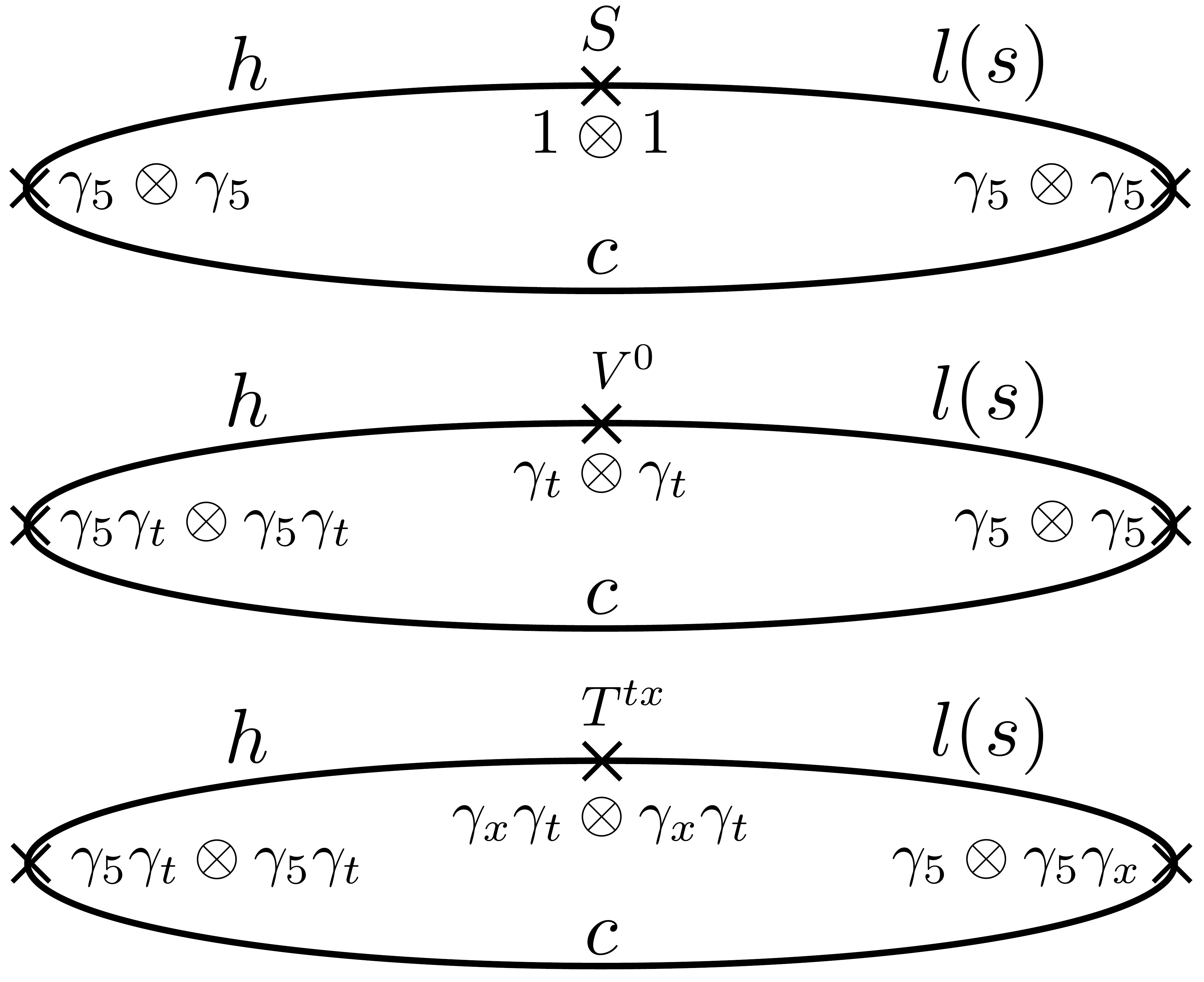}
		\caption{Diagrammatic representations of the three-point functions we calculate on the lattice. The top two diagrams are relevant for extracting matrix elements of the scalar density and temporal vector current, and the bottom diagram is calculated for the case $B_c \to D_s$ and the tensor current. Each operator insertion is shown by a cross and is labelled by its description given in the spin-taste basis, whilst the lines represent lattice quark propagators. The heavy quark propagator is represented by the line, labelled by the flavour $h$, between the left-most operator and the insertion. The daughter quark propagator is represented by the line, labelled by the flavour $l(s)$, between the insertion and the right-most operator. The remaining quark propagator is the spectator quark, labelled by the flavour $c$.}
		\label{fig:3pt_diags}
	\end{center}
\end{figure}

Our choice of operators used in the 3-point correlation functions that we compute are given in Table~\ref{tab:st_interpolators_3pt} and shown in Fig.~\ref{fig:3pt_diags}.
The operators are expressed in the staggered spin-taste basis.
Note that the scalar density, temporal vector current and tensor operator all take the form $\Gamma \otimes \Gamma$ for some combination of gamma matrices $\Gamma$, thus they are all local operators as discussed in Section~\ref{sec:matElts_to_ffs}.

%For the three-point correlation function with an insertion of the scalar density, we use, in the spin-taste basis, $\gamma_5 \otimes \gamma_5$ for $H_c$ and $D_{(s)}$ interpolators at the sink and source, and $I \otimes I$ at the operator insertion.
%For the temporal component of the vector current matrix element, the operators $\gamma_5 \gamma_t \otimes \gamma_5\gamma_t$ for $H_c$, $\gamma_5 \otimes \gamma_5$ at $D_{(s)}$, and $\gamma_t \otimes \gamma_t$ at the current insertion have been used.
%Finally, for the tensor operator, we use the operators $\gamma_5 \gamma_t \otimes \gamma_5\gamma_t$ for $H_c$, $\gamma_5 \otimes \gamma_5 \gamma_x$ at $D_{(s)}$, and $\gamma_x \gamma_t \otimes \gamma_x \gamma_t$ at the insertion.
To extract the overlaps of the $H_c$ and $D_{(s)}$ interpolators used in the 3-point functions onto the low-lying pseudoscalar meson states, we compute the relevant 2-point functions, namely $H_c$ with $\gamma_5 \otimes \gamma_5$ and $\gamma_5 \gamma_t \otimes \gamma_5\gamma_t$ at both the source and sink, and $D_{(s)}$ with $\gamma_5 \otimes \gamma_5$ and $\gamma_5 \otimes \gamma_5\gamma_x$  at both the source and sink.
The $D_{(s)}$ interpolator $\gamma_5 \otimes \gamma_5\gamma_x$ is the only non-local interpolator that we use.
%These choices are summarised in in Table~\ref{tab:st_interpolators_3pt}.
%Diagrammatic representations of the three-point correlation functions are shown in Fig.~\ref{fig:3pt_diags}.
%
\begin{table}%[ht]
	\centering
	\caption{Summary of the interpolators used in the all-HISQ three-point correlation functions. The interpolators are given in the spin-taste basis. Matrix elements of the scalar density, vector current and tensor operator are extracted from the correlation functions constructed from the first, second and third rows of interpolators respectively. The relevant form factor is given in the first column. The tensor form factor is calculated for $B_c \to D_s$ only here.}
	\begin{tabular}{c | c c c} 
		\hline\hline
		& $H_c$ & $D_{l(s)}$ & insertion  \\ [0.1ex] 
		\hline
		$f_0$ & $\gamma_5 \otimes \gamma_5$  & $\gamma_5 \otimes \gamma_5$ & $I \otimes I$ \\
		$f_+$ & $\gamma_5 \gamma_t \otimes \gamma_5 \gamma_t$  & $\gamma_5 \otimes \gamma_5$ & $\gamma_t \otimes \gamma_t$ \\
		$f_T$ & $\gamma_5 \gamma_t \otimes \gamma_5 \gamma_t$  & $\gamma_5 \otimes \gamma_5 \gamma_x$ & $\gamma_x \gamma_t \otimes \gamma_x \gamma_t$ \\
		\hline\hline
	\end{tabular}
	\label{tab:st_interpolators_3pt}
\end{table}

We calculate the correlation functions needed to study the form factors for $B_c \to D_l$ and $B_c \to D_s$ together since the calculations share gluon field configurations and other lattice objects.
From a computational perspective, these processes are similar since they both involve a charm quark which spectates a bottom quark that changes flavour.
Hence, we are able to construct lattice correlation functions such that sequential $b$ quark propagators, i.e. the combined bottom and charm propagator object, can be utilised in both calculations, thus saving us computational expense.

\subsection{Fitting Correlation Functions} \label{sec:fit_correls}

The correlation functions are fit to the following forms using the \textit{corrfitter} package \cite{corrfitter}.
The fit seeks to minimise an augmented $\chi^2$ as described in \cite{Lepage:2001ym,Hornbostel:2011hu,Bouchard:2014ypa}\footnote{In the limit of high statistics the results from this method are equivalent to those from Bayesian inference.}.
We simultaneously fit all of the 2-point  and 3-point correlation functions at all twists and heavy quark masses to account for all possible correlations between the fit parameters.
We use singular value decomposition (SVD) cuts in our fits, thus the $\chi^2/\mathrm{d.o.f.}$ values from our fits of correlation functions do not have a straightforward interpretation in the sense of frequentist statistics.
More discussion and details can be found in Appendix~\ref{sec:corrfit_method}.
This includes details of our priors and a variety of tests of the stability of our fits.

The 2-point correlator data is fit to the functional form
\begin{align} \label{corrfitform_2pt}
C_{\text{2pt}} (t) & = \sum_i^{N_{\mathrm{n}}} (a_{\mathrm{n},i})^2 f (E_{\mathrm{n},i},t) - \sum_i^{N_{\mathrm{o}}}  (a_{\mathrm{o},i})^2  (-1)^t f (E_{\mathrm{o},i},t) \nonumber \\
\end{align}
where
\begin{align} 
f(E,t)=e^{-E t} + e^{-E (N_t - t)}.
\end{align}
This follows from the spectral decomposition of the Euclidean correlation functions.
The sums over $i$ enumerate the tower of states that have non-vanishing overlap with the interpolators such that $E_{\mathrm{n}, i} \leq E_{\mathrm{n}, i+1}$ and $E_{\mathrm{o}, i} \leq E_{\mathrm{o}, i+1}$.
As is characteristic of staggered quarks, we find contributions to the correlation functions that switch sign between adjacent timeslices.
%The states responsible are understood to  have opposite parity to naive expectations from the spin structure of the fermion bilinear interpolator.
These contributions that oscillate with time are accounted for by the second piece in Eq.~\eqref{corrfitform_2pt} where the subscript \lq o\rq \space is shorthand for \emph{\lq oscillating\rq}.
Similarly, the subscript \lq n\rq \space in the first piece in Eq.~\eqref{corrfitform_2pt} is shorthand for \emph{\lq non-oscillating\rq}.
The function $f (E,t)$ accounts for the periodicity of the correlator data in the temporal direction.
The amplitude $a_{\mathrm{n},0}$ is normalised such that
\begin{align}
a_{\mathrm{n},0} = \frac{\braket{0| \mathcal{O}| P}}{\sqrt{2E_P}}
\end{align}
where $\mathcal{O}$ is the pseudoscalar meson interpolator, $P$ is the low-lying pseudoscalar state, and $E_P = E_{\mathrm{n}, 0}$ is its energy.

The 3-point data is fit to the functional form
\begin{align} \label{corrfitform_3pt}
C_{\text{3pt}} (t,T) & = \sum_{i,j}^{N_{\mathrm{n}}, N_{\mathrm{n}}}  a_{\mathrm{n},i} e^{-E_{\mathrm{n},i} t}  V_{\mathrm{nn},ij} b_{\mathrm{n},j} e^{-E_{\mathrm{n},j} (T-t)} \nonumber \\
- \sum_{i,j}^{N_{\mathrm{n}}, N_{\mathrm{o}}}  & (-1)^{T-t} a_{\mathrm{n},i} e^{-E_{\mathrm{n},i} t}  V_{\mathrm{no},ij} b_{\mathrm{o},j}e^{-E_{\mathrm{o},j}(T-t)} \nonumber \\
- \sum_{i,j}^{N_{\mathrm{o}}, N_{\mathrm{n}}}  & (-1)^t a_{\mathrm{o},i} e^{-E_{\mathrm{o},i} t}  V_{\mathrm{on},ij} b_{\mathrm{n},j} e^{-E_{\mathrm{n},j }(T-t)} \nonumber \\
+ \sum_{i,j}^{N_{\mathrm{o}}, N_{\mathrm{o}}}  & (-1)^T a_{\mathrm{o},i} e^{-E_{\mathrm{o},i} t}  V_{\mathrm{oo},ij} b_{\mathrm{o},j} e^{-E_{\mathrm{o},j}(T-t)} \nonumber \\
\end{align}
where the amplitudes $a$ and $b$ are the amplitudes in Eq.~\eqref{corrfitform_2pt} corresponding to the initial and final pseudoscalar meson states in the 3-point correlator.

For an insertion of the local scalar density, both source and sink operators are $\gamma_5 \otimes \gamma_5$.
For an insertion of the temporal component of the local vector current, the $D_{l(s)}$ and $H_c$ mesons are interpolated by $\gamma_5 \otimes \gamma_5$ and $\gamma_0 \gamma_5 \otimes \gamma_0 \gamma_5$ respectively.
The matrix elements of the vector current and tensor operators are related to the fit parameters $V_{\mathrm{nn},ij}$ of the 3-point functions through
\begin{align} \label{Jopinsert}
\langle D_{l(s)} | J | H_c \rangle &= Z V_{\mathrm{nn},00} \sqrt{2E_{D_{l(s)}} 2E_{H_c}},
%a[0] &=& \frac{\braket{0| \phi_{B_s} (\mathbf{q})| B_s} }{\sqrt{2E_{B_s}}},\label{fitforms}
\end{align}
where $J$ here is the insertion that facilitates the $h \to l$ or $s$ flavour transition and $Z$ is the corresponding multiplicative renormalisation factor for $V$ or $T$.
%where $\phi_{B_s} (\mathbf{q})$ is the pseudoscalar interpolator $\overline{s} \gamma_5 b$, and similarly for $b[0]$. The state $\ket{B_s}$ is defined as the state $\ket{n}$ in the spectrum of the system with lowest energy $E_n$ such that $a[0] \neq 0$, i.e. it is the lowest energy state with quantum numbers consistent with the operator $\phi_{B_s}$, and similarly for $\ket{B_c}$ and $\ket{B_d}$.
The pseudoscalar mesons of interest are the lowest lying states consistent with their quark content and the gamma matrix structure of the interpolators, so we only require extraction of the matrix elements for $i=j=0$.
The presence of $i,j>0$ terms are necessary to give a good fit and allows for the full systematic uncertainty from the presence of excited states to be included in the extracted $V_{\mathrm{nn},00}$.
%Firstly, the 2-point data is fit to the oscillating and non-oscillating exponentials in equation (\ref{fitform}).
%The values for the energies and amplitudes are then used as priors for the same parameters in the fit of the 3-point data.
%In addition to this chaining, the 3-point data is marginalised to include only 1 or 2 oscillating and non-oscillating exponentials.

%The decay constant $f_P$ for the low-lying pseudoscalar meson $P$ is defined as
%%
%\begin{align} \label{eqn:defn:meson_decay_constant}
%\braket{0 | A_{\mu} | P(q)} = i q_{\mu} f_P
%\end{align}
%%
%where $q$ is the four-momentum of the particle.
%The Partially Conserved Axial Current Ward identity (PCAC) is
%%
%\begin{align}\label{PCAClatt}
%q_{\mu} \langle A_{\text{lattice}}^{\mu} \rangle Z_A = (m_f + m_{f'}) \langle  P_{\text{lattice}} \rangle  \hspace{1mm},
%\end{align}
%%
%where $f$ and $f'$ are the flavours of valence quark.
%This allows the extraction of the decay constant, a physical observable through combining Equations~\ref{PCAClatt} and ~\ref{eqn:defn:meson_decay_constant} to give
%%
%\begin{align}
%M_P^2 f_P &= (m_f + m_{f'}) \braket{0 | \overline{\psi} \gamma_5 \psi | P(q)}
%\end{align}
%%
%where $m_f$ and $m_{f'}$ are the bare quark masses and $\braket{0 | \overline{\psi} \gamma_5 \psi | P(q)}$ can be found from the amplitude $a[0]$ in Equation (\ref{corrfitform_2pt}) from a 2-point correlator describing a pseudoscalar particle by
%%
%\begin{align}
%a[0] &= \frac{\braket{0  \overline{\psi} \gamma_5 \psi | P(q)}}{\sqrt{E_{P(q)}}}.
%\end{align}
%%

\subsection{Fitting the Form Factors} \label{sec:fit_ffs}

From the parameters $V_{\mathrm{nn},00}$ in the fit form of the 3-point correlation functions in Eq.~\eqref{corrfitform_3pt}, matrix elements are found using Eq.~\eqref{Jopinsert}.
The values of the form factors are then obtained by using Eqs.~\eqref{f0scalar},~\eqref{fplusextract} and~\eqref{fTextract}.
%For each of the processes $B_c \to D_l$ and $B_c \to D_s$, data for the form factors is derived from the $V_{\mathrm{nn},00}$ fit parameters via Eq.~\eqref{Jopinsert}.

The form factor data at all momenta and heavy quark masses on all sets in Table~\ref{LattDesc1} are then fit simultaneously to a functional form that allows for discretisation effects, dependence on the heavy meson mass, and any residual mistuning of the light, strange and charm quark bare mass parameters.
The fit is carried out using the \textit{lsqfit} package \cite{lsqfit} which implements a least-squares fitting procedure.

\subsubsection{$z$-expansion} \label{sec:BcBd_z}

It is convenient, and now standard, to map the semileptonic region $m_{\ell}^2 <q^2< t_- = (M_{H_c} - M_{D_{l(s)}})^2$ to a region on the real axis within the unit circle through
\begin{align} \label{eqn:BcD_littlez}
z(q^2) &= \frac{\sqrt{t_+ - q^2} - \sqrt{t_+ - t_0}}{\sqrt{t_+ - q^2} + \sqrt{t_+ - t_0}}.
\end{align}
{\tiny {\tiny }}The parameter $t_+$ is chosen to be the threshold in $q^2$ for meson pair production with quantum numbers of the current~\cite{Boyd:1997qw}, i.e. $(M_{H} + M_{\pi (K)})^2$.
%We take PDG~\cite{PDG} values for $M_K$ and $M_{\pi}$.
Any quark mass mistunings in our calculations are allowed for by the fit function of the form factor data.
In our $B_c \to D_l$ calculation, we determine the $M_H$ value for evaluating $t_+$ from heavy-light 2-point correlation functions that we fit simultaneously with the correlation functions described in Section~\ref{sec:latt_corr_funcs}.
% and use this value in $t_+$.
In our $B_c \to D_s$ calculation, which we analyse separately from $B_c \to D_l$, we estimate $M_H$ by taking $M_H  = M_{H_s} - (M_{B_s} - M_B)$.
A similar approximation was taken in~\cite{Harrison:2020gvo}, a calculation of the form factors for $B_c \to J/\psi$.
%Should the form factors at the physical-continuum limit be required to be expressed in terms of an alternative definition of $t_+$, then this $z$-transform can be easily undone.
Also, we choose the parameter $t_0$ to be $0$ so that the points $q^2 = 0$ and $z = 0$ coincide.
%This is a particularly convenient choice for imposing the kinematic constraint $f_0 (0) = f_+(0)$ within the fit (discussed in the next section).
The form factors can be approximated by a truncated power series in $z$.
The validity of this truncation is scrutinised in Appendix~\ref{sec:ff_stability}.
	
\subsubsection{Fit form} \label{sec:ff_fit_form}
	
Form factor data from our heavy-HISQ calculation is obtained, as described in Section~\ref{sec:matElts_to_ffs}, from matrix elements extracted from the fits detailed in Section~\ref{sec:fit_correls}.
Data for each of the form factors is fit to the functional form
\begin{align}
&P(q^2) f (q^2) = \nonumber \\
& \mathcal{L} \sum\limits_{n = 0 \vphantom{j = 0}}^{N_n \vphantom{N_j}} \sum\limits_{r = 0 \vphantom{j = 0}}^{N_r \vphantom{N_j}} \sum\limits_{j = 0}^{N_j} \sum\limits_{k = 0 \vphantom{j = 0}}^{N_k \vphantom{N_j}} A^{(nrjk)} \hat{z}^{(n, N_n)} \left( \frac{\Lambda}{M_{H_{l(s)}}}\right)^r \Omega^{(n)}  \nonumber \\
& \hspace{26mm} \times \left(\frac{am_h}{\pi}\right)^{2j} \left(\frac{am_c}{\pi}\right)^{2k} \mathcal{N}_\text{mis}^{(n)}. \label{hhisqffff}
\end{align}
The dominant pole structure is represented by the factor $P(q^2)$ given by $1-q^2/M^2_{\text{res}}$.
The values we use for $M^2_{\text{res}}$ are discussed in Section~\ref{sec:Mres}.
The combination $P(q^2) f (q^2)$ is fitted to a truncated series, or polynomial, in $z(q^2)$ given in the RHS of Eq.~\eqref{hhisqffff}.
We use the Bourreley-Caprini-Lellouch (BCL) parametrisation~\cite{Bourrely:2008za} where
\begin{align}
\hat{z}_0^{(n, N_n)} &= z^n, \nonumber \\
 \hat{z}_{+,T}^{(n, N_n)} &= z^{n}-\frac{n (-1)^{N_n + 1 - n}}{N_n + 1} z^{N_n + 1} \label{eq:zhat}
\end{align}
in Eq.~\eqref{hhisqffff}.
We defined $z(q^2)$ in Eq.~\eqref{eqn:BcD_littlez}.
The priors for $A^{(nrjk)}$ are taken to be $0(2)$ except for $j+k = 1$ where the prior is $0.0(3)$ to account for the removal of $a^{2}$ errors in the HISQ action at tree-level~\cite{Follana:2006rc}.
In Appendix~\ref{sec:ff_fit_results}, we show plots of the lattice data for $P(q^2) f(q^2)$ plotted against $z$ in Figs.~\ref{fig:Pfl} and~\ref{fig:Pfs}.

The factor $\mathcal{L}$ contains a chiral logarithm for the case $B_c \to D$, and we take $\mathcal{L} = 1$ for the case $B_c \to D_s$.
For the case $B_c \to D$, then
\begin{align}
\mathcal{L} = 1 + \left( \zeta^{(0)} + \zeta^{(1)} \frac{\Lambda}{M_{H_{l}}} + \zeta^{(2)}  \frac{\Lambda^2}{M_{H_{l}}^2} \right) x_{\pi} \log x_{\pi} \label{eq:curlyL}
\end{align}
where we take $\Lambda = 500\text{ MeV}$ for the QCD energy scale, $x_{\pi} = M_{\pi}^{2} / \Lambda_{\chi}^{2}$, and $\Lambda_{\chi} = 4 \pi f_{\pi}$ is the chiral scale.
It is convenient for us to write $x_{\pi}$ in terms of quark masses.
By using $M_{\pi}^{2} \approx m_{l} M_{\eta_{s}}^{2} /m_{s}$ and approximating the ratio $M_{\eta_s} / 4\pi f_{\pi}$, we take $x_{\pi} = m_l / 5.63 m_s^{\mathrm{tuned}}$ as in~\cite{Chakraborty:2021qav}.
We give the coefficients $\zeta$, common to all form factors, priors of $0(1)$.

The $\left( \Lambda /M_{H_{l(s)}}\right)^r$ factors in Eq.~\eqref{hhisqffff} account for the dependence of the form factors on the heavy quark mass.
This dependence is given by an HQET-inspired series in $\Lambda / M_{H_{l(s)}}$ which we truncate.

The $\Omega^{(n)}$ factors are given by
\begin{align}
	\Omega^{(n)} = 1+\rho^{(n)} \log \left(\frac{M_{H_{l(s)}}}{M_{D_{l(s)}}}\right).  \label{eq:Omegadefn}
\end{align}
$\Omega^{(n)}$ allows for heavy quark mass dependence that appears as a prefactor to the expansion in inverse powers of the heavy mass given in Eq.~\eqref{hhisqffff}.
From HQET this prefactor could include fractional powers of the heavy quark mass and/or logarithmic terms which vary in different regions of $q^2$~\cite{Charles:1998dr}. 
We allow for this with a logarithmic term with a variable coefficient that depends on the form factor and the power of $z$ in the $z$-expansion.
We take priors for the $\rho^{(n)}$ of $0(1)$. 

The kinematic constraint $f_0 (0) = f_+(0)$ follows since the vector current matrix element must be finite at $q^2 = 0$.
This constraint holds in the  continuum limit for all $M_{H_c}$.
Recalling that we choose $t_0 = 0$, which gives $z(0) = 0$, then this constraint is imposed on the fit by insisting that $(A_0)^{(0r00)} = (A_+)^{(0r00)}$ for all $r$ and $\rho_0^{(0)} = \rho_+^{(0)}$.

The mistuning terms are given by
\begin{align}
\mathcal{N}_\text{mis}^{(n)} &= 1 + \frac{\delta m_c^{\mathrm{sea}}}{m_c^\text{tuned}} \kappa^{(n)}_1 + \frac{\delta m_c^{\text{val}}}{m_c^\text{tuned}} \kappa^{(n)}_2 + \frac{\delta m_l }{10 m_s^\text{tuned}}\kappa^{(n)}_3 \nonumber \\ & \hspace{6mm}
+ \frac{\delta m_s^{\text{sea}}}{10m_s^\text{tuned}}\kappa^{(n)}_4 + \frac{\delta{m_s^{\mathrm{val}}}}{10m_s^\text{tuned}}\kappa^{(n)}_5. \label{eqn:curlyN}
\end{align}
%
%where the term proportional to $\delta m_s^\text{val}$ is included only for the $B_c\rightarrow D_s$ case.
The parameters $\kappa^{(n)}_j$ allow for errors associated with mistunings of both sea and valence quark masses.
For each of the sea and valence quark flavours, $\delta m^{\text{sea}}$ and $\delta m^{\text{val}}$ are given by
\begin{align} \label{eqn:defn_delta_m}
	\delta m^{\text{sea}} = & \hspace{1mm}  m^{\text{sea}} - m^{\text{tuned}} \nonumber \\
	\delta m^{\text{val}} = & \hspace{1mm}  m^{\text{val}} - m^{\text{tuned}},
\end{align}
giving estimates of the extent that the quark masses deviate from the ideal choices in which physical masses of hadrons are exactly reproduced.
The $\delta m_s^{\mathrm{val}}$ term in $\mathcal{N}_\text{mis}^{(n)}$ is not included for the $B_c \to D_l$ form factors since no valence strange quark is present in this case.
For priors, we take $0(1)$ for those $\kappa$ associated with valence quark mass mistunings, and $0.0(5)$ for sea quark mass mistunings which are expected to have a smaller effect.

We now explain the specific values used for $m^{\mathrm{tuned}}$ for each flavour of quark.
The tuned mass $m_s^{\text{tuned}}$ is an estimate of the valence strange quark mass that would reproduce the \lq physical\rq \: $\eta_s$ meson mass on the gauge field configurations we use.
The $\eta_s$ is a fictitious $s \overline{s}$ pseudoscalar meson where the valence strange quarks are prohibited from annihilating.
It is not a particle that is realised in nature, though its mass can be determined in lattice QCD by ignoring disconnected diagrams.
Hence, we use it as a tool to evaluate the extent to which the strange quark mass in simulation has been mistuned.
We construct a \lq physical\rq\ value for the mass of the  $\eta_s$ meson ($M_{\eta_s}^{\mathrm{phys}}$) based on masses of pions and kaons~\cite{Dowdall:2013rya}.
We find $a m_{s}^{\text {tuned }}$ through
\begin{align}
a m_{s}^{\text {tuned }}=a m_{s}^{\text {val }}\left(\frac{M_{\eta_{s}}^{\text {phys }}}{M_{\eta_{s}}}\right)^{2} \label{eqn:amstuned}
\end{align}
where $a m_{s}^{\text {val }}$ is the valence strange quark mass given in Table~\ref{LattDesc1}, $aM_{\eta_s}$ is taken from Table III of~\cite{McLean:2019qcx} (which also used our $a m_{s}^{\text {val }}$ values), and finally we use $M_{\eta_s}^{\mathrm{phys}} = 688.5(2.2) \; \mathrm{MeV}$ from ~\cite{Dowdall:2013rya}.
The value $m_l^{\text{tuned}}$ is fixed by multiplying $m_s^{\text{tuned}}$ from Eq.~\eqref{eqn:amstuned} by the physical ratio~\cite{Bazavov:2017lyh}
\begin{align} 
\frac{m_l}{m_s} = & \hspace{1mm}  \frac{1}{27.18(10)}. \label{eq:mlms_ratio}
\end{align}
%
%\rough{For the $b$ quark, we take tuned values of the quark mass in lattice units from Table XII in~\cite{Dowdall:2011wh}.
%For consistency, we convert these values to physical units by using the lattice spacing $a_{\Upsilon}$, where $a_{\Upsilon}$ is the lattice spacing used in~\cite{Dowdall:2011wh} obtained from the $\Upsilon(2S-1S)$ splitting.}
We take $am_c^{\mathrm{tuned}}$ to be
\begin{align}
	a m_{c}^{\text {tuned }}=a m_{c}^{\mathrm{val}}\left(\frac{M_{J / \psi}^{\mathrm{expt}}}{M_{J / \psi}}\right),
\end{align}
where $M_{J / \psi}^{\mathrm{expt}}= 3.0969 \: \mathrm{GeV}$ (ignoring the negligible uncertainty) from PDG~\cite{Tanabashi:2018oca}, and lattice values for $aM_{J / \psi}$ are obtained from Table III in~\cite{Hatton:2020qhk} (which also used our $a m_{c}^{\text {val }}$ values).
Thus, the tuned valence charm mass is designed to closely reproduce the physical mass of the $J/\psi$ meson.
Detailed discussion of tuning the valence charm quark mass can be found in~\cite{Hatton:2020qhk}.

\subsubsection{Heavy quark mass dependence of $M_{\mathrm{res}}$} \label{sec:Mres}

\begin{table}
	\centering
	\caption{Masses of lightest mesons with $J^{P}$ quantum numbers (given without error) in GeV~\cite{PDG,Bardeen:2003kt,Lang:2015hza} used for approximating the leading order dependence of the heavy quark mass on the location of the vector and scalar poles (see the text in Section \ref{sec:ff_fit_form}). These values are also discussed in Appendix~\ref{app:reconstruct_ff}. The parameter $x$ is defined in Eq.~\eqref{eqn:xls}, and the parameter $\Delta (m_b)$ is defined in Eq.~\eqref{eqn:Deltals}.}
	\begin{tabular}{c c c c c c}
		\hline\hline
		& $0^-$ & $0^+$ & $1^-$ & $\Delta (m_b) \; \mathrm{GeV}$ & $x \; \mathrm{GeV}^2$ \\ [0.1ex]
		\hline
		$B_c \to D_l$ & $5.27964$ & $5.627$ & $5.324$ & $0.34736$ & $0.9368$ \\
		$B_c \to D_s$ & $5.36684$ & $5.711$ & $5.4158$ & $0.34416$ & $1.0510$ \\
		\hline\hline
	\end{tabular}
	\label{tab:res_pole_masses}
\end{table}

For the $f_0$ and $f_{+,T}$ form factors, the relevant poles are the masses of the scalar and vector heavy-light(strange) mesons respectively.
Since these particles have a valence heavy quark, their masses vary with $m_h$.
Determination of these meson masses at comparable precision to the energies of the pseudoscalar mesons is unnecessary.
For the $J^P = 1^-$ mesons, this would require the set of correlation functions described in Section~\ref{sec:fit_correls} to be augmented by 2-point correlation functions with propagators from different sources.
%random wall sources that have. been patterned differently.
Hence, additional propagators would need to be calculated.
%It is most convenient for us to approximate these masses.
% here since the poles do not need to be calculated as precisely for the form factor fits to proceed.
Instead,
%of constructing and fitting the relevant 2-point correlation functions for the $J^P = 0^+, 1^-$ mesons, 
we approximate these meson masses similarly to the estimation of the $J^P = 0^+, 1^-$ heavy-charm mesons in~\cite{McLean:2019qcx} and~\cite{Harrison:2020gvo} 
%(heavy-HISQ calculations of $B_s \to D_s$ and $B_c \to J/\psi$ form factors respectively)
and the estimation of the $J^P = 0^+, 1^-$ heavy-strange mesons in~\cite{Parrott:2020vbe}.
% (a heavy-HISQ calculation of $B_s \to \eta_s$ form factors).

Here, for $B_c \to D_{l(s)}$, we take the extra step in scrutinising this method of approximating the masses of the $J^P = 0^+, 1^-$ mesons by demonstrating that our fits of the form factors are insensitive to
%conservatively large
shifts in these estimates.
These checks are particularly important for processes facilitated by $b \to u$ or $b \to s$ since $q^2_{\mathrm{max}}$ is close to $M_{\mathrm{res}}^2$, and so we expect the $z$-coefficients in the fit form at Eq.~\eqref{hhisqffff} to be more sensitive to the position of nearest pole.
For example, $B_s \to D_s$ has $q^2_{\mathrm{max}} / M_{B_c^*}^2 = 0.29$ whilst $B_c \to D_s$ has $q^2_{\mathrm{max}} / M_{B_s^*}^2 = 0.63$ (errors ignored).
We show this analysis in Appendix~\ref{sec:ffs_fit_analysis} which is summarised by Fig.~\ref{fig:ff_stability}.
	
We now show how we approximate masses of the heavy-light(strange) $J^P = 1^-$ and $J^P = 0^+$ mesons.
We denote these mesons as $H_{l(s)} (1^-)$ and $H_{l(s)} (0^+)$.
%, $H_{s} (1^-)$, and $H_{s} (0^+)$.
Similarly, in this section we will refer to the pseudoscalar meson as $H_{l(s)} (0^-)$.
The nearest pole for $f_+$ is the vector heavy-light(strange) vector meson.
We use the fact that the hyperfine splittings
	\begin{align}
	\Delta_{H_{l(s)} (1^-)} &= M_{H_{l(s)} (1^-)} - M_{H_{l(s)} (0^-)}
	\end{align}
	are expected to vanish as $\Lambda/m_h$ in the limit $m_h \to \infty$~\cite{Falk:1992ws} since, by HQET~\cite{Georgi:1990um}, there is a spin symmetry in this limit meaning that the vector and pseudoscalar mesons become degenerate.
	We model the leading order dependence on $m_h$ through
	\begin{align}
	M_{H_{l(s)} (1^-)} &\approx M_{H_{l(s)} (0^-)} + \frac{x_{l(s)}}{M_{H_{l(s)} (0^-)}} \label{eqn:approx_vector_meson}
	\end{align}
	where $M_{H_{l(s)}}$ are proxies for $m_h$ and the parameters $x_{l(s)}$ are set at $m_h = m_b$ using values from~\cite{PDG}: we take
	\begin{align}
    x_{l(s)} &= (M_{B_{l(s)} (1^-)} - M_{H_{l(s)} (0^-)}) M_{B_{l(s)} (0^-)} \label{eqn:xls}
	\end{align}
	so that the approximation in Eq.~\eqref{eqn:approx_vector_meson} yields $M_{H_{l(s)} (1^-)}$ equal to $M_{B_{l(s)} (1^-)}$ at $m_h = m_b$.
	%Away from the physical point, I take values of $M_{\eta_h}$ from~\cite{McLean:2019qcx}.
	
	Regarding the pole for $f_0$, the differences between the pseudoscalar and scalar mesons
	\begin{align}
	\Delta_{l(s)} (m_h) &= M_{H_{l(s)} (0^+)} - M_{H_{l(s)} (0^-)} \label{eqn:Deltals}
	\end{align}
	are expected to be largely independent of the heavy quark mass because the scalar meson is simply an orbital excitation of the pseudoscalar meson.
	For example, note that $\Delta_s (m_b) = 0.344 \; \mathrm{GeV}$ and $\Delta_s (m_c) = 0.3490 \; \mathrm{GeV}$ (ignoring errors) are very similar ($B_s$, $D_{s0}$ and $D_s$ masses taken from~\cite{PDG} and $B_{s0}$ mass taken from~\cite{Lang:2015hza} (predicted)), providing qualitative support of this statement.
	% what about for h->l?
	Therefore, we approximate $M_{H_{l(s)} (0^+)}$ as
	\begin{align}
	M_{H_{l(s)} (0^+)} \approx M_{H_{l(s)} (0^-)} + \Delta_{l(s)} (m_b).
	\end{align}
	The errors on $\Delta_{l(s)} (m_b)$ are ignored.
	
	In Table~\ref{tab:res_pole_masses}, we summarise the values of the masses that we use and subsequent values for $x_l$ and $x_s$ from Eq.~\eqref{eqn:xls}.
	By construction, all of the heavy-light(strange) meson masses match the physical values (observed or predicted) at the point $m_h = m_b$.
	%Should the approximations above not be sufficiently accurate for any $m_h < m_b$, then this could, be exposed by unreasonably large values of $A^{(n)}_{ijk}$ for $k > 0$ in the form factor fitting, or even a poor fit in which data at the largest and smallest values $m_h$ are at tension.
	
	In Eq.~\eqref{hhisqffff}, the pole factor $P(q^2)^{-1}$ multiples a polynomial in $z$ with degree $N_n$.
	For our final results, we use $N_n=3$, i.e. a cubic polynomial in $z$.
	We demonstrate in Appendix~\ref{sec:ffs_fit_analysis} that results with $N_n=4$ are in good agreement, and hence the truncation of the $z$-series is justified.

%\subsubsection{Alternative: Fitting $R_{0,+}$} \label{sec:fitting_R}
%
%As well as a direct fit of the form factors (see Section \ref{sec:ff_fit_form}), I also fit the quantity
%%
%\begin{align} \label{eqn:R0plus}
%R_{0,+} (q^2) &= \frac{f_{0,+}(q^2)}{f_{H_c} \sqrt{M_{H_c}}}
%\end{align}
%%
%to the same fit function as $f_{0,+}$ at Equation (\ref{hhisqffff}).
%It is expected that $R_{0,+}$ has less discretisation effects that the form factors, hence aiding the fit, and the continuum form factors can be obtained from $R_{0,+}$ by multiplying through by the physical-continuum value for $f_{H_c} \sqrt{M_{H_c}}$.
%	
%\subsection{Fitting the Decay Constants}
%
%Data for the decay constants of the pseudoscalar mesons $H_q$ for $q=c,s$ (see Section \ref{sec:fit_correls}) can be fit similarly to the form factor data (described in Section~\ref{sec:fit_correls}).
%For example, data for the decay constant $f_{H_c}$ from my heavy-HISQ calculation is fit to
%%
%\begin{align}
%\Big(\frac{M_{H_c}}{M_{B_c}} \Big)^{\frac{1}{2}} f_{H_c} = &\sum_{n,r,i,j,k = 0}^{2} A^{(n)}_{rijk} \mathcal{N}_\text{mis}^{(n)} \Delta_{H_c}^{(r)} \nonumber \\
%& \times \left(\frac{a\Lambda_{\text{QCD}}}{\pi}\right)^{2i} \left(\frac{am_h}{\pi}\right)^{2j} \left(\frac{am_c}{\pi}\right)^{2k}. \label{hhisqdcff}
%\end{align}
%%
%For $H_s$, I use a similar form but without the terms allowing for discretisation effects associated with a valence charm quark.
%The parameter $A^{(0)}_{0000}$ takes a prior value of $0.5(5)$ and the other $A^{(n)}_{rijk}$ take prior values of $0(2)$ and $0(1)$ for $H_c$ and $H_s$ respectively.

\section{Results} \label{sec:results}

\subsection{Form factors} \label{sec:results_ffs}

We use the correlation function fits on each set indicated in Table~\ref{tab:ff_stab} of Appendix~\ref{sec:stab_corrfit}.
The energies and matrix elements on each set are stored (with all correlations) in the ancillary file \texttt{corrfit\_results.tar}.
We fit the subsequent form factor data to the form described in Section~\ref{sec:ff_fit_form}.
%update
%We find that the fits gives $\chi^2/\text{d.o.f.} = 0.25$ and $\chi^2/\text{d.o.f.} = 0.35$ for the cases $B_c \to D_l$ and $B_c \to D_s$ respectively.
Fitting with noise added to both the data and priors, as demonstrated in~\cite{Dowdall:2019bea} to compensate for the reduced $\chi^2/\text{d.o.f.}$ from fitting with an SVD cut, we find $\chi^2/\text{d.o.f.} = 0.65$ and $\chi^2/\text{d.o.f.} = 0.43$ for the cases $B_c \to D_l$ and $B_c \to D_s$ respectively.
%
%In the limit of physical quark masses and vanishing lattice spacing, the fit form in Eq.~\eqref{hhisqffff} reduces to the form in Eq.~\eqref{eqn:physcont_ffs} with the coefficients $c^{(n)}$ defined in Eq.~\eqref{eqn:def_cn}.
%In Tables~\ref{tab:ff_coeffs_BcDu} and~\ref{tab:ff_coeffs_BcDs} of Appendix~\ref{app:reconstruct_ff}, we give the meson masses and coefficients $c^{(n)}$ required to reconstruct our form factors in the continuum limit with physical quark masses.

We check that our priors are sensible and conservative by performing Empirical Bayes analyses~\cite{Lepage:2001ym}.
We use the \texttt{lsqfit.empbayes\_fit} function to test the width of the parameters in the following two sets: $\rho^{(n)}$ and $A^{(nr00)}$, and $A^{(nrjk)}$ for $j+k >0$.
The widths of each parameter in these sets are varied simultaneously by a common multiplicative factor $w$.
The Empirical Bayes analyses show that the values for $w$ are around 0.5, so our priors are moderately conservative.
 
\begin{figure}
	\centering
	\includegraphics[width=0.5\textwidth]{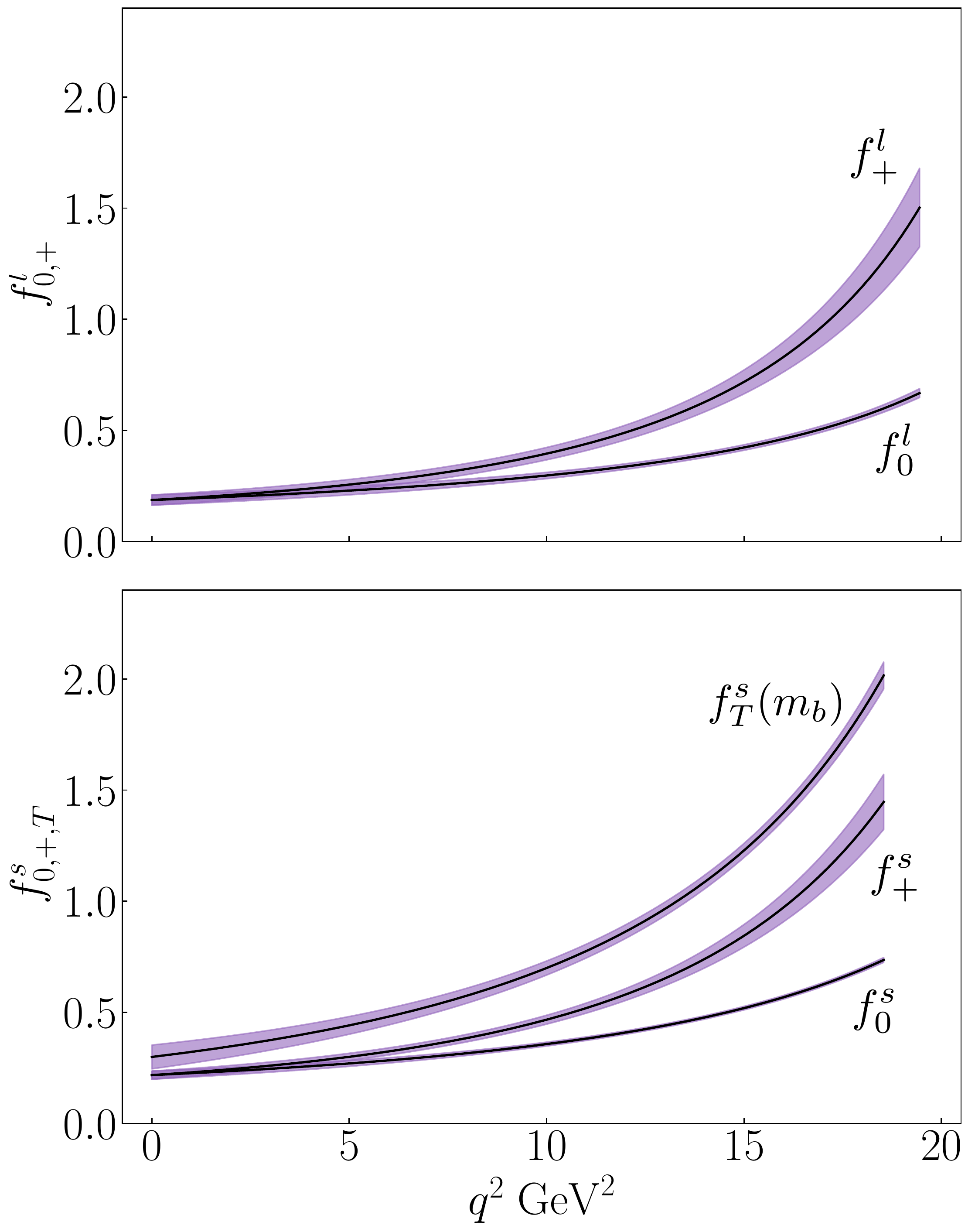}
	\caption{The fit functions for the $B_c \to D_l$ and $B_c \to D_s$ form factors $f_{0,+}^l$ and $f_{0,+,T}^s$ respectively tuned to the continuum limit with physical quark masses. The tensor form factor is at scale $4.8 \; \mathrm{GeV}$.}
	\label{fig:final_ffs}
\end{figure}

In Fig.~\ref{fig:final_ffs}, we present our form factors in the limit of vanishing lattice spacing and physical quark masses across the entire physical range of $q^2$.
Details of the fits of the correlation functions and lattice form factors from which Fig.~\ref{fig:final_ffs} is derived from are given in Appendices~\ref{sec:fitting_analysis} and~\ref{sec:ffs_fit_analysis}.
Appendix~\ref{app:reconstruct_ff} provides details of our form factors in the limit of vanishing lattice spacing and physical quark masses.
%from the fit form described in Section~\ref{sec:ff_fit_form} and the correlation function fits indicated by the boldened entries of Table~\ref{tab:ff_stab} in Appendix~\ref{sec:stab_corrfit}.
%update1this
%
%
%Values taken by each form factor at $q^2 = 0$ and zero-recoil (maximum $q^2$) can be found in Table~\ref{tab:ff_extrema}.

\begin{figure}
	\centering
	\includegraphics[width=0.5\textwidth]{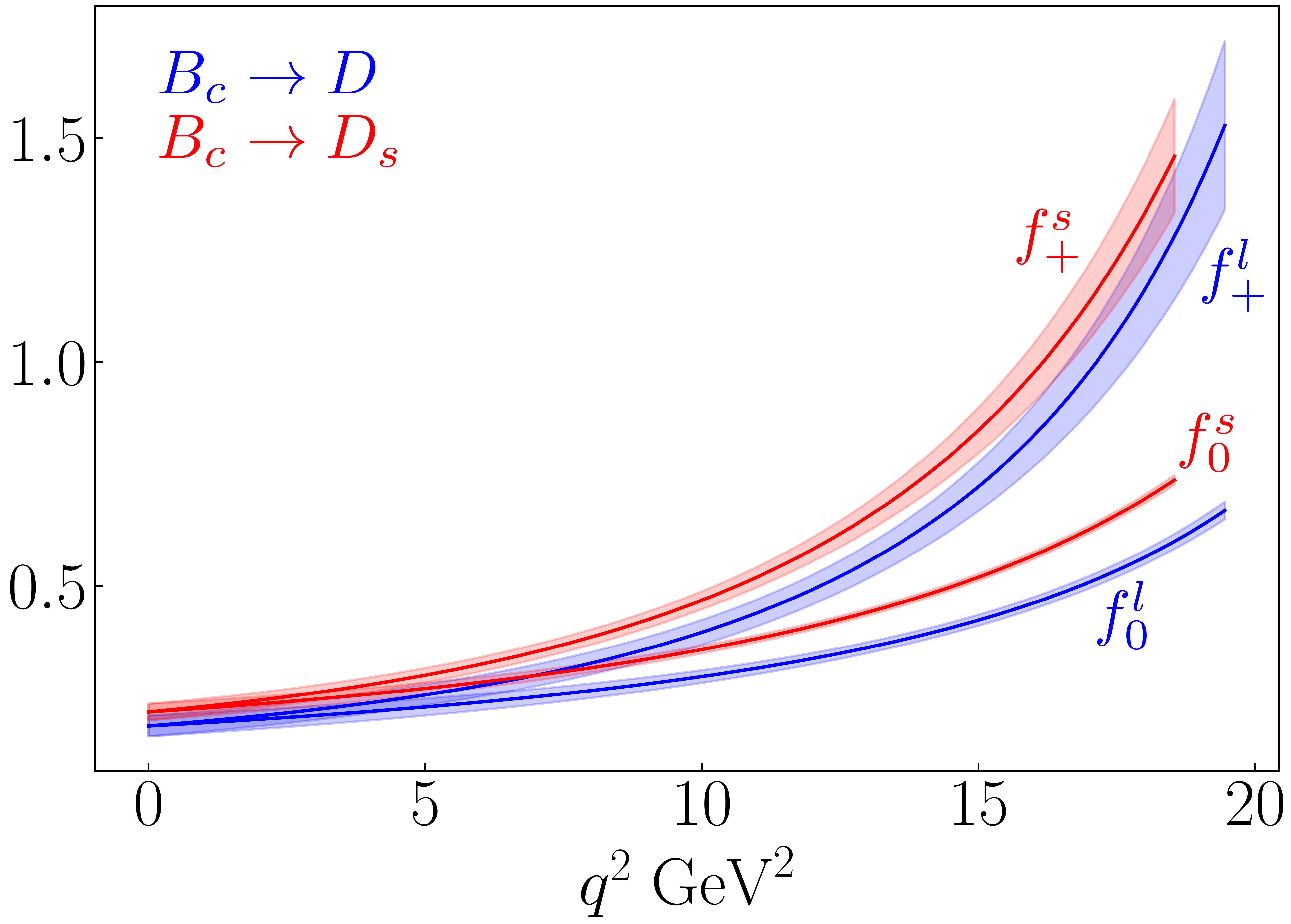}
	\caption{The fit functions for the four form factors $f_{0,+}^{l,s}$ tuned to the continuum limit with physical quark masses.}
	\label{fig:daughter_comparison}
\end{figure}

Fig.~\ref{fig:daughter_comparison} shows the form factors $f_{0,+}^{l,s}$ on the same plot.
This figure shows how the form factors varies as the daughter quark mass changes from $m_s$ to $m_l = m_s/27.4$.
We plot each form factor from $q^2 = 0$ up to the zero-recoil point where $q^2 = (M_{B_c} - M_{D_{(s)}})$ which depends on the daughter quark mass.
The form factors for the strange daughter quark are larger than those for the light daughter quark at all $q^2$ values.
This mirrors what is seen, for example, in the comparison of $D \to \pi$ and $D \to K$ form factors~\cite{Lubicz:2017syv}.

\begin{figure}
	\centering
	\includegraphics[width=0.5\textwidth]{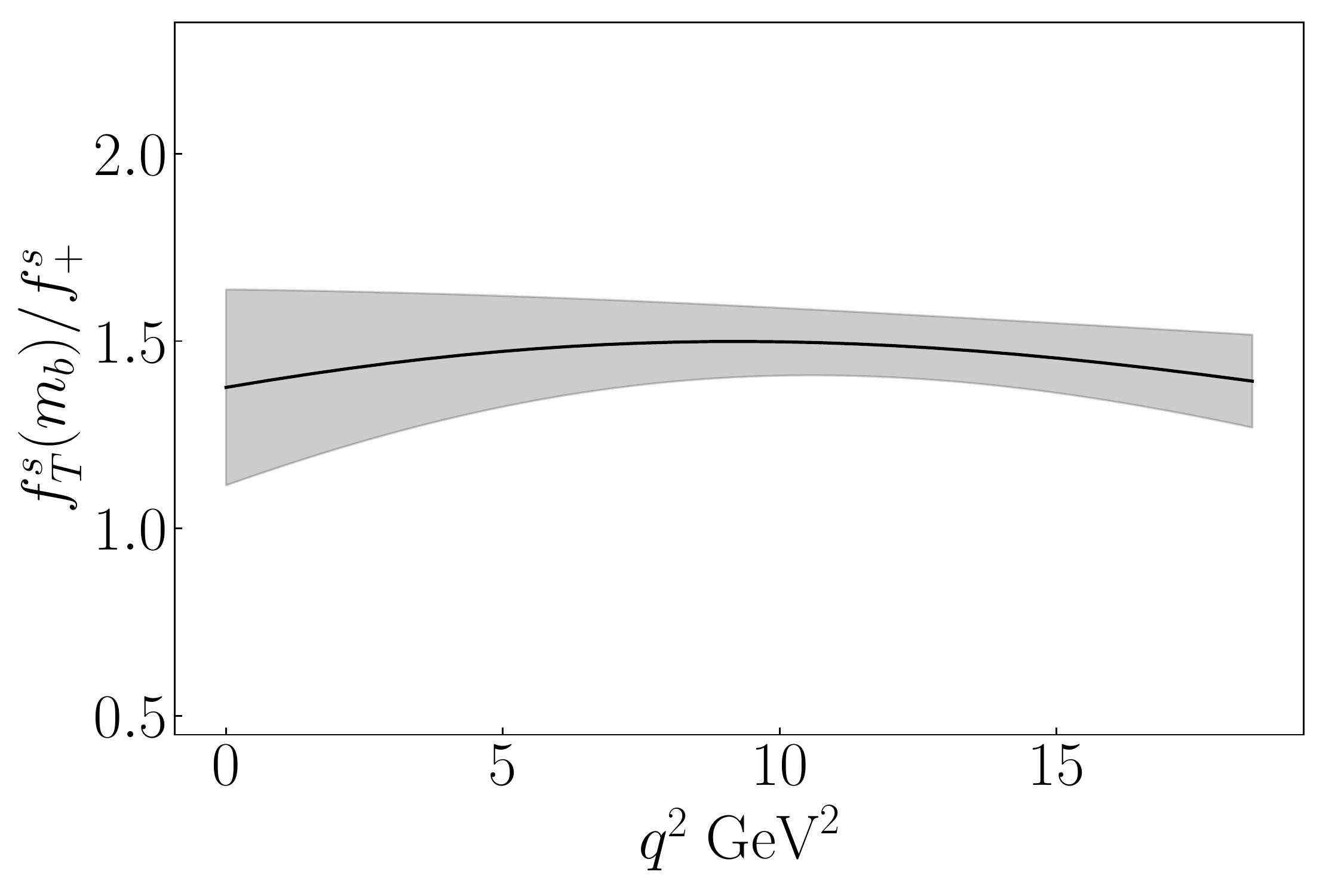}
	\caption{Ratio of the tensor and vector form factors of $B_c \to D_s$ across the entire range of physical $q^2$. The behaviour is in agreement with LEET~\cite{Charles:1998dr} which predicts a constant ratio $(M_{B_c} + M_{D_s})/ M_{B_c}$.}
	\label{fig:fToverFplus}
\end{figure}

For the case $B_c \to D_s$, we show in Fig.~\ref{fig:fToverFplus} the ratio $f_T (m_b) / f_+$ across the entire range of $q^2$.
%This quantity appears very flat as $q^2$ is varied.
Large Energy Effective Theory (LEET)~\cite{Charles:1998dr} expects this ratio near $q^2 = 0$ to take the value $(M_{B_c} + M_{D_s})/ M_{B_c} = 1.31$~\cite{PDG} in the limit $m_b \to \infty$ and ignoring renormalisation corrections.
This follows from the spatial-temporal tensor and spatial vector matrix elements coinciding in the limits $m_b \to \infty$ and $q^2 \to 0$, and the definitions of $f_+$ and $f_T$ in Eqs.~\eqref{fplusextract} and~\eqref{fTextract}.
We find that the ratio $f_T / f_+$ near $q^2=0$ is consistent with LEET, and that this ratio does not change significantly as $q^2$ is varied.
%update1this
%

\begin{figure}
	\centering
	\includegraphics[width=0.5\textwidth]{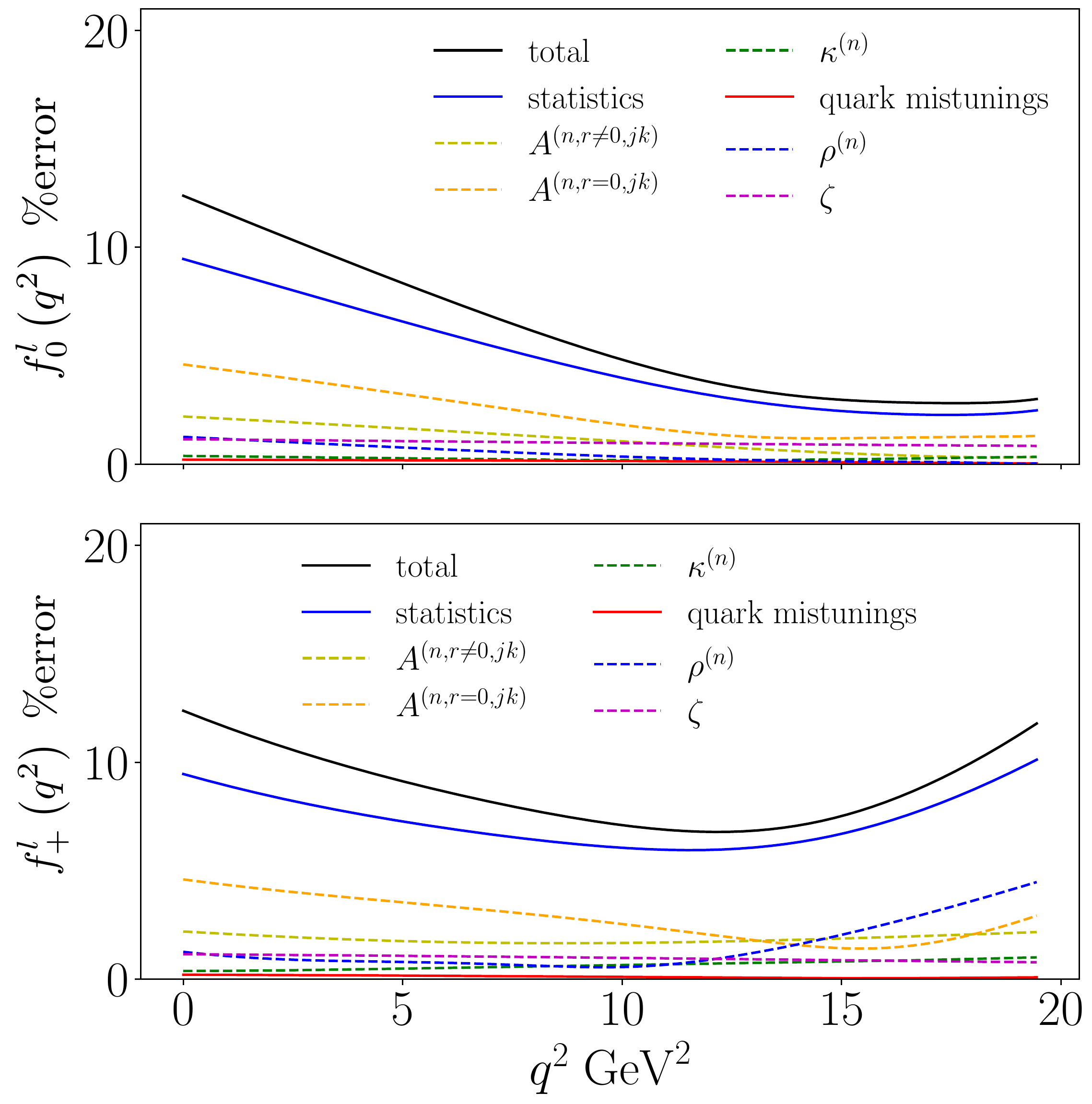}
	\caption{Errors on the form factors $f_{0,+}^l$. The black curve shows the total error and the other lines show a particular partition of the error. When added in quadrature, these contributions yield the black curve. The dashed curves show uncertainties from the fit coefficients in Eq.~\eqref{hhisqffff}. The solid blue curve shows the statistical errors resulting from our fits of correlation functions. The solid red curve represents the contribution to the final error from the determinations of the quark mass mistunings on each lattice (see Eq.~\eqref{eqn:curlyN}).}
	\label{fig:errorbudgetbcdl}
\end{figure}

\begin{figure}
	\centering
	\includegraphics[width=0.5\textwidth]{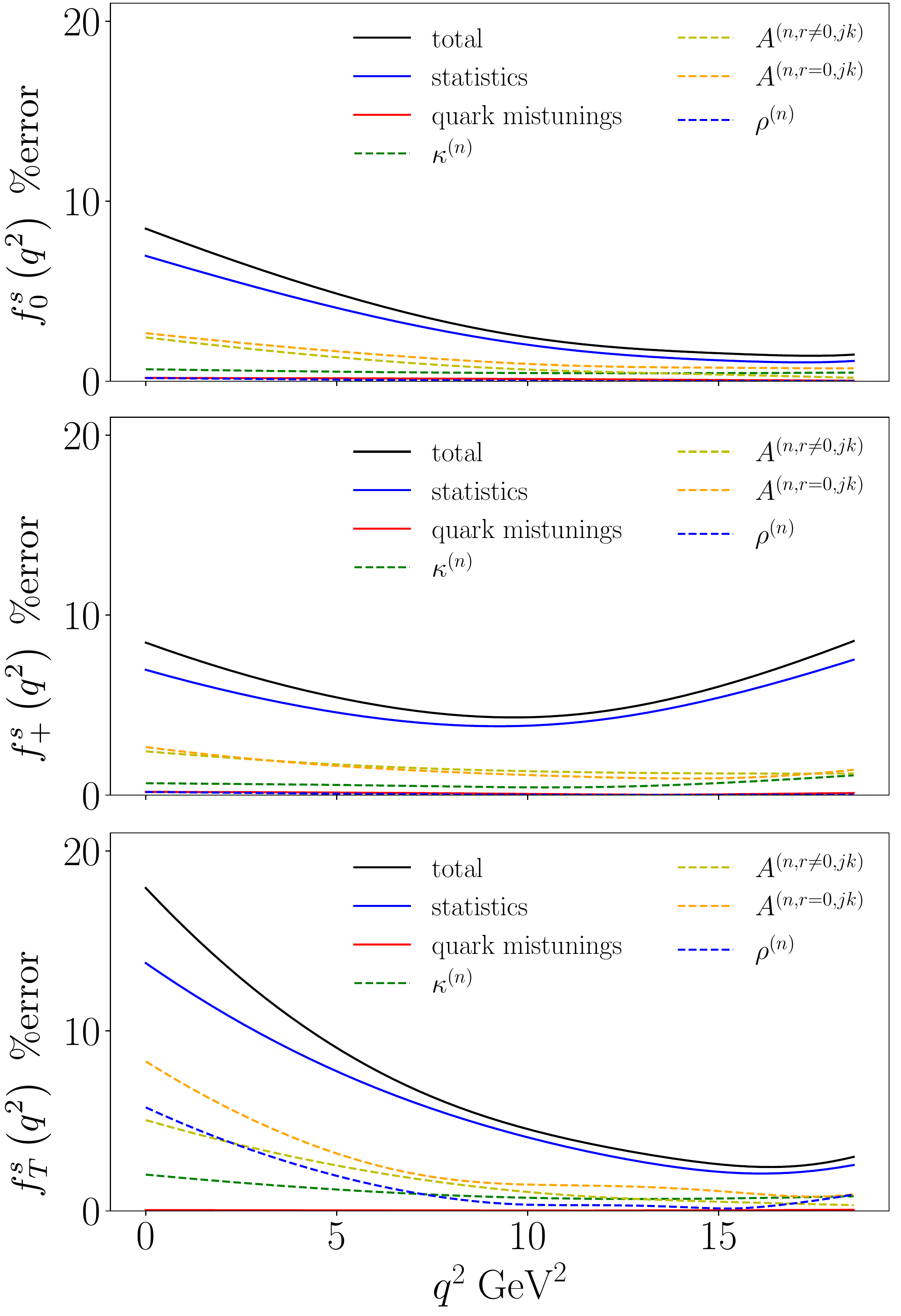}
	\caption{Errors on the form factors $f_{0,+,T}^s$. The curves are labelled similarly to Fig.~\ref{fig:errorbudgetbcdl}.}
	\label{fig:errorbudgetbcds}
\end{figure}

We use the \emph{gvar} package~\cite{gvar} to propagate correlations throughout our calculation.
The package also allows us to decompose the uncertainty on the form factors and resulting branching fractions to create an error budget.
We plot a particular breakdown of the errors in Figs.~\ref{fig:errorbudgetbcdl} and~\ref{fig:errorbudgetbcds} for the form factors $f_{0,+}^l$ and $f_{0,+}^s$ respectively.
%update1this
%
We find that statistical errors contribute substantially to the final error.
Of a similar size are the uncertainties from the coefficients $A^{(n0jk)}$ in the fit form in Eq.~\eqref{hhisqffff}.
The  fit function in Eq.~\eqref{hhisqffff} is complicated since the coefficients $A^{(nrjk)}$ responsible for the extrapolations $am_h \to 0$, $am_c \to 0$ and $\Lambda / M_{H_{l(s)}}  \to \Lambda / M_{B_{l(s)}} $ are mixed to allow for all possible effects.
Terms in the fit form with $r=0$ are associated with discretisation effects of the leading order term in the HQET expansion.
%The largest contribution to the error comes from the parameters associated with resolving the discretisation effects of the heavy quark, i.e. taking $am_h \to 0$, and accessing the point of the physical $b$ quark, i.e. $M_{H_c} \to M_{B_c}$.
This error could be decreased by including the exafine lattice ($a \approx 0.03 \; \mathrm{fm}$) so that $am_h$ can be taken smaller to further constrain the limit $am_h \to 0$.
Also, $b$ quarks, at their physical mass, can be directly simulated on the exafine lattice since $am_b$ is well below $1$.
%Hence, it is possible simulate with physically massive $b$ quarks on the exafine lattice with discretisation effects under control.
%This data should greatly inform the heavy meson mass dependence on the form factors which we parametrise by the $(\Lambda / M_{H_{l(s)}})^r$ factors in Eq.~\eqref{hhisqffff}.
We investigate the impact of adding the exafine lattice in Section~\ref{sec:ex5}.

Regarding the $\zeta$ and $\rho$ parameters in Eq.~\eqref{hhisqffff}, only $\zeta^{(0)}$ and $(\rho^{l,s})^{(0)}$ are determined accurately by the fit.
We find $\zeta^{(0)} = -0.66(24)$, $(\rho_{0,+}^l)^{(0)} = -0.544 (76)$, $(\rho_{0,+}^s)^{(0)} = -0.579 (64)$ and $(\rho_T^s)^{(0)} =-0.676 (92)$.

\subsection{Observables for $B_c^+ \to D^0 \ell^+ \nu_{\ell}$} \label{sec:BcD_obs}

\begin{figure}
	\centering
	\includegraphics[width=0.5\textwidth]{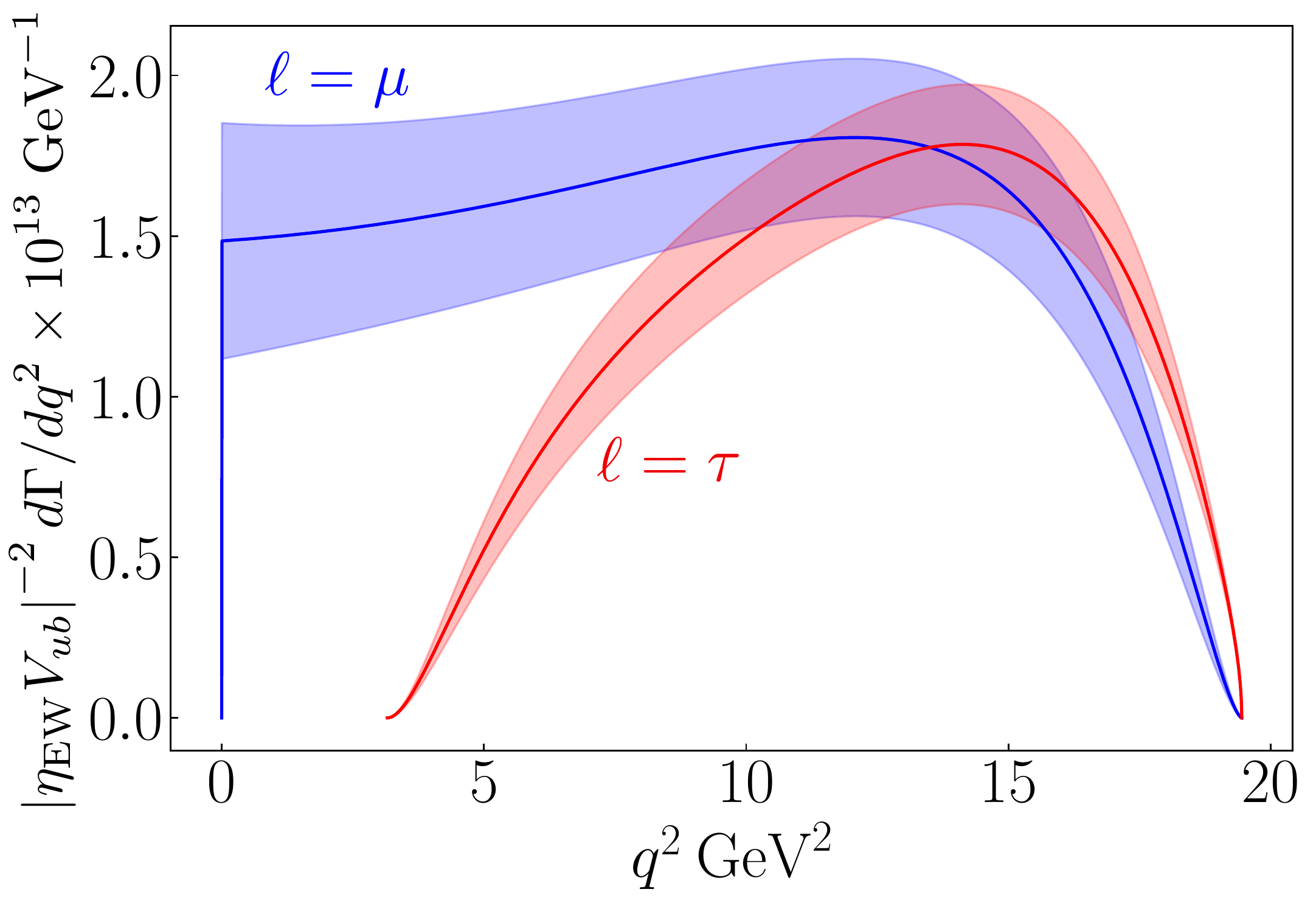}
	\caption{The differential decay rate $\eta_{\mathrm{EW}}^{-2}|V^{ub}|^{-2} d \Gamma (B_c^+ \to D^0 \ell^+ \nu_{\ell}) / dq^2$ as a function of $q^2$ for the cases $\ell = \mu$ in blue and  $\ell = \tau$ in red.}
	\label{fig:diff_decay_rate_BcD}
\end{figure}

We plot the differential decay rate $\eta_{\mathrm{EW}}^{-2} |V_{ub}|^{-2} d \Gamma (B_c^+ \to D^0 \ell^+ \nu_{\ell}) / dq^2$ derived from our form factors as a function of $q^2$ in Fig.~\ref{fig:diff_decay_rate_BcD}.
%update1this
%
The form of the decay rate is given in Eq.~\eqref{eqn:BcD_decay_rate}.
We integrate this function (using \texttt{gvar.ode.integral} in the \emph{gvar} package~\cite{gvar}) to find $\eta_{\mathrm{EW}}^{-2} |V^{ub}|^{-2} \Gamma$.
This is then combined with $\eta_{\mathrm{EW}}$, the CKM matrix element $V_{ub} = 3.82(24) \times 10^{-3}$~\cite{Zyla:2020zbs} (an average of inclusive and exclusive determinations), and the lifetime of the $B_c$ meson to obtain the branching ratios in Table~\ref{tab:integrated_Gamma_BcDl}.
%update1this
%
\begin{table}
	\centering
	\caption{For $B_c^+ \to D^0 \ell^+ \nu_{\ell}$, we give values for the branching ratios (BR) for each of the cases $\ell = e,\mu,\tau$.
		We take the lifetime of $B_c$ meson to be $513.49(12.4) \: \mathrm{fs}$~\cite{Aaij:2014gka}.
		The errors from the lifetime and the CKM matrix element $V_{ub}$ are shown explicitly.
		The error from $\eta_{\mathrm{EW}}$ is negligible.
		We ignore uncertainties from long-distance QED contributions since the meson $D^0$ in the final state is neutral.}
	\begin{tabular}{c | c}
		\hline\hline
		decay mode & BR$\times 10^{5}$ \\ [0.1ex]
		\hline
%		$B_c^+ \to D^0 e^+ \nu_e$  & $3.366(477)_{\mathrm{lattice}}(81)_{\tau_{B_c}} (423)_{\mathrm{CKM}}$ \\
%		$B_c^+ \to D^0 \mu^+ \nu_\mu$  & $3.362(473)_{\mathrm{lattice}}(82)_{\tau_{B_c}}(422)_{\mathrm{CKM}} $  \\
%		$B_c^+ \to D^0 \tau^+ \nu_\tau$  & $2.293(234)_{\mathrm{lattice}}(55)_{\tau_{B_c}}(288)_{\mathrm{CKM}} $  \\
		$B_c^+ \to D^0 e^+ \nu_e$  & $3.37(48)_{\mathrm{lattice}}(8)_{\tau_{B_c}} (42)_{\mathrm{CKM}}$ \\
		$B_c^+ \to D^0 \mu^+ \nu_\mu$  & $3.36(47)_{\mathrm{lattice}}(8)_{\tau_{B_c}}(42)_{\mathrm{CKM}} $  \\
		$B_c^+ \to D^0 \tau^+ \nu_\tau$  & $2.29(23)_{\mathrm{lattice}}(6)_{\tau_{B_c}}(29)_{\mathrm{CKM}} $  \\
		\hline\hline
	\end{tabular}
	\label{tab:integrated_Gamma_BcDl}
\end{table}
%
%
%%
%\begin{table}
%	\centering
%	\caption{Integrated quantities. The second uncertainty in the final column is due to the CKM matrix elements. \rough{AUTOMATE}}
%	\begin{tabular}{c | c c}
%		\hline\hline
%		decay mode & $|V^{ub}|^{-2} \Gamma$ & $\Gamma$ \\ [0.1ex]
%		\hline
%		$B_c^+ \to D^0 e^+ \nu_{e}$ & & \\
%		$B_c^+ \to D^0 \mu^+ \nu_{\mu}$ & & \\
%		$B_c^+ \to D^0 \tau^+ \nu_{\tau}$ & & \\
%		\hline\hline
%	\end{tabular}
%	\label{tab:integrated_Gamma}
%\end{table}
%%
At present, errors from our lattice calculation dominate those associated with the lifetime of the $B_c$ meson and are comparable with those from the CKM element $V_{ub}$. 
For the ratio of widths with $\tau$ and $\mu$ in the final state, we find that
\begin{align}
	\frac{\Gamma(B_c^+ \to D^0 \tau^+ \nu_\tau)}{\Gamma(B_c^+ \to D^0 \mu^+ \nu_\mu)} =  0.682(37).
\end{align}
Much of the error on our form factors cancels in this ratio, and we achieve an uncertainty of $7\%$.

\begin{figure}
	\centering
	\includegraphics[width=0.5\textwidth]{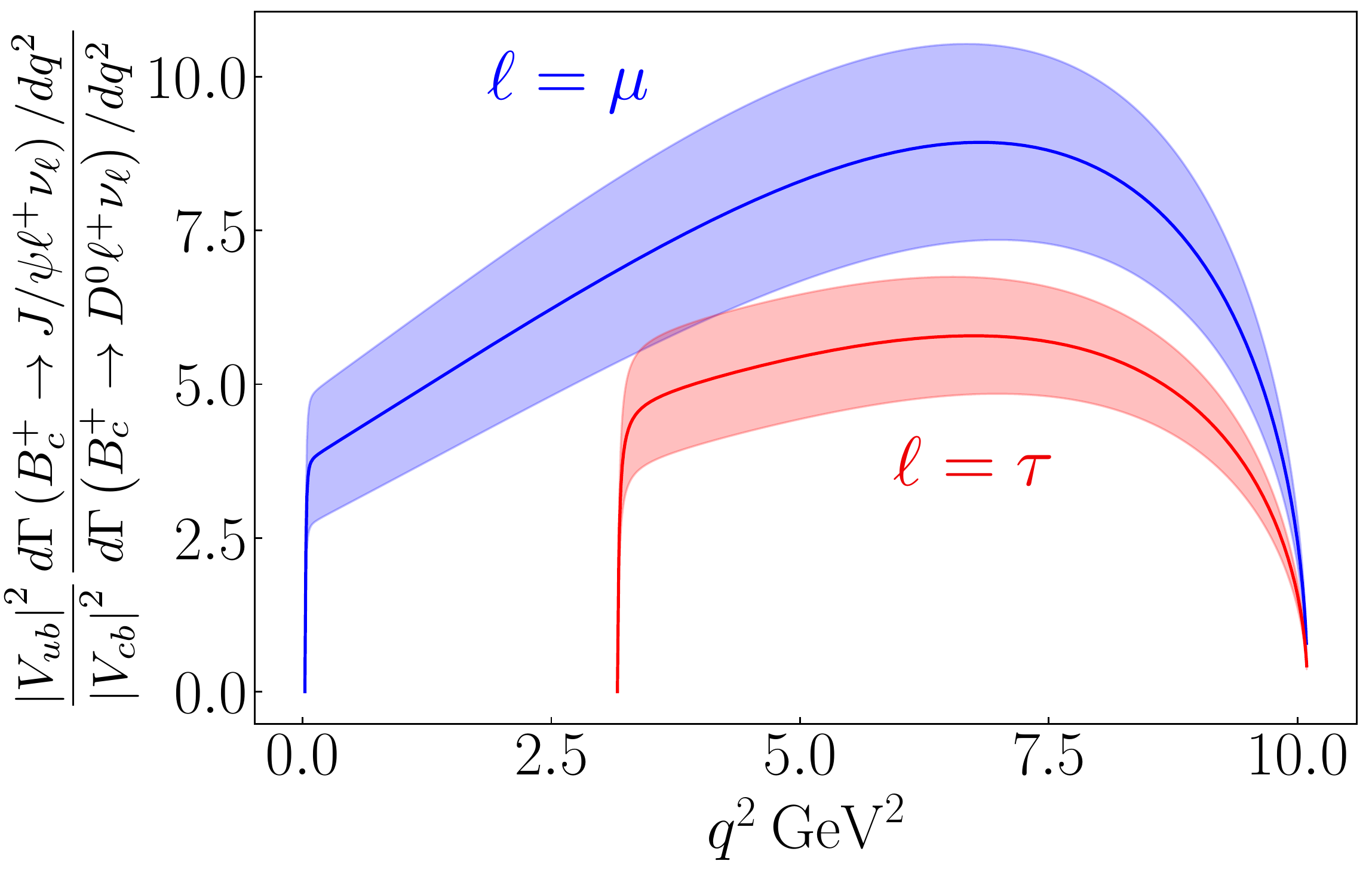}
	\caption{We plot the ratio of $d \Gamma / dq^2$ for  each of the processes $B_{c}^{+} \rightarrow J / \psi \ell^{+} \nu_{\ell}$ and $B_{c}^{+} \rightarrow D^0 \ell^+ \nu_{\ell}$ for the $q^2$ range of the $B_{c}^{+} \rightarrow J / \psi \ell^{+} \nu_{\ell}$ decay. The decay width for the former process is derived from form factors found in~\cite{Harrison:2020gvo}, and the decay width of the latter process is derived from form factors determined in this study. The case $\ell = \mu$ in shown in blue and  the case $\ell = \tau$ in shown red.}
	\label{fig:diff_decay_rate_ratio_BcJpsi_BcD}
\end{figure}

We compare our results with those for the decay mode $B_{c}^{+} \rightarrow J / \psi \ell^{+} \nu_{\ell}$.
We take the form factors for this decay from HPQCD's lattice QCD calculation in~\cite{Harrison:2020gvo}.
We combine these form factors with those for $B_{c}^{+} \rightarrow D^{0} \ell^{+} \nu_{\ell}$ computed in this study to find the ratios
%update1this
%muon
%BcD = gv.gvar(0.0214433955815, 0.00303294499528)
%BcJpsi = gv.gvar(0.0835226606774, 0.00597057275314)
%tau
%BcD = gv.gvar(0.0146252637893, 0.00148956771255)
%BcJpsi = gv.gvar(0.021569207261, 0.00143295647006)
%
\begin{align}
\left|\frac{V_{cb}}{V_{ub}} \right|^2 \frac{\Gamma( B_c^+ \to D^0 \mu^+ \nu_\mu )}{\Gamma(B_{c}^{+} \rightarrow J / \psi \mu^{+} \nu_{\mu})} = 0.257(36)(18), \nonumber \\
\left|\frac{V_{cb}}{V_{ub}} \right|^2 \frac{\Gamma( B_c^+ \to D^0 \tau^+ \nu_\tau )}{\Gamma(B_{c}^{+} \rightarrow J / \psi \tau^{+} \nu_{\tau})} = 0.678(69)(45). \label{eq:BcD_BcJpsi_ratio}
\end{align}
The first error comes from our form factors for $B_{c}^{+} \rightarrow D^0 \mu^+ \nu_{\mu}$, and the second error comes from the form factors for $B_{c}^{+} \rightarrow J / \psi \mu^{+} \nu_{\mu}$ in~\cite{Harrison:2020gvo}.
We treat the form factors for $B_{c}^{+} \rightarrow J / \psi \mu^{+} \nu_{\mu}$ as uncorrelated to the $B_{c}^{+} \rightarrow D^{0} \ell^{+} \nu_{\ell}$ form factors (a conservative strategy).
In Fig.~\ref{fig:diff_decay_rate_ratio_BcJpsi_BcD}, we plot the ratio of $d \Gamma / dq^2$ for the two processes for $m_{\ell}^2 < q^2 < (M_{B_c} - M_{J/\psi})^2$ and each of the cases $\ell = \mu, \tau$.
Note that the ratio plotted is the inverse of the one used in Eq.~\eqref{eq:BcD_BcJpsi_ratio}.
%

%In~\cite{Ghahramany:2010zz}, the authors give values for BR$\times 10^{-5}$ of $3.8(1.0)$ and $1.5(4)$ for $B_c^+ \to D^0 e^+ \nu_e$ and $B_c^+ \to D^0 \tau^+ \nu_\tau$ respectively from 3PSR, and $5.2(2.1)$ and $7.4(3.3)$ for $B_c^+ \to D^0 e^+ \nu_e$ and $B_c^+ \to D^0 \tau^+ \nu_\tau$ respectively from HQET.
%These values are roughly inline with what we see from lattice QCD in Table~\ref{tab:integrated_Gamma}.
A possible method for determining the ratio of $|V_{cs}|/|V_{ub}|$ is to determine the ratio of branching fractions for $B_c$ decay to $D^0 e^+ \nu_e$ and $B_s e^+ \nu_e$. 
Using our form factors for $B_c \to D$ and the form factors for $B_c \to B_s$ from~\cite{Cooper:2020wnj}, we find
%BcBs = gv.gvar(1.73789565323e-14 5.52145858598e-16)
%BcD = gv.gvar(1.13718705194e-10, 6.82456852611e-12)
%
\begin{align}
	\frac{\left|V_{ub}\right|^{2}}{\left|V_{cs}\right|^{2}} \frac{\mathcal{B}\left(B_{c}^{+} \rightarrow B_s^{0} e^{+} \nu_{e}\right)}{\mathcal{B}\left(B_{c}^{+} \rightarrow D^0 e^{+} \nu_{e}\right)} = 5.95(84)(19) \times 10^{-3}.  \label{eq:ratio_with_BcBs}
\end{align}
%
%This can readily be done with the form factors we give here and those in~\cite{Cooper:2020wnj} for $B_c \to B_s$.

Refs.~\cite{Jenkins:1992nb} and~\cite{Leljak:2019fqa} point out that the weak matrix elements for $B_c \to D$ and $B_c \to B_s$ have a simple ratio at the zero recoil point in the limit of $m_b \gg m_c \gg \Lambda_{\mathrm{QCD}}$. 
In this limit, the $B_c$ meson is a point-like particle and the weak matrix elements factorise into a factor that depends on the daughter meson decay constant and a factor that depends on the $B_c$ wave function which is the same in both processes.
Thus, the ratio of weak matrix elements becomes
\begin{align}
\frac{ \braket{ D| V_{\mu}|B_c} }{ \braket{B_s|V_{\mu}|B_c} } \Bigg|_{\mathrm{zero-recoil}} = \frac{ M_D f_D} {M_{B_s} f_{B_s}}. 
\end{align}
Using the decay constants from~\cite{Bazavov:2017lyh}, the RHS evaluates to $0.32$.
We expect an uncertainty on this value of size $\Lambda_{\mathrm{QCD}} / m_c$ ($\sim 30\%$) since the HQET result relies on $m_c \gg \Lambda_{\mathrm{QCD}}$.
By using our form factors for $B_c \to D$ and those for $B_c \to B_s$ from~\cite{Cooper:2020wnj}, we find that the LHS evaluates to $0.571(17)(8)$, much larger than the prediction from HQET.
We conclude that HQET is not a reliable guide here.
Calculations from three-point sum rules~\cite{Leljak:2019fqa} give $0.5(2)$.

\begin{figure}
	\centering
	\includegraphics[width=0.5\textwidth]{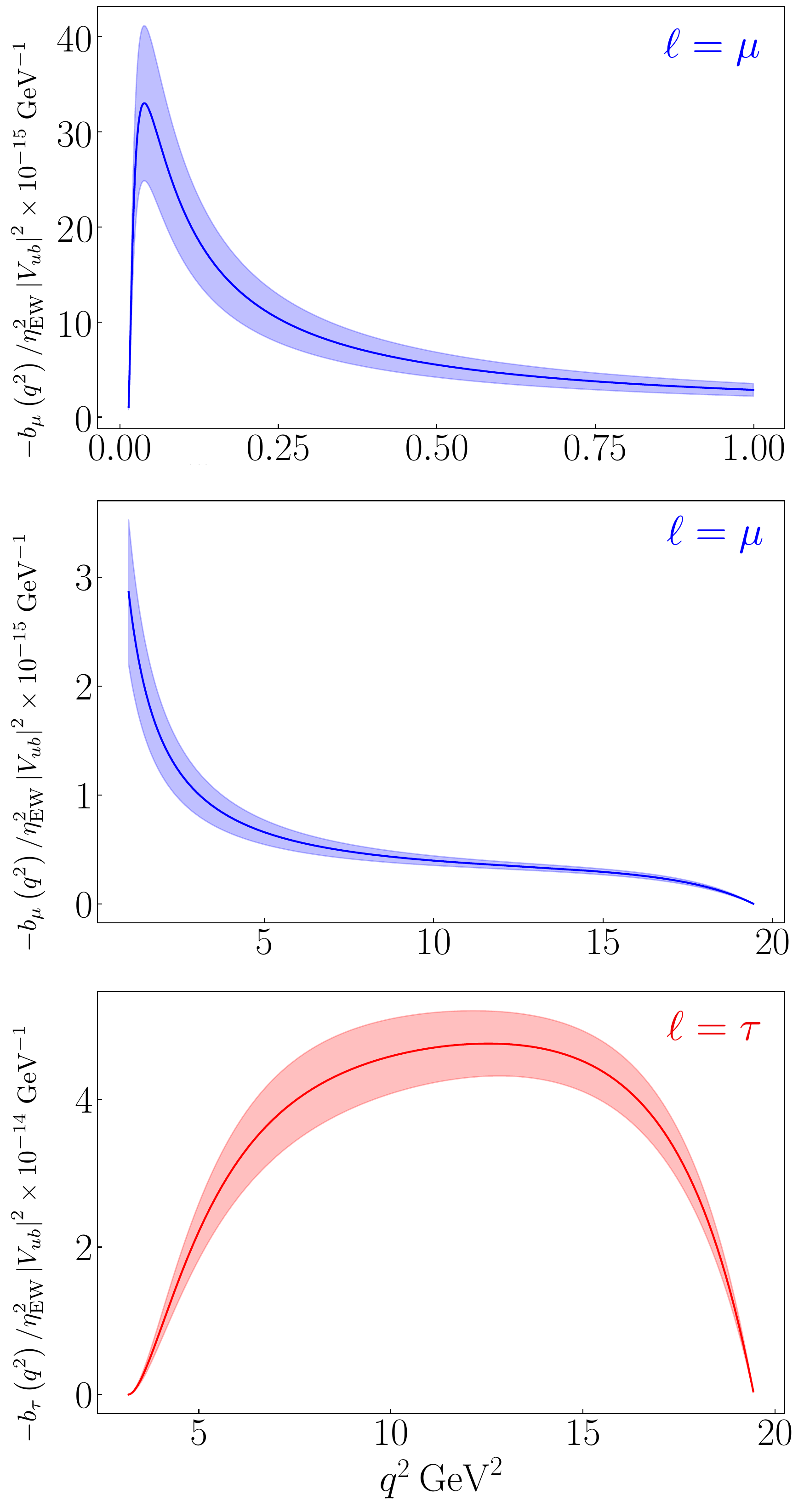}
	\caption{Plot of the $b_{\ell}\left(q^{2}\right)$, as defined in Eqs.~\eqref{eqn:formula_b_ell} and ~\eqref{eqn:dGamma_dqSqdcostheta}, for $B_c^+ \to D_s^+ \ell^+ \ell^-$. The top plot shows the case $\ell = \mu$ (blue) for the region $m_{\mu}^2 < q^2 < 1 \; \mathrm{GeV}^2$. The middle plot shows the case $\ell = \mu$ (blue) for the region $1 \; \mathrm{GeV}^2 < q^2 < q^2_{\mathrm{max}}$. Finally, the lower plot shows the case $\ell = \tau$ (red).}
	\label{fig:AFB}
\end{figure}

We now give the angular dependence of the differential decay rate.
Let $\theta$ be the angle between the direction of flight of the lepton $\overline{\ell}$ and the $D^0$ meson in the centre of mass frame of $\overline{\ell} \nu$.
Then, we have
\begin{align}
\frac{d^{2} \Gamma_{\ell}\left(q^{2}, \cos \theta\right)}{d q^{2} d \cos \theta}=a_{\ell}\left(q^{2}\right)+b_{\ell}\left(q^{2}\right) \cos \theta+c_{\ell}\left(q^{2}\right) \cos ^{2} \theta. \label{eqn:dGamma_dqSqdcostheta}
\end{align}
On performing the integration with respect to $\theta$, the piece linear in $\cos \theta$ vanishes though it is of interest when studying the angular dependence of the decay width.
This forward-backward asymmetric piece, $b_{\ell}\left(q^{2}\right)$, is sensitive to the lepton mass.
It is given by
\begin{align}
b_{\ell}\left(q^{2}\right)&=-\frac{\eta_{\mathrm{EW}}^2G_{F}^{2}\left|V_{ub}\right|^{2}|\mathbf{q}|}{64 \pi^{3} M_{B_c}^{2}}\left(1-\frac{m_{\ell}^{2}}{q^{2}}\right)^{2} \frac{m_{\ell}^{2}}{q^{2}} \times \nonumber \\
&\lambda (M_{B_c}^2, M_D^2, q^2)^{1/2} (M_{B_c}^2 - M_{D}^2) f_0 (q^2) f_+ (q^2) \label{eqn:formula_b_ell}
\end{align}
where $\lambda\left(x^{2}, y^{2}, z^{2}\right)=\left[x^{2}-(y-z)^{2}\right]\left[x^{2}-(y+z)^{2}\right]$.
In Fig.~\ref{fig:AFB}, we plot $b_{\ell}\left(q^{2}\right)$ for the cases $\ell = \mu, \tau$.
The shape of $b_{\ell}\left(q^{2}\right)$ differs between the two cases.
To exhibit in more detail the low-$q^2$ behaviour of $b_{\mu}\left(q^{2}\right)$, we plot separately the regions $q^2 < 1 \; \mathrm{GeV}^2$ and $1 \; \mathrm{GeV}^2 \leq q^2$.

\subsection{Observables for $B_c^+ \to D_s^+ \ell^+ \ell^-$ and $B_c^+ \to D_s^+ \nu \overline{\nu}$} \label{sec:BcDs_obs}

Like $B \to K \ell^+ \ell^-$, the process $B_c^+ \to D_s^+ \ell^+ \ell^-$ is a rare decay mediated by the loop-induced $b \to s$ transition.
Here, we follow nomenclature commonly used for $B \to K \ell^+ \ell^-$ as in~\cite{Becirevic:2012fy} and replace the initial and final mesons in the $B \to K$ formulae with $B_c$ and $D_s$ respectively.
% contains the details which are conveniently summarised in~\cite{Bouchard:2013eph}.
We calculate observables for $B_c^+ \to D_s^+ \ell^+ \ell^-$ from our form factors $f_{0,+,T}^s$ ignoring small non-factorisable contributions at low $q^2$~\cite{Khodjamirian:2012rm,Hambrock:2015wka}.

We use the same value for $\left|V_{t b} V_{t s}^{*}\right| = 0.0405(8)$~\cite{Charles:2004jd} and the Wilson coefficients in~\cite{Bouchard:2013eph}.
The Wilson coefficients used in~\cite{Bouchard:2013eph} are quoted st scale $4.8 \; \mathrm{GeV}$.
%Therefore, we need to run our tensor form factor, shown in Fig.~\ref{fig:final_ffs} at scale $4.8 \; \mathrm{GeV}$, down to scale $4.2 \; \mathrm{GeV}$.
%We achieve this by following~\cite{Hatton:2020vzp}, in effect multiplying our form factor $f_T$ (at scale $4.8 \; \mathrm{GeV}$) by a factor of $1.00933(16)$.

The determination of the branching fraction includes effective Wilson coefficients expressed in terms of the functions $h(q^2, m_c)$ and $h(q^2, m_b)$ that depend on the $c$ and $b$
 pole masses. 
We take $m_c$ and $m_b$ in the $\overline{\mathrm{MS}}$ scheme to be $1.2757(84) \; \mathrm{GeV}$~\cite{Lytle:2018evc} and $4.209(21) \; \mathrm{GeV}$~\cite{Hatton:2021syc} respectively each at their own scale.
%For the pole masses of the charm and bottom quark, we use the relationship between the pole mass and the mass in the $\overline{\mathrm{MS}}$ scheme given in~\cite{Melnikov:2000qh}.
Using the 3-loop expression in Eq.~(12) of~\cite{Melnikov:2000qh} that relates the pole mass to the mass in the $\overline{\mathrm{MS}}$ scheme, we find the values $1.68 \; \mathrm{GeV}$ and $4.87 \; \mathrm{GeV}$ for the pole mass of the charm and bottom quarks respectively, each taken with an uncertainty of $200 \; \mathrm{MeV}$ to account for the presence of a renormalon in the pole mass~\cite{Beneke:1994sw} suffered by the perturbation series in the expression in~\cite{Melnikov:2000qh}.

%We take $m_c$ and $m_b$ in the $\overline{\mathrm{MS}}$ scheme to be $1.2757(84) \; \mathrm{GeV}$~\cite{Lytle:2018evc} and $4.209(21) \; \mathrm{GeV}$~\cite{Hatton:2021syc} respectively.
%%For the pole masses of the charm and bottom quark, we use the relationship between the pole mass and the mass in the $\overline{\mathrm{MS}}$ scheme given in~\cite{Melnikov:2000qh}.
%Using the 3-loop expression in~\cite{Melnikov:2000qh} that relates the pole mass to the mass in the $\overline{\mathrm{MS}}$ scheme, we find the values $1.684 \; \mathrm{GeV}$ and $4.874 \; \mathrm{GeV}$ for the pole mass of the charm and bottom quarks respectively, each taken with an uncertainty of $200 \; \mathrm{MeV}$ to account for the presence of a renormalon in the pole mass~\cite{Beneke:1994sw} suffered by the perturbation series in the expression in~\cite{Melnikov:2000qh}.
%The scale appearing in the function $h$ in~\cite{Becirevic:2012fy} is taken as $m_b$ in the $\overline{\mathrm{MS}}$ scheme.

\begin{figure}
	\centering
	\includegraphics[width=0.5\textwidth]{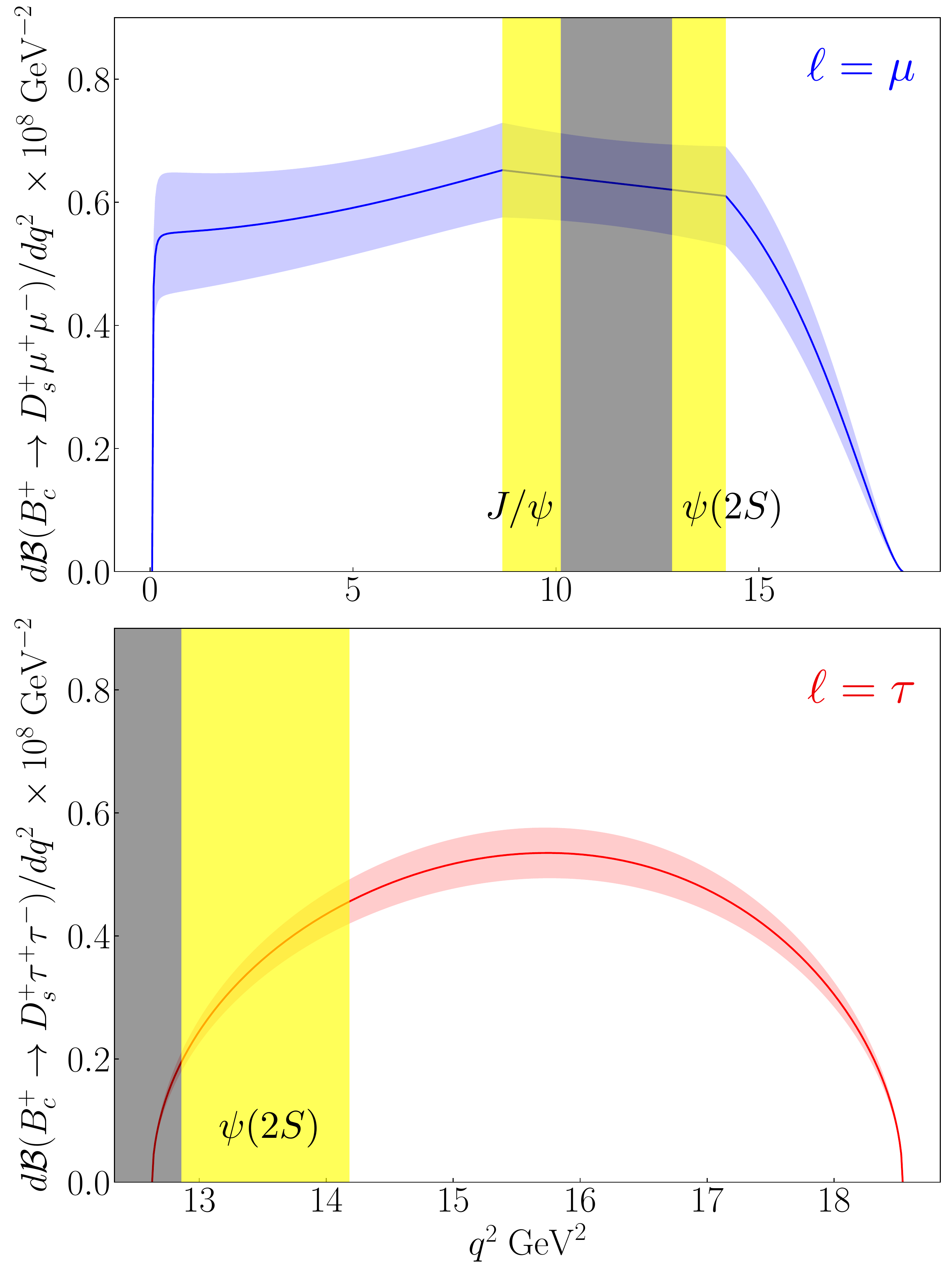}
	\caption{Plot of the $B_c^+ \to D_s^+ \ell^+ \ell^-$ differential branching ratio for $\ell = \mu$ (top) and $\ell = \tau$ (bottom) in the final state. The yellow bands show regions where charmonium resonances (not included in our calculation) could have an impact. The grey band is between the two yellow regions labelling the charmonium resonances. Through the yellow and grey bands, we interpolate the function $d\mathcal{B}_{\mu}/dq^2$ linearly when integrating to find the branching fraction and related quantities.} 
	\label{fig:dbdqsqbcdslepton}
\end{figure}

In Fig.~\ref{fig:dbdqsqbcdslepton} we plot the differential branching fractions for the cases $\ell = \mu, \tau$ for the physical range $4m_{\ell}^2 < q^2 < (M_{B_c} - M_{D_s})^2$.
These are constructed from the expressions in~\cite{Becirevic:2012fy} for $B\to K$.
%The discontinuous derivative in the curve at $q^2 = 4m_c^2$ originates from the $h(q^2, m_c)$ contribution to $Y(q^2)$ where $C^9_{\mathrm{eff}} = C^9 + Y(q^2)$.
%The function $h(q^2, m)$ is a piecewise-defined function with a different form for $q^2 < 4m^2$ and $q^2 \geq 4m^2$.
The yellow bands span across $\sqrt{q^2} = 2.956-3.181 \; \mathrm{GeV}$ and $3.586-3.766 \; \mathrm{GeV}$.
These regions are the same as in~\cite{Aaij:2012cq} and they represent veto regions which largely remove contributions from charmonium resonances via intermediate $J/\psi$ and $\psi(2S)$ states.
The effects of charmonium resonances are not included in our differential branching fractions.
For $d \mathcal{B}_{\mu} / dq^2$  between $\sqrt{q^2} = 2.956$ and $\sqrt{q^2} = 3.766$, we interpolate the function linearly as performed in~\cite{Du:2015tda} for the $B \to K$ branching fraction.

On integrating with respect to $q^2$, we report the ratio
\begin{align}
R_{\ell_2}^{\ell_1}\left(q_{\text {low }}^{2}, q_{\text {high }}^{2}\right) = \frac{\int_{q_{\text {low }}^{2}}^{q_{\text {high }}^{2}} d q^{2} d \mathcal{B}_{\ell_1} / d q^{2}}{\int_{q_{\text {low }}^{2}}^{q^2_{\text {high }}}{d q^{2}} d \mathcal{B}_{\ell_2} / d q^{2}}
\end{align}
for different choices of final-state lepton $\ell_{1,2}$ and integration limits $q^2_{\mathrm{low}}, q^2_{\mathrm{high}}$.
We find that
\begin{align}
	R_{e}^{\mu}\left(4 m_{\mu}^{2}, q_{\max }^{2}\right) &=1.00203(47) \\
	R_{e}^{\mu}\left(1 \; \mathrm{GeV}^2, 6 \; \mathrm{GeV}^2\right) &= 1.00157(52) \label{eqn:mu_e_lower_ratio}\\
	R_{e}^{\mu}\left(14.18 \;\mathrm{GeV}^{2}, q_{\max }^{2}\right) &= 1.0064(12) \\
	R_{e}^{\tau}\left(14.18 \;\mathrm{GeV}^{2}, q_{\max }^{2}\right) &= 1.34(13) \\
	R_{\mu}^{\tau}\left(14.18 \; \mathrm{GeV}^{2}, q_{\max }^{2}\right) &= 1.33(13)
\end{align}
where $q^2_{\mathrm{max}} = (M_{B_c} - M_{D_s})^2$.
The latter three ratios above involve the differential decay widths above the veto region associated with the resonance from $\psi(2S)$.
The ratio in Eq.~\eqref{eqn:mu_e_lower_ratio} lies beneath the $J/\psi$ veto region and above $q^{2} \lesssim 1 \; \mathrm{GeV}^{2}$ where effects from $u \bar{u}$ resonances could have an impact; these are not included in our calculation.
We tabulate in Table~\ref{tab:integrated_Gamma_BcDs} integrals of differential branching fractions for these ranges of $q^2$.
%Since we do not consider effects of charmonium and $u \bar{u}$ resonances, our branching fractions here may differ from values measured in experiment.
As in the case $B_c^+ \to D^0 \ell^+ \nu_{\ell}$, the ratio of widths with $\ell = \tau$ and $\ell = \mu$ in the final state
\begin{align}
	\frac{\Gamma(B_c^+ \to D_s^+ \tau^+ \tau^-)}{\Gamma(B_c^+ \to D_s^+ \mu^+ \mu^-)} = 0.245(20)
\end{align}
has reduced error.
%update1this
%
\begin{table}
	\centering
	\caption{For $B_c^+ \to D_s^+ \ell^+ \ell^-$, we give values for $d\mathcal{B}/dq^2 \times 10^{7}$ integrated with respect to $q^2$ over the given ranges $(q^2_{\mathrm{low}}, q^2_{\mathrm{high}})$ in $\mathrm{GeV}^2$ for each of the cases $\ell = e,\mu,\tau$.
		We take the lifetime of $B_c$ meson to be $513.49(12.4) \: \mathrm{fs}$~\cite{Aaij:2014gka}. Note that these results do not include effects from charmonium or $u \overline{u}$ resonances. }
	\begin{tabular}{c | c c c }
		\hline\hline
		decay mode & $(4 m_{\ell}^2, q^2_{\mathrm{max}})$ & $(1, 6)$ & $(14.18, q^2_{\mathrm{max}})$  \\ [0.1ex]
		\hline
		$B_c^+ \to D_s^+ e^+ e^-$ & $1.00(11)$ & $0.285(41)$ & $0.146(22)$\\
		$B_c^+ \to D_s^+ \mu^+ \mu^-$ & $1.00(11)$ & $0.286(41)$ & $0.147(22)$ \\
		$B_c^+ \to D_s^+ \tau^+ \tau^-$ & $0.245(18)$ & --- & $0.195(14)$ \\
		\hline\hline
	\end{tabular}
	\label{tab:integrated_Gamma_BcDs}
\end{table}
%

%update1this
%

In the low-$q^2$ region $1 \; \mathrm{GeV}^2$ to $6 \; \mathrm{GeV}^2$, we find that the ratio of integrated branching fractions for $B_c^+ \to D_s^+ \mu^+ \mu^-$ and $B_{c}^{+} \rightarrow J / \psi \mu^{+} \nu_{\mu}$ to be
%update1this
%muon
%BcDs = gv.gvar(2.86195941544e-08, 4.0685749854e-09)
%integrated_F_1to6.mean = 0.0252928644133
%integrated_F_1to6.sdev = 0.00190463266656
%BcJpsi = gv.gvar(0.00453751347727, 0.000468397774132)
%
\begin{align}
\frac{ \int_{1 \; \mathrm{GeV}^2}^{6 \; \mathrm{GeV}^2}d q^{2} \frac{ d \mathcal{B} (B_c^+ \to D_s^+ \mu^+ \mu^-)}{ d q^{2}}}{ \int_{1 \; \mathrm{GeV}^2}^{6 \; \mathrm{GeV}^2}d q^{2} \frac{d \mathcal{B} (B_{c}^{+} \rightarrow J / \psi \mu^{+} \nu_{\mu})}{ d q^{2}}} = 6.31(90)(65) \times 10^{-6}.
\end{align}
The first error is from the numerator and the second error is from the denominator which we compute using the form factors for $B_{c}^{+} \rightarrow J / \psi \mu^{+} \nu_{\mu}$ from~\cite{Harrison:2020gvo}.
As in~\cite{Harrison:2020gvo}, we take $\left|V_{c b}\right|=41.0(1.4) \times 10^{-3}$ ~\cite{pdg20} from an average of inclusive and exclusive determinations, scaling the uncertainty by $2.4$ to allow for their inconsistency.

\begin{figure}
	\centering
	\includegraphics[width=0.5\textwidth]{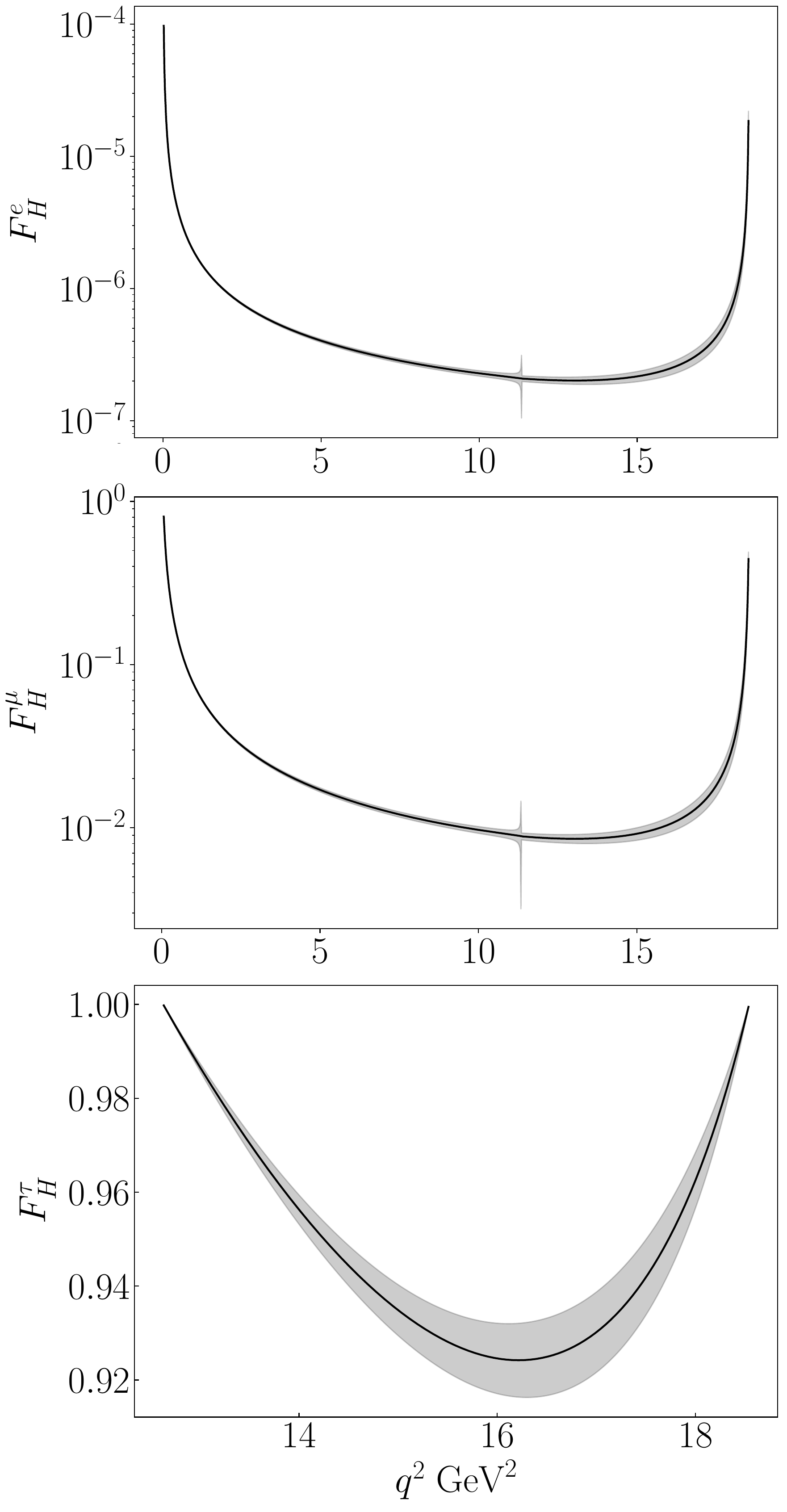}
	\caption{From top to bottom, we show plots of the flat-terms $F_H^{\ell}$ for each of $\ell = e, \mu, \tau$ respectively. We use a log-scale for the cases $\ell = e, \mu$. Error bands are presented though the errors are small due to the correlations in the construction of the flat-term.}
	\label{fig:flatTerms}
\end{figure}

Next, we show in Fig.~\ref{fig:flatTerms} the \lq flat-term\rq \:$F_H^{\ell}$, first introduced in~\cite{Bobeth:2007dw} in the context of $B \to K$.
This term appears as a constant in the angular distribution of the decay width.
Taking the same parametrisation of the decay width as in Eq.~\eqref{eqn:dGamma_dqSqdcostheta}, then performing the integration with respect to $q^2$, we have
\begin{align}
	\frac{1}{\Gamma_{\ell}} \frac{d \Gamma_{\ell} (\cos \theta)}{d \cos \theta}=\frac{3}{4}\left(1-F_{H}^{l}\right)\left(1-\cos ^{2} \theta\right)+\frac{1}{2} F_{H}^{\ell}+A_{\mathrm{FB}}^{\ell} \cos \theta
\end{align}
where
\begin{align}
	A_{\mathrm{FB}}^{\ell} &=  \frac{1}{\Gamma_{\ell}} \int_{q^2_{\mathrm{min}}}^{q^2_{\mathrm{max}}}  dq^2 \hspace{1mm} b_{\ell} (q^2),  \\
	F_{H}^{\ell } &= \frac{2}{\Gamma_{\ell}} \int_{q^2_{\mathrm{min}}}^{q^2_{\mathrm{max}}}  dq^2 \hspace{1mm} \left( a_{\ell} (q^2) + c_{\ell} (q^2) \right)
\end{align}
and we define
\begin{align}
F_{H}^{\ell} (q^2) = \frac{2\left(a_{\ell} (q^2) + c_{\ell} (q^2) \right)}{2a_{\ell} (q^2) + \frac{2}{3}c_{\ell} (q^2)}.
\end{align}
%
 %where $\Gamma_{l}$ is the decay width with $l^+ l^-$ in the final state, $\theta$ is the angle measured in the dilepton rest frame between the initial $B_c$ meson and the lepton $\ell$ in the final state, and $A_{\mathrm{FB}}^l$ is a constant parametrising the forward-backward asymmetry.
 The flat-term $F_{H}^{\ell }$ may be sensitive to contributions from new physics since it is small according to the Standard Model.
This quantity is a ratio of combinations of the form factors and uncertainties are much less than those exhibited by the raw form factors or branching fractions.
%The flat-term is expected to be sensitive to new physics.

%update1this
%
\begin{figure}
	\centering
	\includegraphics[width=0.5\textwidth]{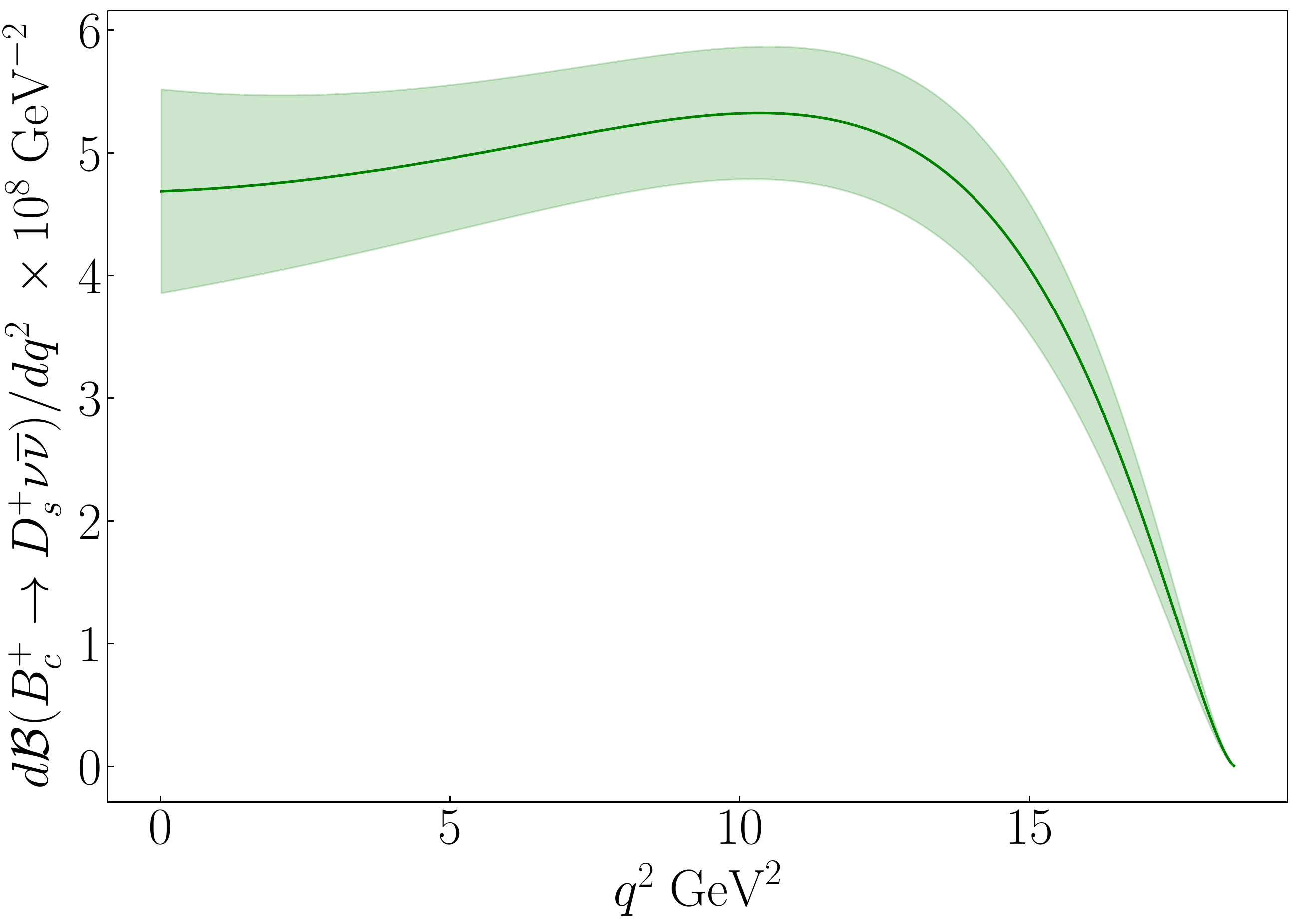}
	\caption{The differential branching fraction for $B_c^+ \to D_s^+ \nu \overline{\nu}$ as a function of $q^2$.}
	\label{fig:diff_bf_BcDs}
\end{figure}

We determine the differential branching fraction for $B_c^+ \to D_s^+ \nu \overline{\nu}$ using the expressions for the $B \to K \nu \overline{\nu}$ case in~\cite{Buras:2014fpa,Altmannshofer:2009ma}.
The differential branching fraction, summing over the three neutrino flavours, is 
\begin{align}
&\frac{d \mathcal{B}(B_c^+ \to D_s^+ \nu \overline{\nu})}{d q^{2}}= \tau_{B_c}\left|V_{t b} V_{t s(d)}^{*}\right|^{2} \frac{G_{F}^{2} \alpha^{2}}{32 \pi^{5}} \frac{X_{t}^{2}}{\sin ^{4} \theta_{W}} \nonumber \\
& \hspace{40mm} \times \left|\boldsymbol{q}\right|^{3} f_{+}^{2}\left(q^{2}\right)
\end{align}
which we plot in Fig.~\ref{fig:diff_bf_BcDs}.
We take $X_t = 1.469(17)$~\cite{Brod:2010hi} and $\alpha^{-1} (M_Z) = 127.952(9)$~\cite{pdg20}.
Integrated from $q^2 =0$ to $q^2_{\mathrm{max}}$, we find the branching fraction
%mean = 8.22548566879e-07
%sdev = 8.51983675664e-08
\begin{align}
	\mathcal{B} \left( B_c^+ \to D_s^+ \nu \overline{\nu} \right) = 8.23(85) \times 10^{-7}.
\end{align}
%
%As for the branching fraction $\mathcal{B}(B_c^+ \to D_s^+ \ell^+ \ell^-)$, we take $\left|V_{t b} V_{t s}^{*}\right| = 0.0405(8)$~\cite{Charles:2004jd}.
There are no issues from charmonium resonances or non-factorisable pieces in this case.
Since $m_{\tau} > M_{D_s}$, there is also no long-distance contribution for the $\tau$ case (unlike for $B \to K \nu_{\tau} \overline{\nu}_{\tau}$). 
We find the ratio of branching fractions
% BcDs = gv.gvar(8.22548566879e-07, 8.51983675664e-08)
% BcJpsi = gv.gvar(0.0149838781519, 0.00150552205378)
%
\begin{align}
	\frac{\mathcal{B} \left( B_c^+ \to D_s^+ \nu \overline{\nu} \right)}{\mathcal{B} \left( B_{c}^{+} \rightarrow J / \psi \mu^{+} \nu_{\mu} \right)} = 5.49(57)(55) \times 10^{-5}.
\end{align}
The first error is from the numerator and the second error is from the denominator which we compute using the form factors for $B_{c}^{+} \rightarrow J / \psi \mu^{+} \nu_{\mu}$ from~\cite{Harrison:2020gvo}.
%again using the form factors for $B_{c}^{+} \rightarrow J / \psi \mu^{+} \nu_{\mu}$ from~\cite{Harrison:2020gvo}.

\section{Future prospects: improving accuracy of the form factors}\label{sec:prospects}

%In this final main section, we present investigations into achieving more accurate determinations of the form factors beyond our calculation is the preceding sections.
We consider two extensions to our current strategy to improve uncertainties in the future: the addition of a finer lattice and the inclusion of the spatial vector current.
%These methods can be applied to other heavy-to-light semi-leptonic weak decays in the future.

\subsection{Simulating with a physically massive $b$ quark on the exafine lattice}  \label{sec:ex5}

%To reduce the error in our form factors associated with the heavy quark mass, we could simulate at heavy quark masses closer to the mass of the physical $b$ quark.
%However, the lattice spacings of the sets in described in Table~\ref{LattDesc1} are such that $am_h = am_b$ is too large for discretisation effects to be reasonably under control.
%On all sets, the largest heavy quark mass that we use is $am_h = 0.8$.

We carry out the first heavy-to-light decay analysis on the exafine gluon field configurations, with size $N_x^3 \times N_t = 96^3 \times 288$ and lattice spacing $a \approx 0.033 \; \mathrm{fm}$.
These configurations are finer than all the sets used in our calculation thus far.
The lattice spacing is such that $am_b \approx 0.625$, therefore we are able to simulate with physically heavy $b$ quarks on this lattice with reasonably small discretisation effects associated with $am_h$.

%We carry out the first heavy-to-light decay analysis on the exafine lattice.
%We proceed with the decay $B_c \to D_s$ only.
Computations on the exafine lattices are expensive due to the large size, $N_x^3 \times N_t$.
Hence, since these investigations are preliminary, we restrict the calculation to $B_c \to D_s$ and compute with a small selection of parameters on 100 configurations, each with 4 different positions of random wall source.
We take $am_h = 0.35, 0.625$ and calculate with three different momenta (including zero-recoil), plus a further larger momentum for $am_h = 0.625$: a three-momentum transfer of roughly $2.8 \; \mathrm{GeV}$.
%We wish to assess the range of $q^2$ that we can cover without suffering from large signal-to-noise degradation that would render the data uninformative on the fit.
%Experience gathered here will guide the choice of momenta in future calculations.
%The propagators at the mass of the strange quark that we generate are saved for future use.
%update1this
%
\begin{figure}
	\begin{center}
		\includegraphics[width=0.5\textwidth]{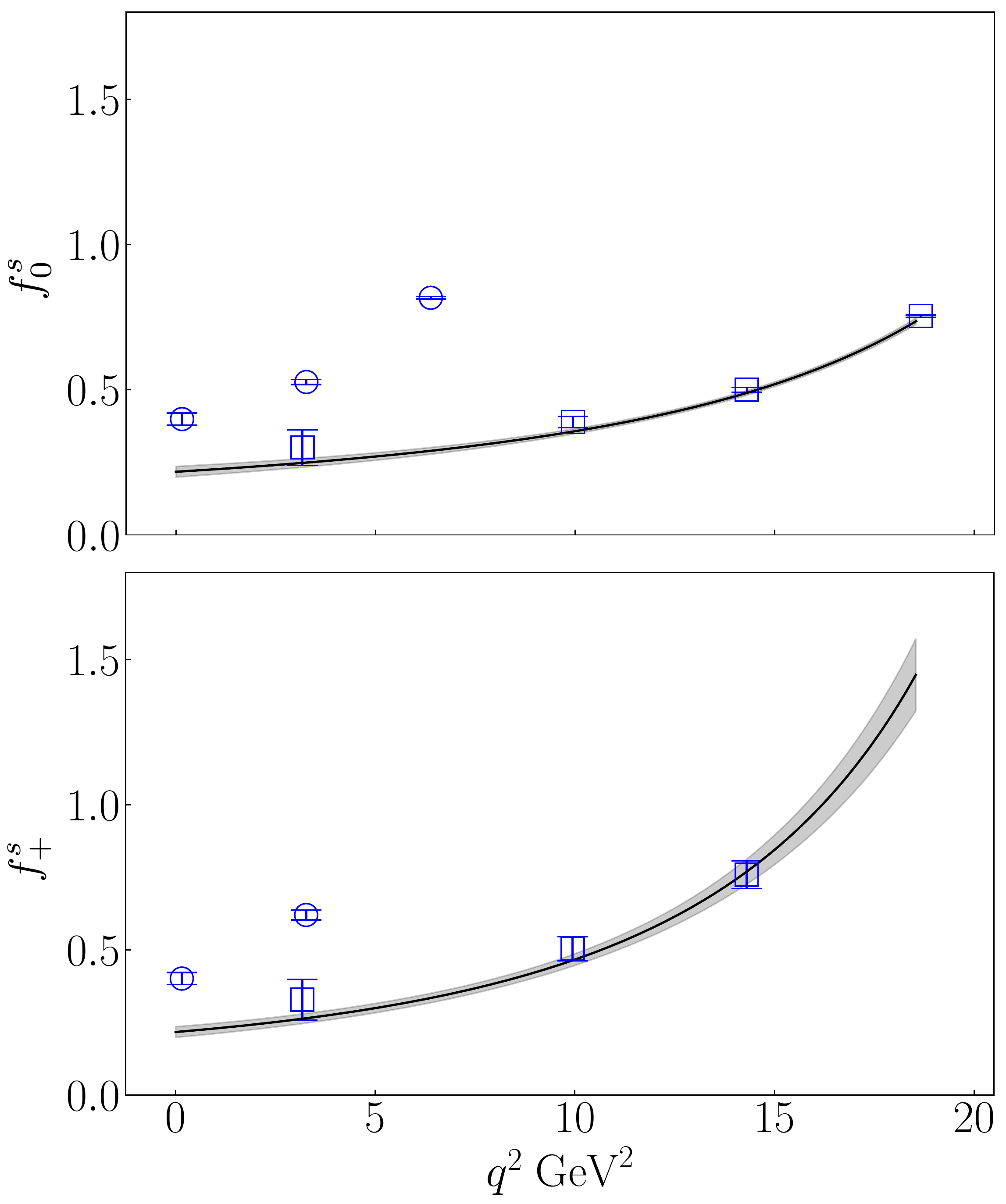}
		\caption{We show data from the exafine lattice shown in blue with squares, denoting $am_h = 0.625$, and circles, denoting $am_h = 0.35$.
		Alongside the exafine data, we show the fits of the form factors $f_0^s$ (top) and $f_+^s$ (bottom) from ultrafine sets to coarser sets as presented in the upper plot of Fig.~\ref{fig:final_ffs}. The lattice data at $am_h = 0.625$ ($\approx am_b$) closely follows the fit curves.}
		\label{fig:ffs_with_ex5}
	\end{center}
\end{figure}

In Fig.~\ref{fig:ffs_with_ex5}, we show form factor results on the exafine lattice with these two masses along with our physical-continuum curve at $m_h = m_b$ derived from the coarser lattices (presented in Section~\ref{sec:results_ffs}).
The exafine data at $am_h = 0.625$ closely follow the physical-continuum curve.

Errors on the physical-continuum form factors from fits with and without the data from the exafine lattice are shown in Table~\ref{tab:ff_errors_ex5}.
%We also show errors with the largest 3-momentum data point on exafine excluded.
From this table, we see that errors are reduced by 15-25\% at zero-recoil on inclusion of data on the exafine lattice.
%We also see that fits with and without the largest 3-momentum data point are very similar.
%This point does not appear to offer much improvement on the errors.
%On inspection of the form factor data in Fig.~\ref{fig:ffs_with_ex5}, the error on the exafine data nearest to $q^2 = 0$ has an error greater than the points closer to $q^2_{\mathrm{max}}$, likely due to signal-to--noise degradation at large 3-momentum transfer.
%The $f_+$ form factor is less adversely affected than $f_0$.
%update1this
%
\begin{table}
	\centering
	\caption{Comparison of extremal values of the form factors in the physical-continuum limit. The second column gives results from our fit without any data points on the exafine lattice. The third column gives results using the same fit form but now including results on the exafine lattice.
	Errors are reduced from the second to third column.}
	\begin{tabular}{c | c c}
		\hline\hline
		 & w/o & with exafine \\ [0.1ex]
		\hline
		$f_{0,+}^s(0)$& $0.217(18)$ & $0.221(16)$ \\
		$f_0^s(q^2_{\mathrm{max}})$& $0.736(11)$ &$0.7383(91)$ \\
		$f_+^s(q^2_{\mathrm{max}})$& $1.45(12)$ & $1.433(97)$ \\
		\hline\hline
	\end{tabular}
	\label{tab:ff_errors_ex5}
\end{table}

Given our present statistics on the exafine lattice, we are able to cover at least half the range of $q^2$ with reasonable errors.
Reducing the uncertainties at lower $q^2$ values will require higher statistics;
% than is presently viable,
however data on exafine with $q^2 > q^2_{\mathrm{max}} / 2$ does give some error reduction at $q^2 = 0$.
%This suggests that the momentum dependence of the form factors is well-resolved by the data on the coarser lattices.

%\rough{\emph{[if we wanted to do the fit/extrap of $Z_T$, may want to include details of that here. I'll need to do this for my heavy-to-light decays where I'll be doing the tensor on ex5]}}

\subsection{Extracting $f_+$ from matrix elements of the spatial vector current} \label{sec:fplus_from_spatVec}

As can be clearly seen in Figs.~\ref{fig:ffs_BcDl},~\ref{fig:ffs_BcDs},~\ref{fig:Pfl} and~\ref{fig:Pfs} in Appendix~\ref{sec:ffs_fit_analysis}, the errors on the lattice data for $f_+$ near zero-recoil (maximum $q^2$) are much larger than the errors seen away from zero-recoil.
This is not because our extraction of the matrix elements $\langle D_{l(s)} | S_{\text{local}} | H_c \rangle $ and $\langle D_{l(s)} | V^{0}_{\text{local}} | H_c \rangle$ is especially imprecise at these momenta, but because we extract the form factor via Eq.~\eqref{fplusextract}.
The denominator in Eq.~\eqref{fplusextract} approaches zero as $q^2$ approaches$q^2_{\mathrm{max}}$.
However, $f_+$ is finite and analytic at $q^2_{\mathrm{max}}$, and so the numerator also vanishes at $q^2_{\mathrm{max}}$.
%\rough{\emph{[is there a better way of saying this?]}}
In practice, the smallness of both the numerator and denominator at large $q^2$ results in a large error for the extracted value of $f_+$.
As a consequence, the error on the final physical-continuum form factor $f_+$ is large, and certainly larger than the error on $f_0$ at zero-recoil.

We now propose and investigate a method to reduce the error on $f_+$ near zero-recoil.
For these purposes, we consider only the process $B_c \to D_s$.
As an alternative to extracting $f_+$ via Eq.~\eqref{fplusextract}, we set $\mu =i \neq 0$ in Eq.~\eqref{form factors} to find
\begin{align} \label{extract_fplus2}
f_{+}^s\left(q^{2}\right)=\frac{- q^2  Z_{V} \langle D_{s} | V^{i} | H_c \rangle / q^{i} +f_{0}^s \left(q^{2}\right) \left(M_{H_{c}}^{2}-M_{D_{s}^{2}}^{2}\right)}{q^2+M_{H_c}^2-M_{D_s}^2}
\end{align}
which, in addition to the matrix elements calculated in our existing setup, involves matrix elements of a spatial component of the vector current.

%We now design appropriate correlation functions for extraction of matrix elements of the spatial vector current.
%To begin this discussion, we refrain from specifying a particular vector current, and denote its multiplicative renormalisation factor by $Z_{V^i}$.
%As is typical when designing correlation functions that make exclusive use of valence quark in a staggered formalism, we have a choice as to the taste structure of each operator provided that the overall taste of the correlation function is conserved.
%In choosing which combination of operators to use, we are mindful of the following.
%Firstly, we wish to use local operators wherever possible since they generally lead to cleaner correlation functions and thus more precise extraction of the matrix elements.
%Secondly, we wish to minimise the cost of computing additional propagators beyond those calculated for the diagrams in Fig.~\ref{fig:3pt_diags}.
%Thirdly, we must have available the means to accurately and non-perturbatively renormalise the operator insertion that is used for the spatial vector current.
%With these guiding principles in mind, we now discuss two different constructions.

\begin{figure}
	\begin{center}
		\includegraphics[width=0.5\textwidth]{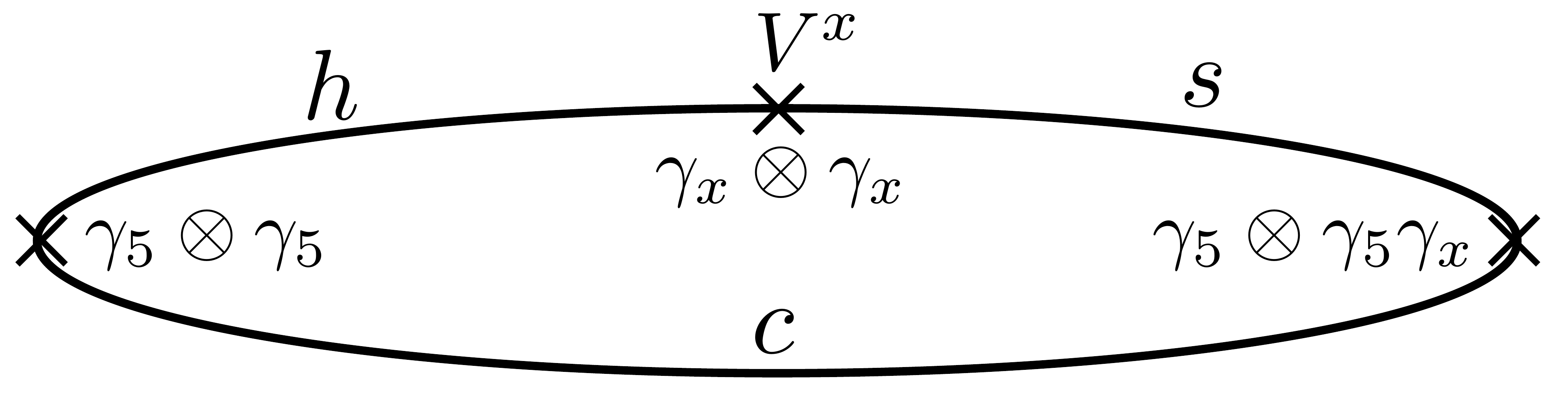}
		\caption{Diagrammatic representation of the three-point functions we calculate on set 1 for insertions of the local spatial vector current $\gamma_x \otimes \gamma_x$ as described in the text in Section~\ref{sec:fplus_from_spatVec}. Each operator is shown by a cross and is labelled by its description given in the spin-taste basis, whilst the lines represent lattice quark propagators as in Fig.~\ref{fig:3pt_diags}.}
		\label{fig:3pt_diag_Vx}
	\end{center}
\end{figure}

To achieve this we include the 3-point function given in Fig.~\ref{fig:3pt_diag_Vx} where the spatial vector current has spin-taste $\gamma_x \otimes \gamma_x$.
This correlation function has the advantage that the spatial vector current has the same multiplicative renormalisation as for the $\gamma_t \otimes \gamma_t$ insertion in the middle diagram of Fig.~\ref{fig:3pt_diags} (up to discretisation effects handled when fitting the form factor data).

\begin{figure}
	\begin{center}
		\includegraphics[width=0.5\textwidth]{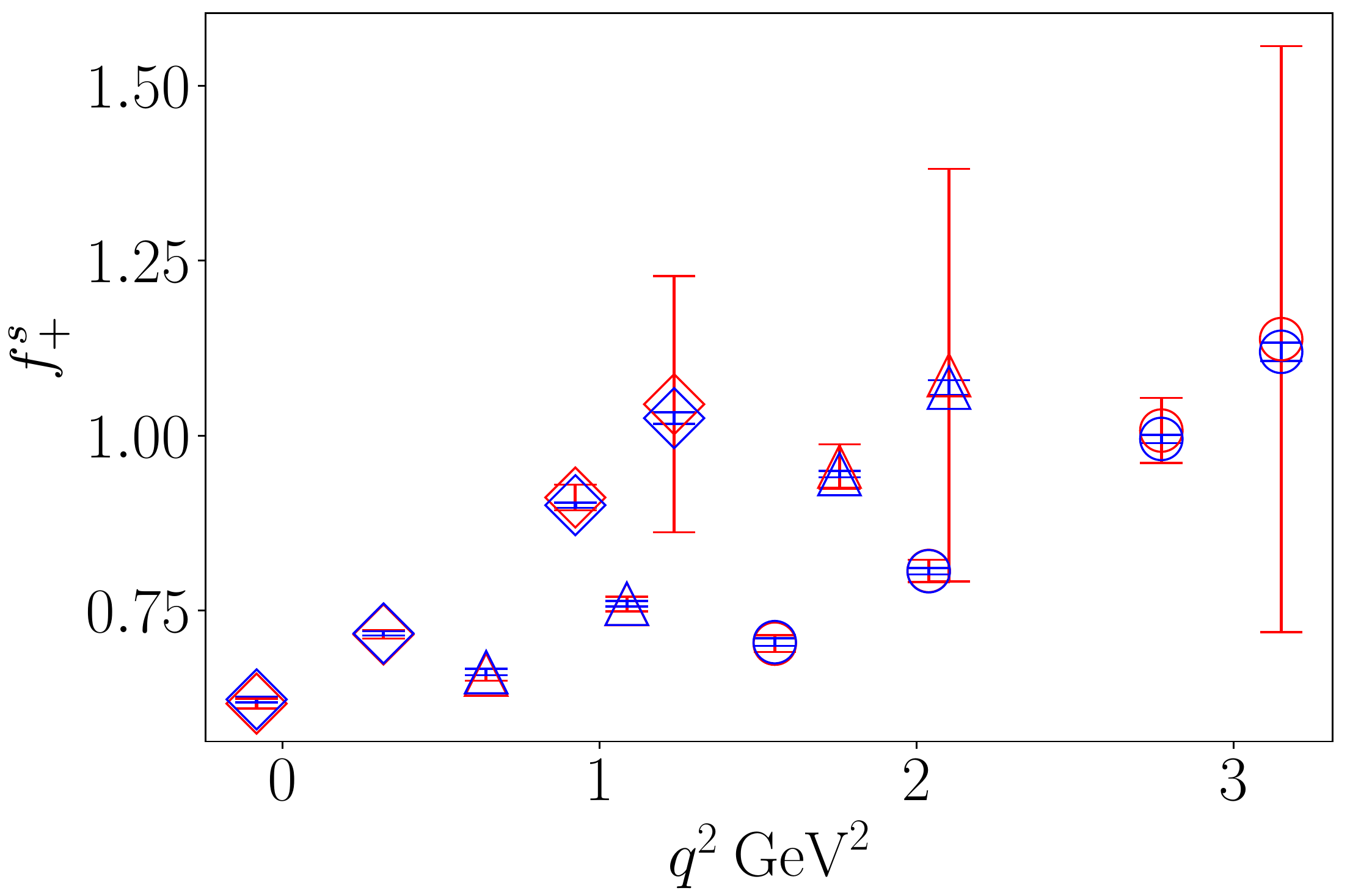}
		\caption{Lattice data on set 1 (see Table~\ref{LattDesc1}) for $f_+^s$ for different methods of extraction which we differentiate by colour.
		The blue points are the $f_+^s$ data extracted via Eq.~\eqref{extract_fplus2} using the local spatial vector current $\gamma_x \otimes \gamma_x$.
		The red points are the $f_+^s$ data extracted via Eq.~\eqref{fplusextract} using the local temporal vector current.
		The blue and red points agree very well at all $q^2$.
		Near zero-recoil, the errors on blue points are much smaller than the red points.}
		\label{fig:lattData_fplus_differentExtractions}
	\end{center}
\end{figure}

To demonstrate the effectiveness of extracting $f_+$ via Eq.~\eqref{extract_fplus2} versus the extraction of $f_+$ via Eq.~\eqref{fplusextract}, we apply the method outlined above to set 1 in Table~\ref{LattDesc1}.
%We take $Z_{V^i}^{\gamma_x \otimes 1}$ from Table VI of~\cite{Hatton:2019gha}.
In Fig.~\ref{fig:lattData_fplus_differentExtractions}, we show lattice data for $f_+$ from the different methods of extraction.
%Means and errors for the lattice data of $f_0$ and $f_T$ do not vary significantly between the different methods.
%

From Fig.~\ref{fig:lattData_fplus_differentExtractions}, we see that the different extractions are in excellent agreement and that the improvement in accuracy of the lattice data for $f_+$ by using Eq.~\eqref{extract_fplus2} is very large close to zero-recoil (maximum $q^2$).
By utilising the spatial vector current, we observe errors near zero-recoil comparable to those seen at momenta further away from maximum $q^2$.
Hence, using this approach on all lattices, we can expect an error on the physical-continuum $f_+$ form factor near zero-recoil comparable to that seen for $f_0$.
Therefore, by including matrix elements of the spatial vector current, we expect errors on our physical-continuum $f_+$ form factor at zero-recoil to reduce roughly by a factor of 2.

\section{Conclusions and Outlook}

For the first time from lattice QCD, we obtain the scalar and vector form factors $f_{0,+}$ for $B_c \to D_l$ and the scalar, vector and tensor form factors $f_{0,+,T}$ for $B_c \to D_s$ across the entire physical ranges of $q^2$ in the continuum limit with physical quark masses.
Our lattice QCD calculation uses four different lattices with three different lattice spacings, unphysically and physically massive light quarks, and a range of heavy quark masses.
Together the lattice data informs the limit of vanishing lattice spacing, physical $b$ quark mass, and physical (equal-mass) up and down quark masses.
The reader should consult Appendix~\ref{app:reconstruct_ff} for instructions on how to reconstruct our form factors.

%Our calculations presented here show the first attempt in calculating the form factors for these processes.
%The errors are under control throughout.
%The calculation of the form factors for $B_c^+ \to D^0 \ell^+ \nu_{\ell}$ demanded the generation of new sets of propagators at small quark masses which we can use for other heavy-to-light transitions in the future.
The error on the decay widths $\Gamma(B_c^+ \to D^0 \ell^+ \nu_{\ell})$ (see Table~\ref{tab:integrated_Gamma_BcDl}) from our form factors is similar to the error on the present determination of $V_{ub}$.
For the cases $\ell = e$ or $\mu$, the lattice error is $13\%$ larger than the error from $V_{ub}$, whereas, for $\ell = \tau$, the lattice error is nearly $20\%$ smaller than the error from $V_{ub}$.
The error on the form factors calculated here for $B_c \to D_s$ is smaller than that for $B_c \to D$ by up to a factor of 2 at small recoil.

Experimental observations are expected from LHC in the near future~\cite{Bediaga:2018lhg}.
In Sections~\ref{sec:BcD_obs} and~\ref{sec:BcDs_obs} we give results for a host of observables that can be compared to experiment.
In Section~\ref{sec:prospects} we demonstrate how the uncertainties in our calculation can be reduced in the future to complement experimental results as they improve.

\section*{ACKNOWLEDGEMENTS}

We thank Jonna Koponen, Andrew Lytle, William Parrott and Andre Zimermmane-Santos for making previously generated lattice propagators available for our use; we thank Daniel Hatton et al. for the calculation of $Z_T$ in~\cite{Hatton:2020vzp}, and Chris Bouchard, Judd Harrison, Peter Lepage and William Parrott for useful discussions.
We are grateful to the MILC collaboration for making publicly available their gauge configurations and their code MILC-7.7.11 \cite{MILCgithub}.
This work was performed using the Cambridge Service for Data Driven Discovery (CSD3), part of which is operated by the University of Cambridge Research Computing on behalf of the STFC DiRAC HPC Facility.
The DiRAC component of CSD3 was funded by BEIS capital funding via STFC capital grants ST/P002307/1 and ST/R002452/1 and STFC operations grant ST/R00689X/1.
DiRAC is part of the National e-Infrastructure.
We are grateful to the CSD3 support staff for assistance.
This work has been partially supported by STFC consolidated grant ST/P000681/1.

%===========================================================================

\appendix

\section{Correlator fitting analysis} \label{sec:fitting_analysis}

\subsection{Method} \label{sec:corrfit_method}

As described in Section~\ref{sec:fit_correls}, we fit our two- and three- point correlations functions to the fit forms given in Eqs.~\eqref{corrfitform_2pt} and~\eqref{corrfitform_3pt}.
We minimise the usual $\chi^2$
\begin{align} \label{eqn:unaug_chi2}
	\chi^2 &= \sum_{i,j} \big(f(x_i;p)-y_i\big) (\sigma^y)^{-2}_{ij}\big(f(x_j;p)-y_j\big)
\end{align}
with the additional piece
\begin{align}
	\chi_{\text{prior}}^2 &= \sum_a \Bigg(\frac{p_a-p_a^{\text{prior}}}{\sigma_{a}}\Bigg)^2
\end{align}
with respect to the fit parameters $p$, where $f (x_i;p)$ is the corresponding fit function with parameters $p$ (functions of the amplitudes, energies and matrix elements), $y$ is the data, and the (estimated) covariance matrix $\sigma^y$ is
\begin{align} \label{eqn:est_cov}
	\sigma_{ij}^y &= \frac{\overline{f (x_i;p) f (x_j;p)} -  \overline{f (x_i;p)} \hspace{2mm} \overline{f (x_j;p)}}{N_s(N_s-1)}.
\end{align}
The prior distribution for the parameter $p_a$ in the fit function $f(x_i; p)$ is the normal distribution $\mathcal{N}(p_a^{\text{prior}}, \sigma_a)$.
Therefore, the function to be minimised is $\chi_{\text{aug}}^2 = \chi^2 + \chi_{\text{prior}}^2$~\cite{Lepage:2001ym,Hornbostel:2011hu,Bouchard:2014ypa}.

The covariance matrix $\sigma^y$ of the correlation function data is very large and so small eigenvalues of the covariance matrix are underestimated~\cite{Michael:1993yj, Dowdall:2019bea}, causing problems when carrying out the inversion of $\sigma^y$ in Eq.~\eqref{eqn:unaug_chi2} to find $\chi^2$.
This is overcome by using an SVD (singular-value decomposition) cut; any eigenvalue of the covariance matrix smaller than some proportion $c$ of the biggest eigenvalue $\lambda_{\text{max}}$ is replaced by $c\lambda_{\text{max}}$.
By carrying out this procedure, the covariance matrix becomes less singular.
These eigenvalue replacements will only inflate our final errors, hence this strategy is conservative.
The $\chi^2 / \mathrm{d.o.f.}$ values are affected by the SVD cut, demonstrated in Appendix D of~\cite{Dowdall:2019bea}.

Priors for ground state energies, amplitudes and matrix elements ($V_{\mathrm{nn},00}$) are motivated by plateaus in plots of effective quantities.
% are broadly identified to motivate a sensible prior.
For example, a straightforward effective energy can be constructed from a 2-point correlation function as
\begin{align}
	%aE_{\text{sim,eff}} &= -\frac{1}{2} \log \Big( \frac{C(t)}{C(t-2)} \Big). \label{effsimenergy}
	aE_{\text{eff}} &= -\log \Big( \frac{C_{\mathrm{2pt}}(t)}{C_{\mathrm{2pt}}(t-1)} \Big). \label{effsimenergy}
\end{align}
and the effective simulation amplitude
\begin{align}
	%a[0]_{\text{sim,eff}} &= \sqrt{\frac{C(t)+C(t-2)}{2\cosh(E)} e^{E(t-1)}}. \label{effsimamplitude}
	a_{\text{eff}} &= \sqrt{C_{\mathrm{2pt}}(t)e^{aE_{\text{eff}}(t)}}. \label{effsimamplitude}
\end{align}
%
%It is sometimes helpful to use alternative expressions to Equations~\ref{effsimenergy} and~\ref{effsimamplitude} in the case where oscillatory contributions are large.
%For example, similar expressions can be derived from the 2-point correlation function data separated by two timeslices in order to suppress the oscillations.

Priors associated with the oscillating and excited states are informed by our previous experiences.
From expectations of QCD, the energy splittings between excited states are taken as $a\Lambda_{\mathrm{QCD}} \times 2(1)$ where $\Lambda_{\text{QCD}}$ is taken to be $500 \text{ MeV}$.
The prior for the energy of the lowest lying oscillating state is given a prior twice as wide as the prior for the energy of the non-oscillating ground state.
The log of the amplitudes for the oscillating states and the remaining non-oscillating states are given priors of $-2.3(4.6)$.
%The log of the energy difference between excited states are given priors of $a\Lambda_{\text{QCD}} \times 2.0(1.5)$ where $\Lambda_{\text{QCD}}$ is taken to be $500 \text{ MeV}$.
Finally, $V_{\mathrm{nn},ij}$ for $i.j$ other than $i=j=0$ are given priors of $0(1)$ for the case of insertions of the scalar density and temporal vector current, and $0.0(5)$ for the tensor current insertion.

A variety of different fits are carried out with different SVD cuts, numbers of exponentials, and trims of correlator data at early and late times.
Results from these fits are inspected in Appendix~\ref{sec:stab_corrfit}.
Insensitivity to these choices is observed thus demonstrating stable and robust determination of the matrix elements.
The SVD cuts considered for each lattice are based around the suggested cut given by the \texttt{svd\_diagnosis} tool within the \textit{corrfitter} package~\cite{corrfitter}.

\subsection{Energies and amplitudes} \label{sec:sim_energies_amps}

As described in Appendix~\ref{sec:corrfit_method}, plots of effective energies and amplitudes from Eqs.~\eqref{effsimenergy} and~\eqref{effsimamplitude} are inspected to guide the selection of suitable priors for the non-oscillating ground states.
The ground state energies from the fit are always within their prior distribution and the error from the fit is always at least considerably smaller than the error on the prior.

\begin{figure*} 
	\centering
	\includegraphics[width=1.0\textwidth]{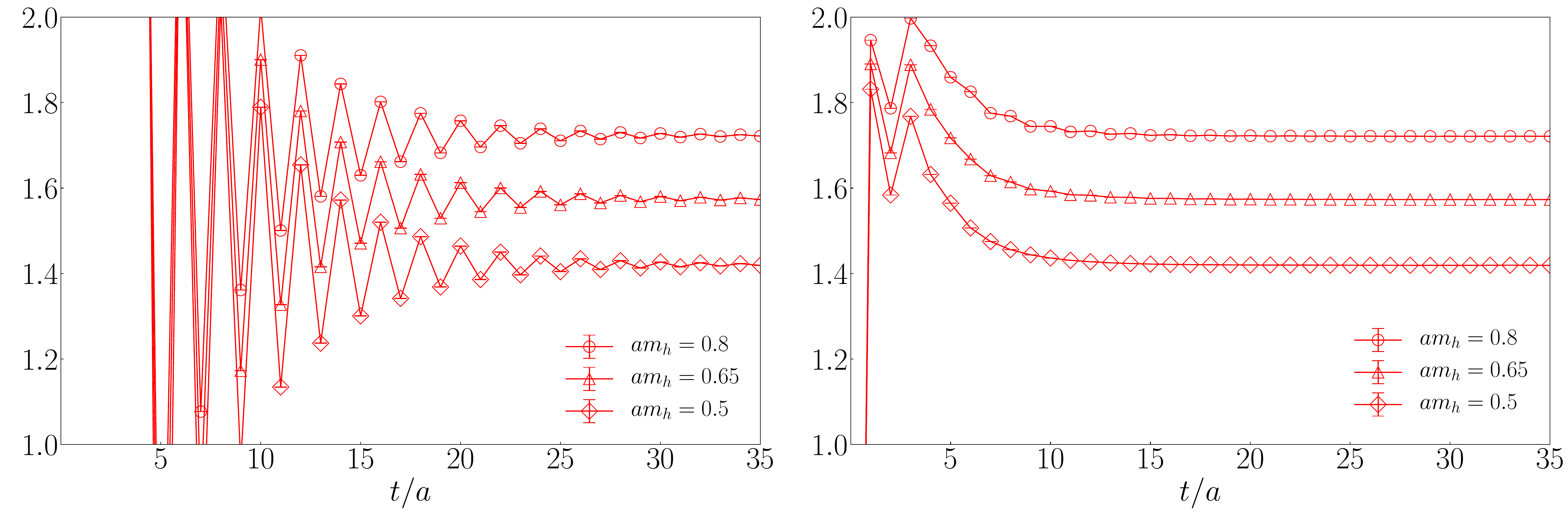}
	\caption{Plots of effective energies for the pseudoscalar heavy-charm meson at each $am_h$ value for set 1. The left plot shows the $\gamma_5 \gamma_t \otimes \gamma_5 \gamma_t$ meson. The right plot shows the $\gamma_5 \otimes \gamma_5$ meson.}
	\label{HsEffEnergy}
\end{figure*}

\begin{figure*} 
	\centering
	\includegraphics[width=1.0\textwidth]{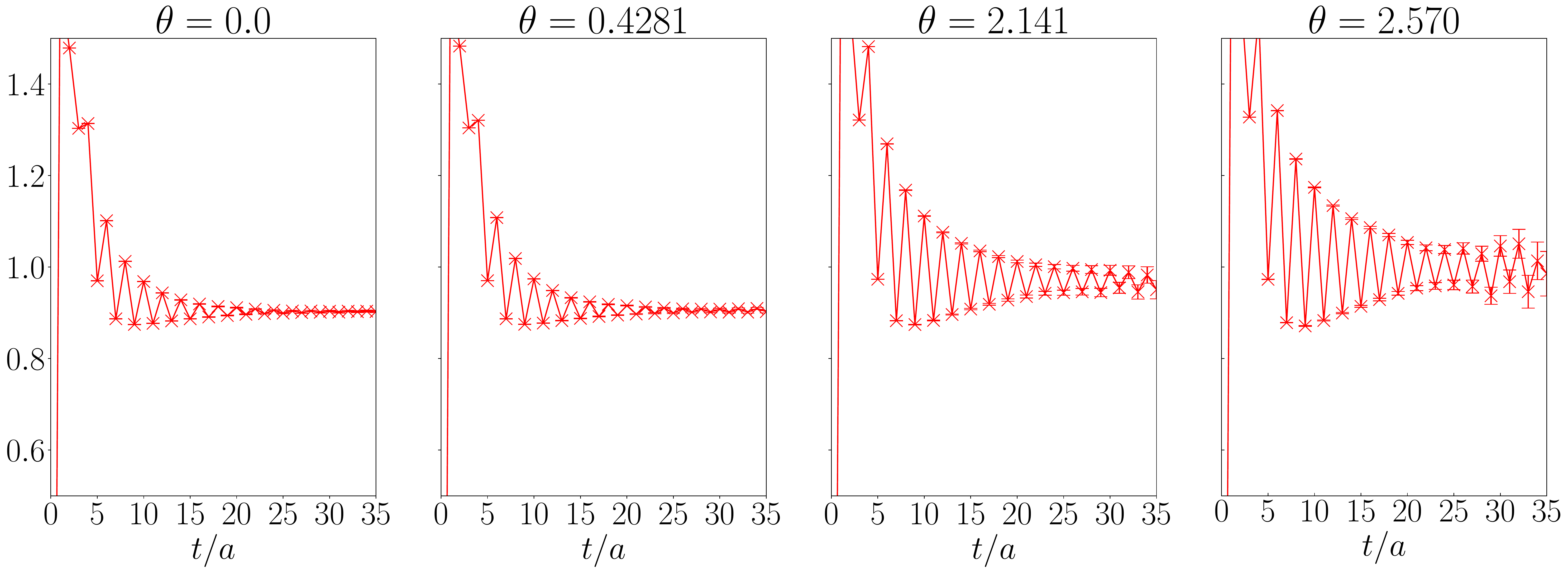}
	\caption{Plots of effective energies for the $D_s$ meson with taste $\gamma_5 \otimes \gamma_5$ at each twist in Table~\ref{LattDescHHISQmom} for set 1.}
	\label{DsG5EffEnergy}
\end{figure*}

\begin{figure*} 
	\centering
	\includegraphics[width=1.0\textwidth]{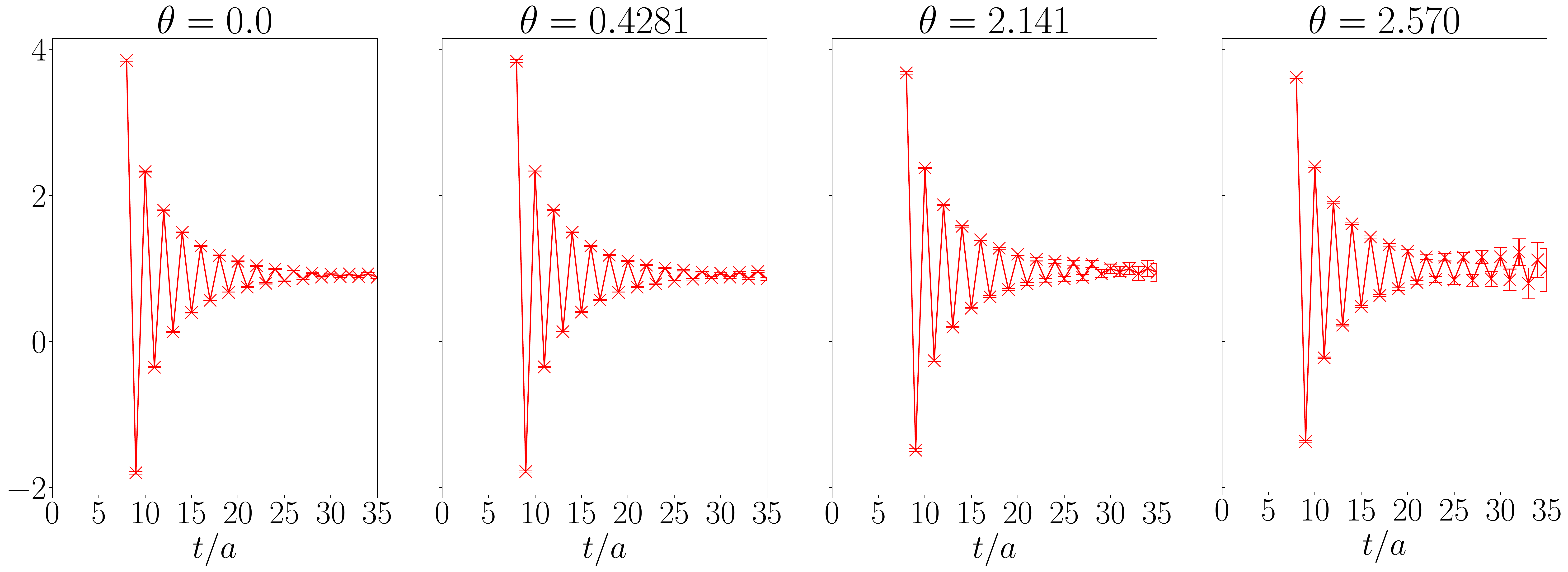}
	\caption{Plots of effective energies for the $D_s$ meson with taste $\gamma_5 \gamma_0 \otimes \gamma_5 \gamma_0$ at each twist in Table~\ref{LattDescHHISQmom} for set 1.}
	\label{DsG5TEffEnergy}
\end{figure*}

For the purposes of demonstration, we consider the effective energies on set 1 (the fine lattice).
Fig.~\ref{HsEffEnergy} shows how the effective energies for the $H_s$ pseudoscalar meson plateau over the first 35 timeslices.
The behaviour is an oscillatory decay towards a plateau whose position is read off and used as the mean of the prior value accompanied with a broad error that comfortably accounts for any misread of the plateau position.
%\rough{The captions of the figures give specific examples of the priors used and the corresponding values determined by the fit.}
Similar behaviour is observed for the other three sets in Table~\ref{LattDesc1}.
The size of the oscillatory behaviour differs according to which 2-point correlation function is being analysed.
The effective energy for the $D_s$ pseudoscalar with interpolator $\gamma_5 \otimes \gamma_5$ in Fig.~\ref{DsG5EffEnergy} shows almost no oscillatory contamination, whereas the effective energy for the $D_s$ meson with taste $\gamma_5 \gamma_0 \otimes \gamma_5 \gamma_0$ in Fig.~\ref{DsG5TEffEnergy} fluctuates strongly between early timeslices, but nevertheless a plateau emerges at later timeslices which indicates a suitable prior.
%

%\begin{figure}
%	\centering
%	\includegraphics[width=0.5\textwidth]{effEnergy_2pt_HcG5.pdf}
%	\caption{Plot of effective energies for the pseudoscalar heavy-charm meson at rest with taste $\gamma_5 \otimes \gamma_5$.}
%	\label{HcG5EffEnergy}
%\end{figure}
%\begin{figure}
%	\centering
%	\includegraphics[width=0.5\textwidth]{effEnergy_2pt_HcG5T.pdf}
%	\caption{Plot of effective energies for the pseudoscalar heavy-charm meson at rest taste $\gamma_5 \gamma_0 \otimes \gamma_5 \gamma_0$.}
%	\label{HcG5TEffEnergy}
%\end{figure}

\subsection{Vector current renormalisation} \label{sec:report_ZV}

\begin{figure}
	\centering
	\includegraphics[width=0.5\textwidth]{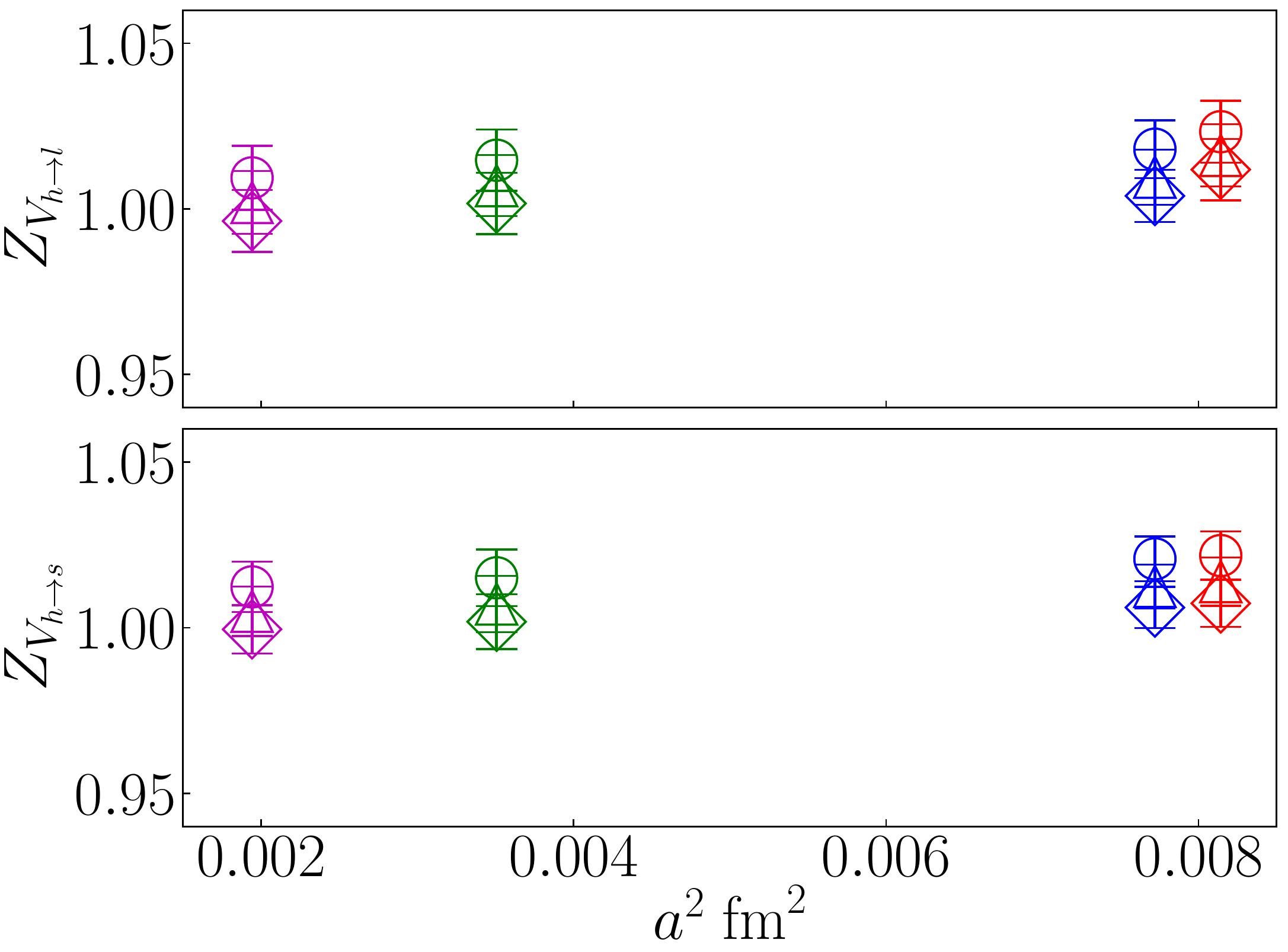}
	\caption{Results for the local vector current renormalisation factor $Z_V$ obtained from Eq.~\eqref{PCVClatt} by the ratio of scalar density and temporal vector current matrix elements at zero-recoil. The top and bottom plots show the results from the calculation of $B_c \to D_l$ and $B_c \to D_s$. The different colours and shapes of markers relate to sets and $am_h$ values as described in Fig.~\ref{hhisq_prop_qSquaredMax}.}
	\label{ZVdata}
\end{figure}

For each heavy-quark mass, the renormalisation factor $Z_V$ is obtained at zero-recoil using Eq.~\eqref{PCVClatt}.
Results are plotted in Fig.~\ref{ZVdata}.
The smallest uncertainties are observed on sets 1 and 2 (red and blue points) which have the best statistics.
%Using more configurations on sets 3 and 4 may result in errors comparable to those achieved for $Z_V$ on sets 1 and 2.
The differences between the top ($B_c \to D_l$) and bottom ($B_c \to D_s$) plots are very small.
This is expected since the $Z_V$ values in the two plots differ only by a discretisation effect associated with the mass of the two daughter quarks, strange and light, which are both small.
%Since the masses of these quarks are very small, the differences between the top and bottom plot of Fig.~\ref{ZVdata}. are tiny.
The figure suggests a mild discretisation effect associated with the bare heavy quark mass $am_h$, though the values are comparable across the four sets.
The central values for $Z_V$ with $am_h = 0.8$ are positioned above the other two values for $am_h$ for each set.
Discretisation errors associated with $am_h$ are taken into consideration when fitting the form factor data obtained on the lattices.

\subsection{Stability of Correlation Function Fits} \label{sec:stab_corrfit}

\begin{figure*}
	\centering
	\includegraphics[width=1.0\textwidth]{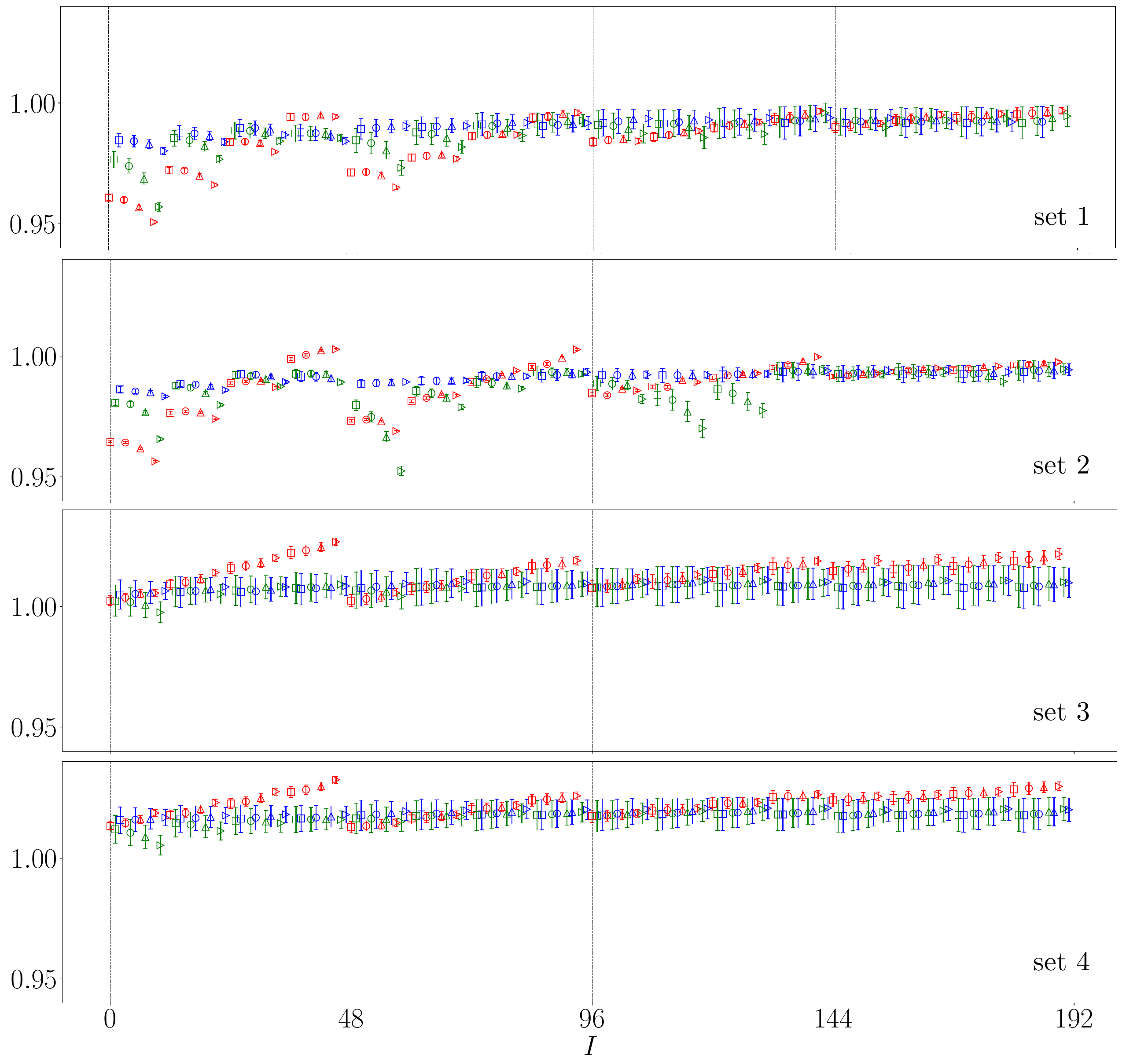} 
	\caption{The parameter $V_{\mathrm{nn},00}$ from Eq.~\eqref{corrfitform_3pt} corresponding to the $H_c \to D_s$ 3-point correlator at zero-recoil with $am_h = 0.65$ is plotted against the fit index $I$ (defined in Eq.~\eqref{eqn:I_defn}). From top to bottom, results on sets 1, 2, 3 and 4 (see Table~\ref{LattDesc1}) are presented respectively. Red, green and blue points indicate that the fit used $N_{\mathrm{n}} + N_{\mathrm{o}}=4,5,6$ exponentials respectively (see Eqs.~\eqref{corrfitform_2pt} and~\eqref{corrfitform_3pt}). The different marker styles reflect the SVD cut chosen: squares, circles, triangles, right and left pointing triangles correspond to SVD cuts of $0.001, 0.025, 0.05, 0.075$ and $0.01$ respectively. The scale of the y axis is shared by the four plots. We scrutinise the form factors associated with correlator fits detailed in Table~\ref{tab:ff_stab}.}
	\label{corrfit_stab}
\end{figure*}

%This section is dedicated to demonstrating the reliability of our extraction of matrix elements from the correlation functions.
We are required to make many choices when fitting the correlation functions to the forms given in Eqs.~\eqref{corrfitform_2pt} and~\eqref{corrfitform_3pt}.
However, we demonstrate in this section that the fit results for the sought-after groundstate quantities are insensitive to the particular strategy of any given fit.
In fact, we explore many different choices to assess robustness.
For the purposes of demonstration, Fig.~\ref{corrfit_stab} shows a selection of matrix elements plotted against $I$ which enumerates different fits which we now describe.
By inspecting this plot, we can identify a region in the space of fitting strategies where the fit results are stable and reasonable.

In Table~\ref{tab:ff_stab}, we tabulate the regimes for each set used in our final determination of the physical-continuum form factors.
\begin{table}
	\centering
	\caption{Input parameters (see text for definition) to the fits of correlation functions for the heavy-HISQ calculation together with fits including variations of the SVD cut, $t^{\text{2pt}}_\text{min}/a$, $t^{\text{3pt}}_\text{min}/a$ and $N$. Bold entries indicate those fits used to obtain the final results. Other values are used in tests 
		of the stability of our form factor fits to be discussed in Appendix~\ref{sec:ff_stability}.}
	\begin{tabular}{c | l c c c | l c c c} 		
		\hline\hline
		& \multicolumn{4}{c |}{$B_c \to D_l$} & \multicolumn{4}{c}{$B_c \to D_s$} \\
		set & SVD & $t_\text{min}^\text{2pt}/a$  & $t_\text{min}^\text{3pt}/a$ & $N$ & SVD & $t_\text{min}^\text{2pt}/a$  & $t_\text{min}^\text{3pt}/a$ & $N$ \\ [0.1ex] 
		\hline
		1 & \textbf{0.005} & \textbf{4} & \textbf{4} & \textbf{8} & \textbf{0.005} & \textbf{6} & \textbf{4} & \textbf{7}\\
		& 0.0025 & 8 & 8 & 8 & 0.0025 & 6 & 2 & 7 \\
		%& 0.0025 & 4 & 4 & 8 & 0.0025 & 8 & 4 & 7 \\
		\hline
		2 & \textbf{0.0025} & \textbf{4} & \textbf{6} & \textbf{8} & \textbf{0.005} & \textbf{6} & \textbf{6} & \textbf{7}\\
		%& 0.0025 & 6 & 8 & 8 & 0.0025 & 4 & 8 & 7 \\
		& 0.005 & 4 & 8 & 8 & 0.0025 & 8 & 6 & 8 \\
		\hline
		3 & \textbf{0.005} & \textbf{8} & \textbf{8} & \textbf{6}  & \textbf{0.005} & \textbf{10} & \textbf{10} & \textbf{6}\\
		& 0.0075 & 10 & 10 & 5 & 0.005 & 10 & 12 & 5 \\
		%& 0.005 & 6 & 12 & 6 & 0.0075 & 8 & 12 & 6 \\
		\hline
		4 & \textbf{0.0075} & \textbf{10} & \textbf{12} & \textbf{6}  & \textbf{0.0075} & \textbf{12} & \textbf{12} & \textbf{6}\\
		& 0.0075 & 10 & 10 & 5 & 0.0075 & 10 & 14 & 6 \\
		%& 0.01 & 8 & 14 & 6 & 0.01 & 8 & 12 & 5 \\
		\hline \hline
	\end{tabular}%}
	\label{tab:ff_stab}
\end{table}
These fits are chosen from a variety of fits that, as explained in Section~\ref{sec:fit_correls}, use different SVD cuts, numbers of exponentials, and trims of the correlator data.
To demonstrate the robustness of the correlation function fits used to extract the form factor data, we show that the fits are stable and are selected among regions in parameter space where the matrix elements are insensitive to these choices of fitting regime.
In Fig.~\ref{corrfit_stab}, as an example, we display results for the $V_{\mathrm{nn},00}$ parameter associated with the scalar density at zero-recoil for $am_h = 0.65$ on each of the four sets in Table~\ref{LattDesc1} (similar behaviour is found for the other currents, momenta and heavy quark masses).
We plot $V_{\mathrm{nn},00}$ against an index $I$ which enumerates the fit.
We define $I$ as
\begin{align} \label{eqn:I_defn}
	I = n_i + 3s_i + 15t_i^{\text{3pt}} + 75t_i^{\text{2pt}}
\end{align}
where $n_i = 0,1,2$ indexes the choice of the number of exponentials $N_{\mathrm{n}} + N_{\mathrm{o}} \in \{4,5,6\}$, $s_i = 0,1,2,3$ indexes the choice of SVD cut in either $\{0.0075, 0.005, 0.0025, 0.001\}$ for sets 1 and 2, or the set $\{0.01, 0.0075, 0.005, 0.0025\}$ for sets 3 and 4.
These ranges of SVD cut cover the recommendation from the \texttt{svd\_diagnosis} tool within the \textit{corrfitter} package~\cite{corrfitter}.
We investigate the effect of trimming the correlator data: $0 \leq t_i^{\text{2pt}},t_i^{\text{3pt}} \leq 3$ indexes the choice of $t_{\text{min}}^{\text{2pt}}/a$ and $t_{\text{min}}^{\text{3pt}}/a$ in $\{2,4,6,8\}$ for sets 1 and 2, in $\{6,8, 10, 12\}$ for set 3, and in $\{8, 10, 12, 14\}$ for set 4.
We are guided by the expectation that we should trim according to some fixed distance in physical units away from the interpolator.
Hence, we generally trim more data points for finer lattices.
Note that $t_{\text{min}}^{\text{2pt}}$ is the slowest running parameter.
To aid the reader's understanding of the organisation of the fits in Fig.~\ref{corrfit_stab}, we separate fits with different values of $t_{\text{min}}^{\text{2pt}}/a$ with black dashed vertical lines.
%update1this
%
\begin{figure}
	\centering
	\includegraphics[width=0.5\textwidth]{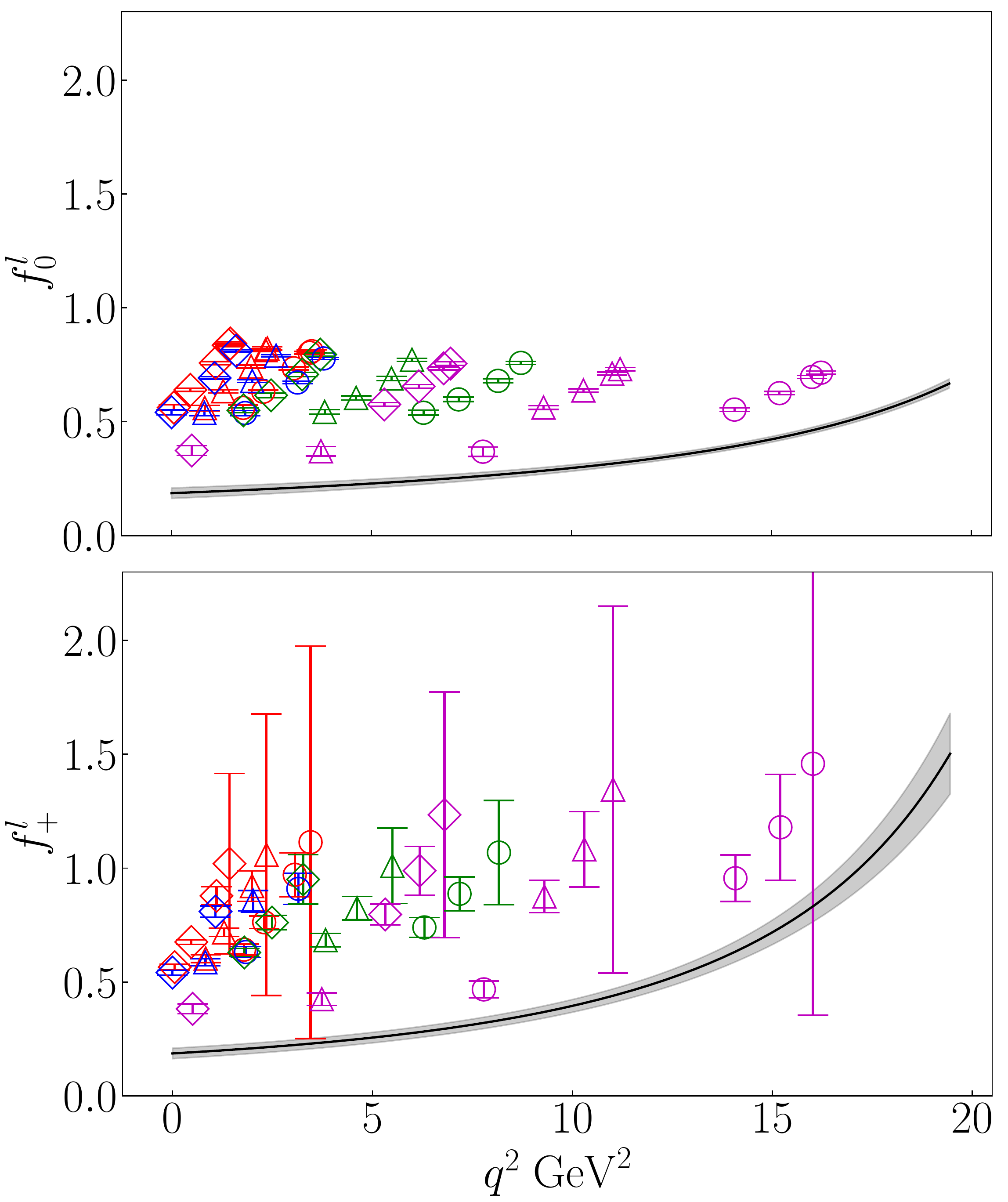}
	\caption{Data and fit for the form factors $f_{0,+}^l$. The scale of the $y$-axis is the same as for Fig.~\ref{fig:ffs_BcDs}. The different colours and shapes of markers relate to sets and $am_h$ values as described in Fig.~\ref{hhisq_prop_qSquaredMax}.}
	\label{fig:ffs_BcDl}
\end{figure}
%
%update1this
%
\begin{figure} 
	\centering
	\includegraphics[width=0.5\textwidth]{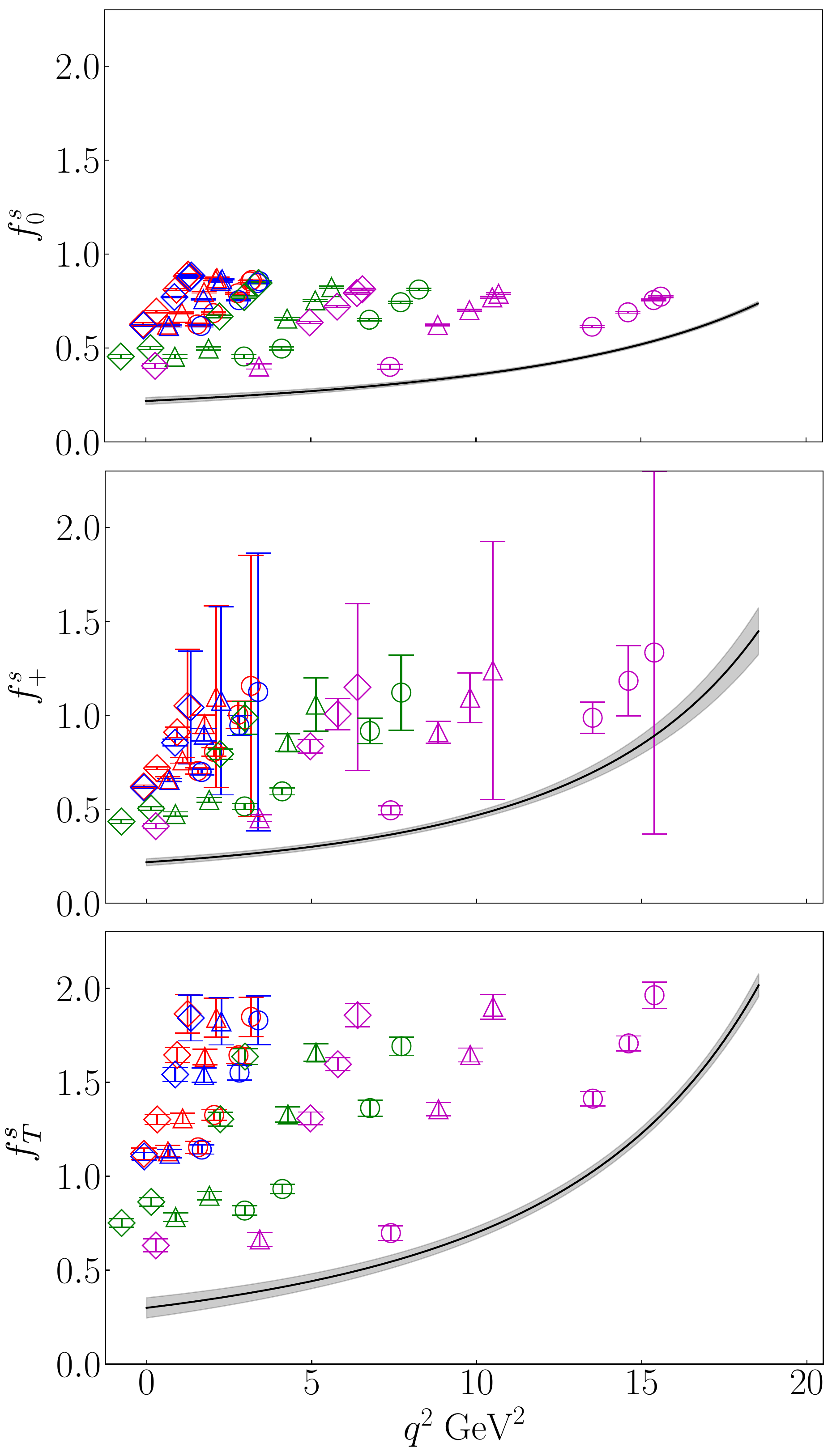}
	\caption{Data and fit for the form factors $f_{0,+,T}^s$. The scale of the $y$-axis is shared with Fig.~\ref{fig:ffs_BcDl}.}
	\label{fig:ffs_BcDs}
\end{figure}

Considering figures such as Fig.~\ref{corrfit_stab} for all matrix elements helps us to identify choices of parameters where the fit is stable, whilst also ensuring that we avoid unnecessarily bloated fit models with more exponentials than required.
The fit takes longer to complete for more exponentials, hence a judicious selection of $N_{\mathrm{n}}$ and $N_{\mathrm{o}}$ allows us to feasibly explore, in reasonable computing time, the parameter landscape in other directions. 
Nevertheless, a variety of fits with $N_{\mathrm{n}} + N_{\mathrm{o}} = 7, 8$ and greater have also been carried out to ensure that the convergence demonstrated in Fig.~\ref{corrfit_stab} is maintained for more exponentials.
Indeed, similar extractions of the groundstate quantities are obtained by these fits.
For the purposes of fitting form factors, it suffices to use fits with $N_{\mathrm{n}} + N_{\mathrm{o}} = 7$ or $8$ on sets 1 and 2, and $N_{\mathrm{n}} + N_{\mathrm{o}} = 5$ or $6$ on sets 3 and 4.
In summary, each plot shows results from 192 different fits ($0 \leq I \leq 191$).
The parameters used for our final fits are shown by the boldened entries in Table~\ref{tab:ff_stab}, and the plots demonstrate that these choices lie within regions of parameter space that admit stable fit results.
%update1this
%
\begin{figure}
	\centering
	\includegraphics[width=0.5\textwidth]{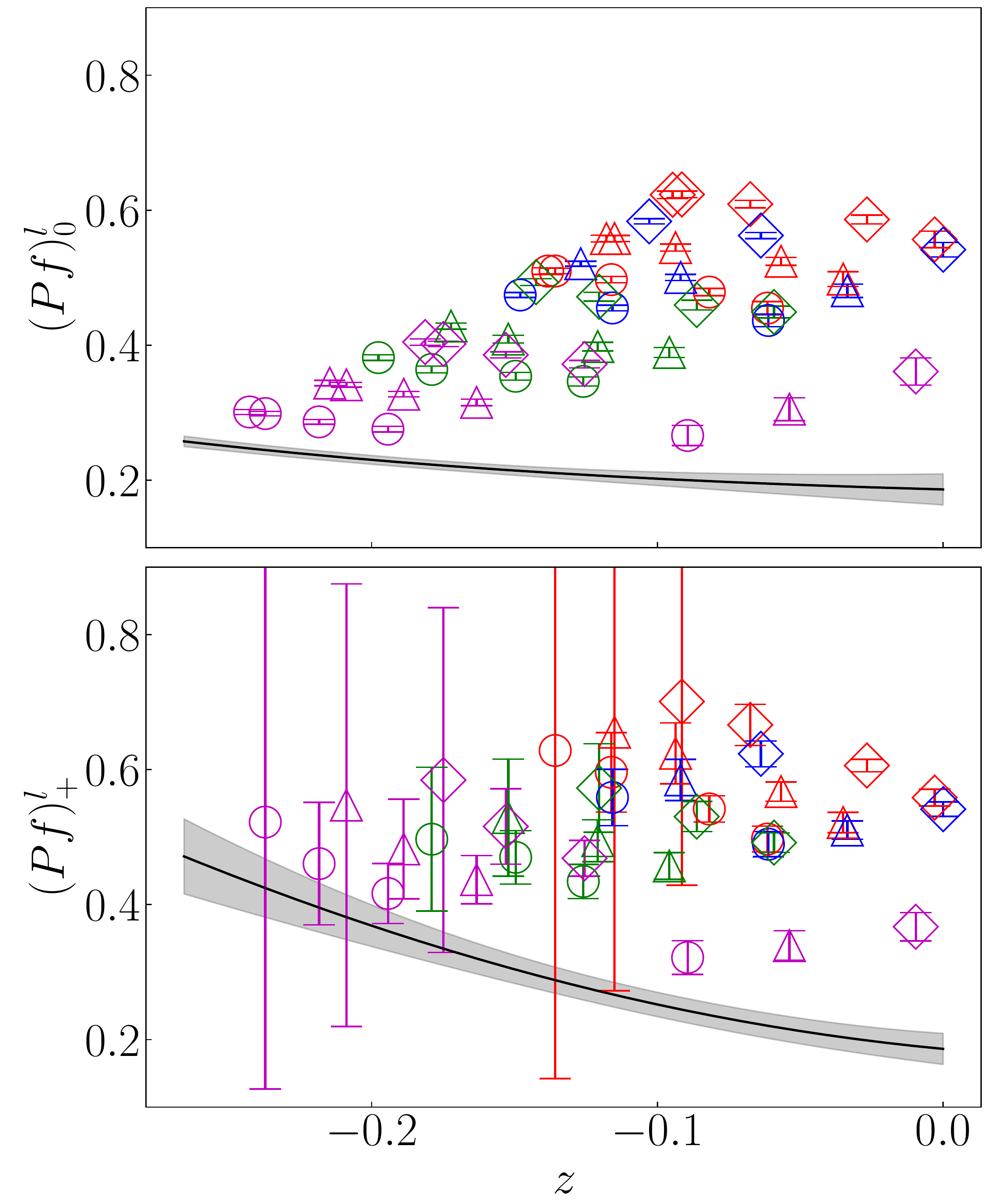}
	\caption{Data and fit for the form factors $f_{0,+}^l$ multiplied by the pole factor $P(q^2)$ (see Eq.~\eqref{hhisqffff}). The fit band is the polynomial $\sum_n c^{(n)} (-z)^n$ (coefficients $c^{(n)}$ are defined in Eq.~\eqref{eqn:def_cn}).}
	\label{fig:Pfl}
\end{figure}
%
%update1this
%
\begin{figure}
	\centering
	\includegraphics[width=0.5\textwidth]{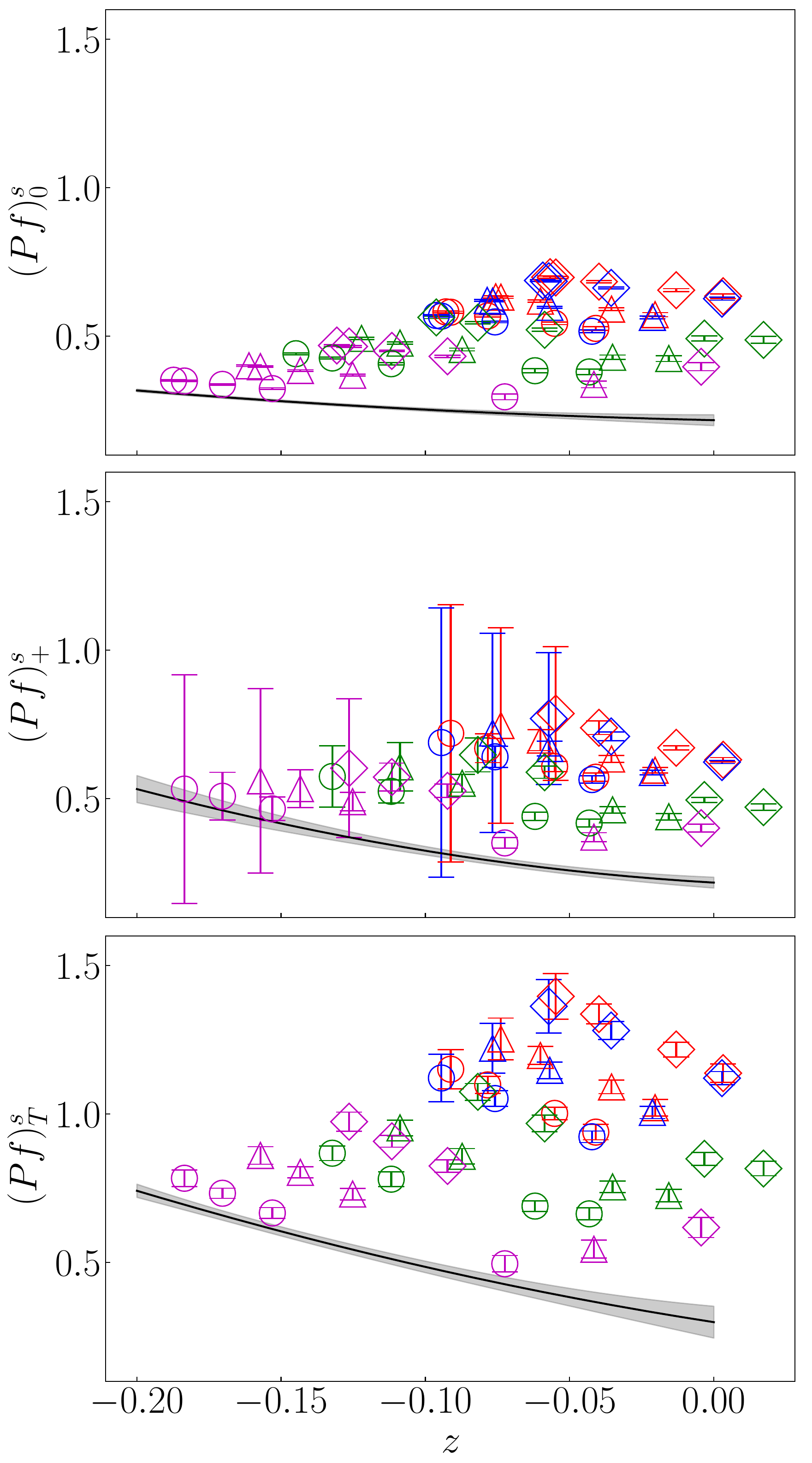}
	\caption{Data and fit for the form factors $f_{0,+,T}^s$ multiplied by the pole factor $P(q^2)$ (see Eq.~\ref{hhisqffff}). The fit band is the polynomial $\sum_n c^{(n)} (-z)^n$ (coefficients $c^{(n)}$ are defined in Eq.~\eqref{eqn:def_cn}).}
	\label{fig:Pfs}
\end{figure}

Firstly, we address the dependence on the number of exponentials.
In Fig.~\ref{corrfit_stab}, we show fits for $ N_{\mathrm{n}} + N_{\mathrm{o}} \in \{4,5,6\}$ with $N_{\mathrm{n}} - N_{\mathrm{o}} = 0$ for $N_{\mathrm{n}} + N_{\mathrm{o}}$ even, and $N_{\mathrm{n}} - N_{\mathrm{o}} = 1$ for $N_{\mathrm{n}} + N_{\mathrm{o}}$ odd.
The fits with $N_{\mathrm{n}} + N_{\mathrm{o}}=4$ show some variation as the other parameters are varied, particularly for smaller $t_{\text{min}}^{\text{2pt}}/a$ and smaller SVD cuts.
In contrast, fits with $N_{\mathrm{n}} + N_{\mathrm{o}}=5$ and $N_{\mathrm{n}} + N_{\mathrm{o}}=6$ are in good agreement with each other for most choices of SVD cut and larger correlator function trims, and there are clear regions where $V_{\mathrm{nn},00}$ appear stable.

Addressing the different extents that correlation function data has been trimmed, the fit results show some mild instability for $t_{\text{min}}^{\text{2pt}}/a = 2$ where the correlation function data to be fit contains the most excited state contamination.
This instability is expected to be better resolved by introducing more exponentials that can absorb more contributions from higher energy states and short-distance effects.
For example, fits with $t_{\text{min}}^{\text{2pt}}/a > 2$ appear more stable than those for $t_{\text{min}}^{\text{2pt}}/a = 2$.
%For the fits with the hardest pruning of correlation functions, the error is larger.
%For example, consider the fits exhibited by the right-most points in Figure \ref{corrfit_stab}.
%Here, $t_{\text{min}}^{\text{2pt}}/a = 10$ and $2 \leq t_{\text{min}}^{\text{3pt}}/a \leq 10$.
%As is especially clear from the fit with $t_{\text{min}}^{\text{2pt}}/a = t_{\text{min}}^{\text{3pt}}/a = 10$ (the \emph{very} right-most point), the parameter $V_{nn}[0,0]$ is determined with much larger error than many other fits.
%This is due to the fact that the severity of the trim is detrimental to the fit: the fitter does not have enough data points to determine $V_{nn}[0,0]$ to the same accuracy as with fits where correlation functions have been pruned less harshly.

Finally, we discuss the behaviour of the fit results as the SVD cut is varied, denoted by different marker styles in Fig.~\ref{corrfit_stab}.
It is consistently apparent throughout the fits on each set that increasing the SVD cut has the effect of increasing the error on the value obtained for the $V_{nn}$ parameter.
The matrix elements extracted are consistent with each other as the SVD cut is increased, so it appears from these plots that using too large an SVD cut is too conservative.
Decreasing the SVD cut substantially below the recommended cut taken from the \texttt{svd\_diagnosis} tool within the \textit{corrfitter} package~\cite{corrfitter} gives unstable and unreliable results.
Hence, we do not deviate far from this recommended cut.
On the finer lattices, sets 3 and 4, fits with an SVD cut of $0.001$ are frequently in tension with the other fits.
Whilst this may be an appropriate SVD cut for some fits on set 1 and 2, this same is not true on sets 3 and 4.
This is unsurprising since sets 3 and 4 have poorer statistics than sets 1 and 2.
Fits on sets 3 and 4 benefit from a larger SVD cut.
Indeed, in Table~\ref{tab:ff_stab}, we show that we take fits with SVD cuts of no smaller than $0.005$ for sets 3 and 4.
SVD cuts for sets 1 and 2 are chosen among $0.0025$ and $0.005$.
Obtaining higher statistics on sets 3 and 4 would enable a smaller SVD cut to be taken, thus achieving a smaller error on the extracted matrix elements.

In conclusion, based on our exploration of different fits, it is clear that fitting with larger trims of the correlation function data are warranted for the finer lattices, reflected by our choice of fits in Table~\ref{tab:ff_stab}.
The finer lattices also require fewer exponentials and slightly larger SVD cuts than the fine and fine-physical sets.
%update1this
%
\begin{figure}
	\centering
	\includegraphics[width=0.5\textwidth]{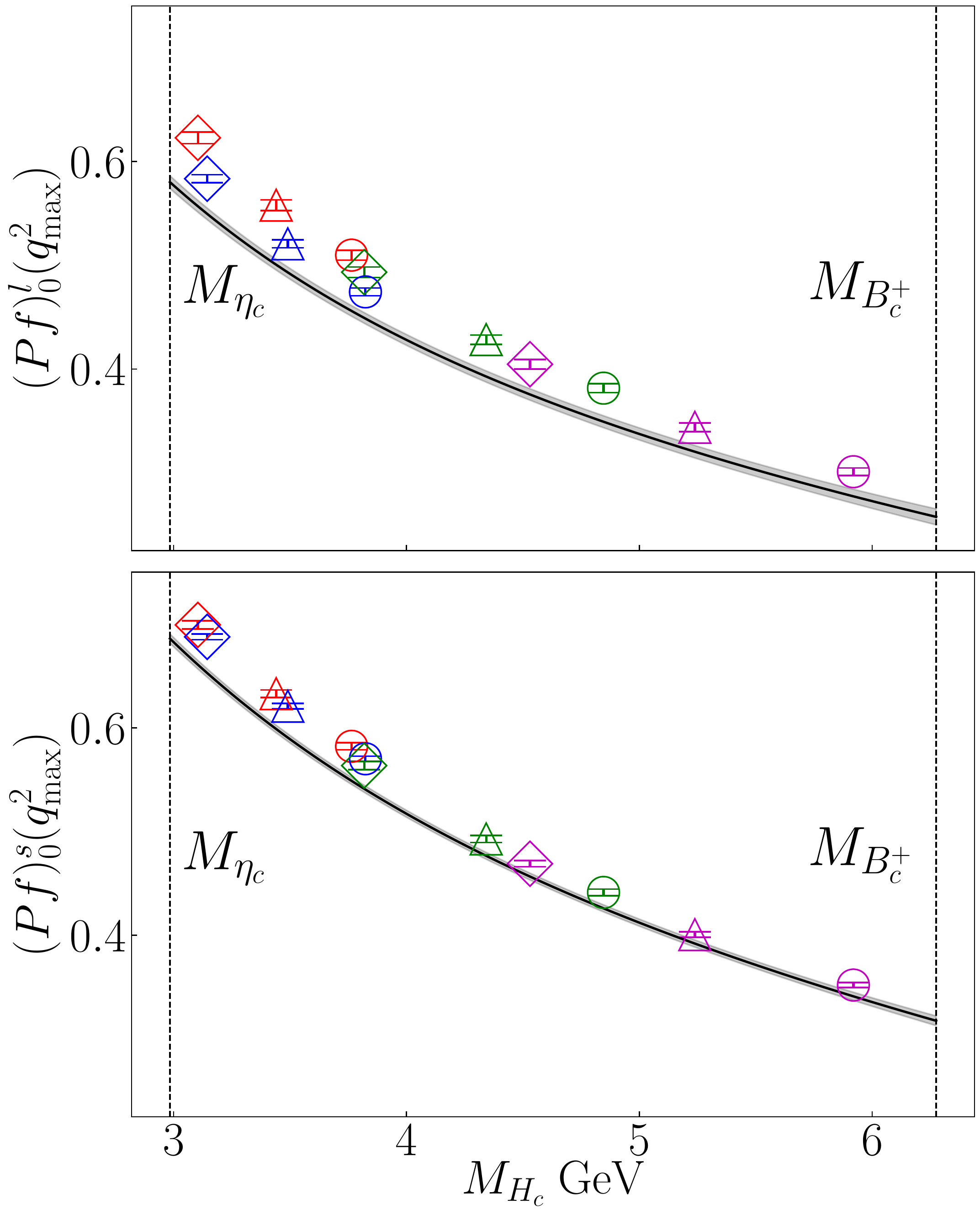}
	\caption{Data and fit for the form factor $f_0$  multiplied by the pole factor (see Eq.~\eqref{hhisqffff}) plotted at zero-recoil as a continuous function of $M_{H_c}$. The vertical dotted lines show the masses of $H_c$ meson for the cases that the heavy quark coincides with the charm and bottom quark.}
	\label{fig:Pf0_against_MHc}
\end{figure}

\section{Form Factor fitting analysis} \label{sec:ffs_fit_analysis}

\subsection{Fit results} \label{sec:ff_fit_results}

In Figs.~\ref{fig:ffs_BcDl} and~\ref{fig:ffs_BcDs}, we show our form factor data alongside the fit functions tuned to the physical-continuum point.
Note that the $q^2$ corresponding to zero-recoil, $q^2_{\text{max}} = (M_{H_c} - M_{D_{l(s)}})^2$, varies as a function of the heavy quark mass.
%(in contrast to $B_c \to B_{s(d)}$~\cite{Cooper:2020wnj} for example).
Hence, the spread over $q^2$ of the form factor data for larger $am_h$ is greater than for smaller $am_h$ on each set.
%This is reminiscent of $B_s \to D_s$ in \cite{McLean:2019qcx} and other heavy-HISQ analyses of decays of the valence heavy quark.
See Fig.~\ref{hhisq_prop_qSquaredMax} for the $q^2$ we access as a proportion of $q^2_{\text{max}}$ on each set and heavy quark mass $am_h$.

Errors on the data for $f_+$ near zero-recoil are large and we exclude points with error in excess of $25\%$ from the fit.
These large errors are a result of the kinematic factors associated with determining $f_+$ from the temporal vector current matrix elements (see Eq.~\eqref{fplusextract}).
%are encountered since the denominator of Eq.~\eqref{fplusextract} increases from zero at zero-recoil.
Further discussion can be found in Section II C of~\cite{Cooper:2020wnj} and Section~\ref{sec:fplus_from_spatVec} here.

Figs.~\ref{fig:Pfl} and~\ref{fig:Pfs} show the same data and fit after multiplying by the pole factor $P(q^2)$ (see Eq.~\eqref{hhisqffff}).
%Recall the function form in Eq.~\eqref{eqn:physcont_ffs}.
The fit function shown in Eq.~\eqref{hhisqffff} is the polynomial in $z$ that gives the residual momentum dependence of the form factors not accounted for by the pole factor $P(q^2)^{-1}$.
Note that the $y$-axis is smaller in Figs.~\ref{fig:Pfl} and~\ref{fig:Pfs} than for Figs.~\ref{fig:ffs_BcDl} and~\ref{fig:ffs_BcDs} since most of the $q^2$ dependence of the form factors has been removed on multiplying by the pole factor $P(q^2)$.
The polynomial for $f_0$ appears linear in $z$-space to a good approximation.
For $f_{+,T}$, the fit curves show a small amount of curvature.
We compare fits with $N_n=3$ and $4$ in Appendix~\ref{sec:ff_stability} to ensure that our truncation of the $z$-expansion is appropriate.

As is standard with heavy-HISQ analyses of decays of a valence $b$ quark, the $q^2$ dependence of the form factors is inferred from data on multiple lattices which each have a different range of $q^2$ since $q^2_{\mathrm{max}}$ varies with $am_h$.
This can make plots shown in Figs.~\ref{fig:ffs_BcDl},~\ref{fig:ffs_BcDs},~\ref{fig:Pfl} and~\ref{fig:Pfs} difficult to interpret since there are several different extrapolations taking place simultaneously to reach the fit curve in the continuum limit with physical quark masses.
Considering just the data at zero-recoil can provide a clearer understanding of how the fit curves shown in figures relate to the lattice data for the form factors.
Fig.~\ref{fig:Pf0_against_MHc} shows, for both the cases $B_c \to D_l$ and $B_c \to D_s$, data for $f_0$ at zero recoil plotted against $M_{H_c}$ alongside the fit function tuned to the continuum limit with physical light, strange and charm quark masses.
This figure shows how the dependence on the heavy quark mass is resolved by the factors $\Omega^{(n)} (\Lambda / M_{H_{l(s)}})^r$ in Eq.~\eqref{hhisqffff}.
For the purposes of presenting the fit as a continuous function of the $M_{H_c}$, we approximate the heavy-light and heavy-strange pseudoscalar mass as $M_{H_q} \approx M_{H_c} - (M_{B_c} - M_{B_q})$ where $q=l$ or $s$.
The lattice data follows the curve closely.
The error band is most narrow at around $4 \; \mathrm{GeV}$ and the error flares slightly as $M_{H_c}$ approaches $M_{B_c}$.
%update1this
%%
%\begin{figure} 
%	\centering
%	\includegraphics[width=0.5\textwidth]{errorratio_withWithoutKC}
%	\caption{For each of the form factors $f_{0,+}$ for $B_c \to D_l$ and $B_c \to D_s$, we plot the ratio $r_{\mathrm{KC}}$ of errors (see text in Appendix~\ref{sec:imposition_KC}) on the form factors found by fitting with and without the kinematic constraint $f_0 (0) = f_+ (0)$ (imposed through the parameter constraint $(A_0)^{(0)}_{r00} = (A_+)^{(0)}_{r00}$ for all $r$ (see Eq.~\eqref{hhisqffff})). }
%	\label{fig:withWithoutKC}
%\end{figure}
%%

%\subsubsection{Truncating the $z$-expansion} \label{sec:z_trunc}
%
%Our results use a $z$-expansion of the form factors (see Eq.~\eqref{hhisqffff}) with $N_n=3$, i.e. the series is truncated after the term $z^3$.
%In Table~\ref{tab:ff_extrema}, we show results from fits instead with $N=2$ and $N=4$ .
%Good agreement is seen among these fits indicating that the the dominant $q^2$ behaviour has been successfully absorbed into the pole factor $P(q^2)^{-1}$ with the choice of pole masses described in Section~\ref{sec:ff_fit_form}.
%The remnant dependence on $q^2$ has been captured by the $z$-polynomial.
%Form factor values and errors at both $q^2 = 0$ and zero-recoil change very little between fits with $N=2,3,4$.
%We use $N=3$ in our final results.

\subsection{Imposition of the kinematic constraints} \label{sec:imposition_KC}
%In Fig.~\ref{fig:withWithoutKC}, we show, as a proportion of the mean value, the difference of errors $r_{\mathrm{KC}}$ of fits carried out without and with impositions of the kinematic constraint 
The form factors must obey $f_0(0) = f_+(0)$ in the continuum limit for all mass of heavy-charm pseudoscalar meson (see Section~\ref{sec:ff_fit_form}).
Since we take $t_0 = 0$ in Eq.~\eqref{eqn:BcD_littlez}, $z=0$ at $q^2 = 0$.
Hence, the kinematic constraint can be straightforwardly applied to our fit: we insist that $(A_0)^{(0r00)} = (A_+)^{(0r00)}$ for all $r$ and $\rho_0^{(0)} = \rho_+^{(0)}$ (see Eq.~\eqref{hhisqffff}) by setting a narrow prior on their differences.
Table~\ref{tab:ff_extrema} compares the errors at the $q^2$ extremes from fitting with and without these parameter constraints.
We also compare integrated quantities.
The two fits are in good agreement.
Uncertainties are reduced very slightly when fitting with the kinematics constraint.
The form factors $f_{0,+}^l$ at $q^2 = 0$ see the most benefit.
%Table~\ref{tab:ff_extrema} compares the errors at the $q^2$ extremes.
%update1this
%
\begin{table}
	\centering
	\caption{We compare fits with and without imposition of the kinematic constraint (KC) $f_0 (0) = f_+ (0)$. Form factors are shown at $q^2 = 0$ and maximum $q^2$. We also present integrated values where we find the variation between the two fits to be especially small. The three uncertainties on the branching fractions are from the lattice, the lifetime of the $B_c$ meson, and $V_{ub}$ respectively.}
	\begin{tabular}{l | c c} 
		\hline\hline
		& final & without KC \\ [0.1ex] 
		\hline
		$f_0^l(0)$ & $0.186(23)$ & $0.191(27)$ \\
		$f_+^l(0)$ & --- & $0.158(34)$  \\
		$f_0^l(q^2_{\text{max}})$ & $0.668(20)$ & $0.669(20)$\\
		$f_+^l(q^2_{\text{max}})$ & $1.50(18)$ & $1.48(17)$ \\
		\hline
		$\mathcal{B}(B_c^+ \to D^0 e^+ \nu_e) \times 10^5$ & $3.37(48)(8)(42)$ & $3.17(51)(8)(40)$ \\
		$\mathcal{B}(B_c^+ \to D^0 \tau^+ \nu_{\tau}) \times 10^5$ & $2.29(23)(6)(29)$ & $2.29(23)(6)(29)$ \\
		\hline
		$f_0^s(0)$ & $0.217(18)$ & $0.224(19)$ \\
		$f_+^s(0)$ & --- & $0.192(23)$ \\
		$f_0^s(q^2_{\text{max}})$ & $0.736(11)$ & $0.736(11)$ \\
		$f_+^s(q^2_{\text{max}})$ & $1.45(12)$ & $1.44(12)$ \\
		\hline
		$\mathcal{B}(B_c^+ \to D_s^+ e^+ e^-) \times 10^7$ & $1.00(11)$ & $0.95(11)$ \\ 
		$\mathcal{B}(B_c^+ \to D_s^+ \tau^+ \tau^-) \times 10^7$ & $0.246(18)$ & $0.246(18)$ \\ 
		\hline \hline
	\end{tabular}%}
	\label{tab:ff_extrema}
\end{table}
%
%The kinematic constraints are a powerful restriction on the vector form factor, where the precision of $f_0$ data is leveraged by $f_+$ to reduce errors throughout the entire range of physical $q^2$.

%update1this
%
\begin{figure*}
	\centering
	\includegraphics[height=0.65\textwidth]{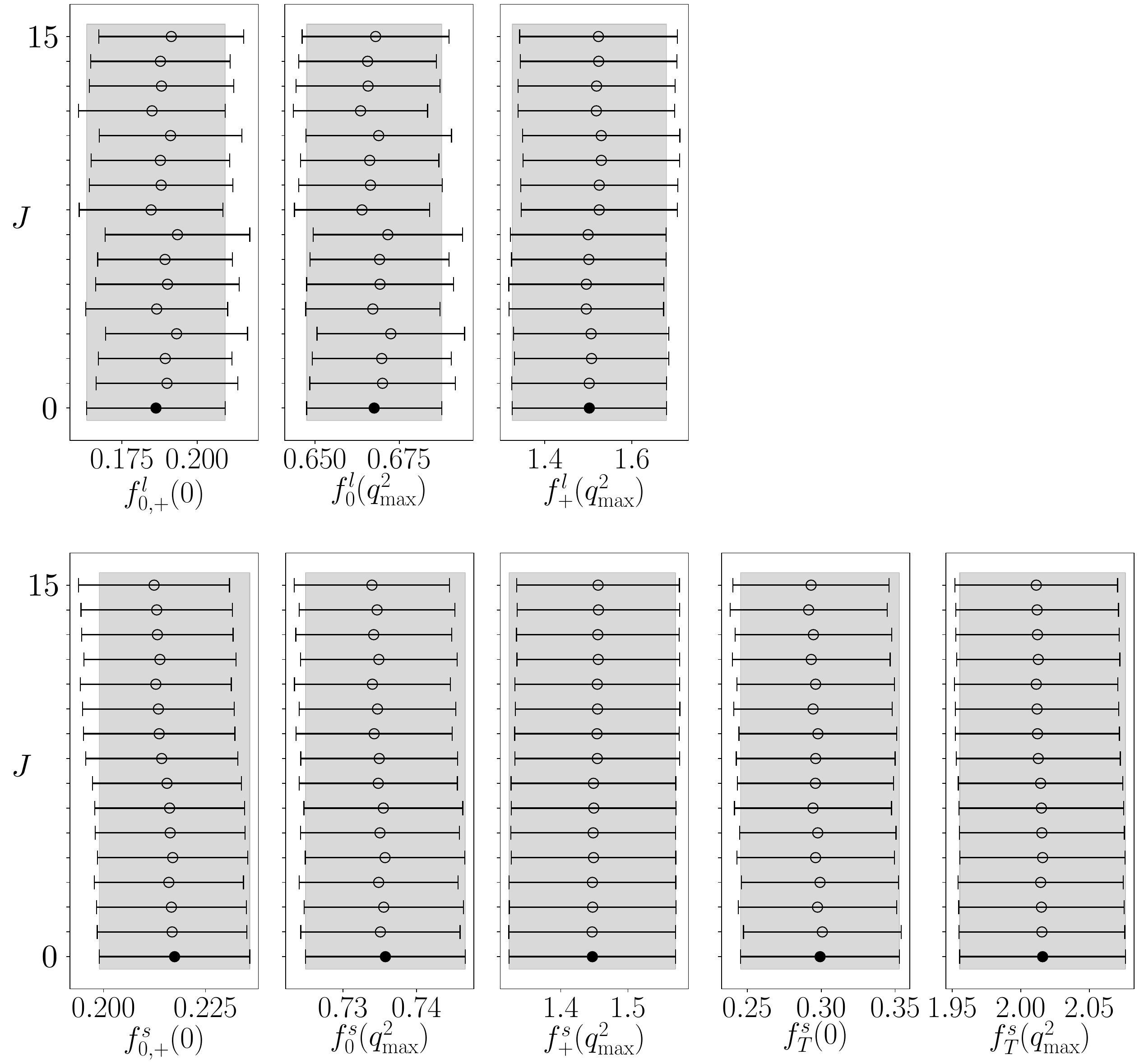}
	\caption{For each of the $16$ different correlator fits indexed by $J$ (see Eq.~\eqref{eqn:defn_J}), we show the fitted values of the physical-continuum form factors for $B_c \to D_l$ (top) and $B_c \to D_s$ (bottom) evaluated at maximum $q^2$ and $q^2 = 0$. These plots show results from all possible combinations of the correlation function fits described in Table~\ref{tab:ff_stab} and demonstrate the stability of our results under these changes. The filled black points show the results from our final fit.}
	\label{fig:ff_stability_with corrf_BcDl}
\end{figure*}
%

%%
%\begin{figure*}
%	\centering
%	\includegraphics[width=0.85\textwidth]{stability_ff_BcDl}
%	\caption{For each of the $81$ different fits indexed by $J$ (see Eq.~\eqref{eqn:defn_J}), we show the parameters $(c_{0,+})^{(0,1)}_{0000}$ of the physical-continuum form factors for $B_c \to D$. These plots show results from all possible combinations of the correlation function fittings described in Table \ref{tab:ff_stab}.}
%	\label{fig:ff_stability_with corrf_BcDl}
%\end{figure*}
%%

\subsection{Fit variations} \label{sec:ff_stability}

In Table~\ref{tab:ff_stab}, we describe two different fits of correlation functions on each set and fit the form factors to each different combination, resulting in $16$ different fits of the form factors.
In Fig.~\ref{fig:ff_stability_with corrf_BcDl}, we show the physical-continuum form factors evaluated at $q^2 = 0$ and $q^2_{\mathrm{max}}$ from each of the fits.
%where $c^{(n)}$ is the coefficient of $(-z)^n$ in the fit form taken to the physical continuum limit (Eq.~\eqref{eqn:def_cn}).
The fits are indexed by $J$ where
\begin{align} \label{eqn:defn_J}
	J &= \sum_{j=1}^4 2^{j-1}n_j
\end{align}
where $n_j \in \{0,1\}$ indexes each of the two fits on set $j$ given in Table~\ref{tab:ff_stab}.
For example, the fit labelled by $J=0$ uses correlation function fit results corresponding to all the boldened entries in the Table \ref{tab:ff_stab}.
The figure shows that the form factors are insensitive to the particular choice of correlator fits.
%As an example, Fig.~\ref{fig:ff_stability} shows the variation of $c_{0,+}^{(0)}$ between different fits is well within the $1$-$\sigma$ error bars.
The fit $J=0$ yields form factors very similar the 15 alternative fits with $J>0$.
All central values lie within the $1$-$\sigma$ error band of those parameters corresponding to the $J=0$ fit from which our final results for the form factors are derived.
We conclude that the form factor fits are robust and stable as the choices of correlation function fits are varied.

\begin{figure*}
	\centering
	\includegraphics[height=0.95\textwidth]{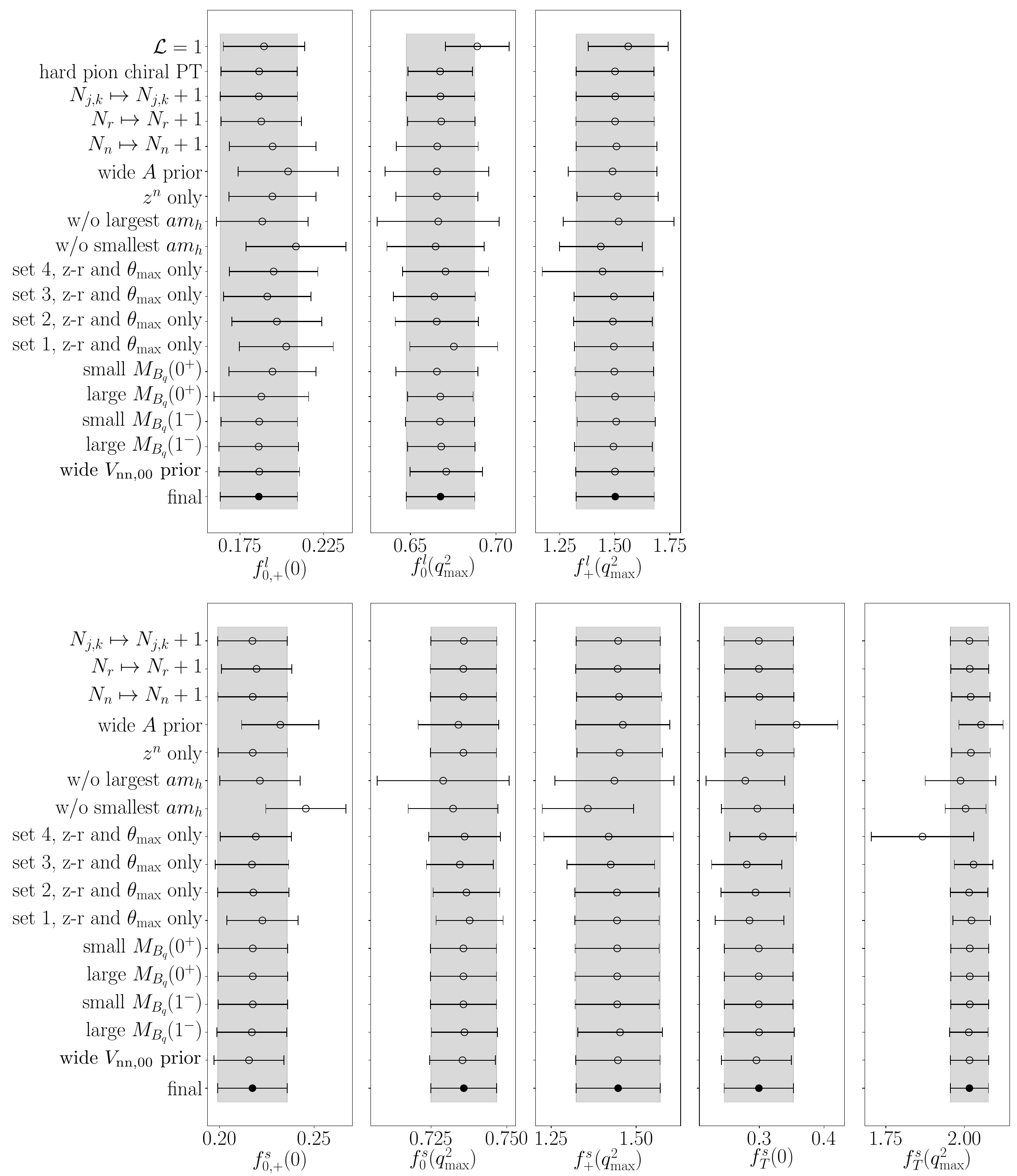}
	\caption{For each of the different form factor fits described in Appendix~\ref{sec:ff_stability}, we show the physical-continuum form factors for $B_c \to D_l$ (top) and $B_c \to D_s$ (bottom) evaluated at maximum $q^2$ and $q^2 = 0$. The filled black points shows the results from our final fit.}
	\label{fig:ff_stability}
\end{figure*}

Next, we consider other variations of form factor fits.
In Fig.~\ref{fig:ff_stability}, we show result from a variety of different fits which we now describe.
The fit variations are labelled on the $y$ axis.
Our final fit, results from which we report in Section~\ref{sec:results}, is labelled \lq final\rq.

Beginning at the top of the plot for $f_{0,+}^l$, we consider removing the chiral log by setting $\mathcal{L} = 1$.
The fit labelled\ \lq hard pion chiral PT\rq \space uses $\mathcal{L} = 1 + \zeta^{(0)} x_{\pi} \log x_{\pi}$ instead of the $\mathcal{L}$ given in Eq.~\eqref{eq:curlyL}.
Similar fit results are achieved with these fit variations indicating that, with the current status of errors, the dependence on the light quark mass can be absorbed into the analytic terms in the $\mathcal{N}_{\mathrm{mis}}^{(n)}$ factor in the fit form at Eq.~\eqref{hhisqffff}.

Next, we consider fits varying $N_{n,r,j,k}$ in the fit form at Eq.~\eqref{hhisqffff}.
Doing so allows us to investigate the impact of truncating our fit form.
Varying $N_n$ tests the truncation $N_n = 3$ of the $z$ series for $P(q^2)f(q^2)$.
%We see good agreement in the form factors obtained from fits with $N_n=3$ and $N_n=4$.
Form factor values and errors at both $q^2 = 0$ and zero-recoil change very little between fits with $N_n=3,4$.
We use $N_n=3$ in our final results.
Similarly, increasing $N_{r,j,k}$ yields consistent fit results.

Results from increasing prior widths of parameters $A^{(nrjk)}$ and $\rho^{(n)}$ are shown next.
The fit results are in agreement with our normal priors.
Recall in Section ~\ref{sec:results_ffs} that we perform an Empirical Bayes analysis to check that our priors are appropriate.

Fits where the $z^{N_n+1}$ terms are removed are shown.
It appears as though these terms make very little difference to the form factors.

We then consider fitting with different subsets of the data.
Firstly, we consider fitting without the smallest and largest $am_h$ values on all sets.
Next, we remove certain twists on the four different sets.
Fitting with these smaller datasets gives form factors consistent with our final results.
It is often the case that fitting with these reduced datasets gives errors larger than those observed when fitting with all of the data.

We also check that the fits are insensitive to the value given for $M_{\mathrm{res}}$ in $P(q^2)$ by perturbing the pole mass.
In Section~\ref{sec:Mres}, we described how we estimate the masses of the heavy-strange(light) vector and scalar mesons used in the pole factor $P(q^2)$.
With the pseudoscalar meson mass fixed, the splitting between the pseudoscalar and vector mesons are changed by $\pm 50\%$, similarly for the splitting of the pseudoscalar and scalar mesons.
The agreement of the fits here suggests that the approximations made in Section~\ref{sec:Mres} are appropriate.
Finally, we show a fit that uses correlation functions in which the priors for $V_{\mathrm{nn}, 00}$ of each insertion are $25\%$ wider.

% replacing $(-z^n)$ with $(-z^n) - (-1)^{N+1-n} (-z)^{N+1}$ in the fit forms for $f_+$ and $f_T$ in Eq.~\eqref{eqn:physcont_ffs} to match the Bourreley- Caprini-Lellouch (BCL) parametrisation~\cite{Bourrely:2008za}.
%These $(-z)^{N+1}$ terms impose the constraint $df_{+,T}/dz = 0$ at $z=-1$ ($q^2 = t_+$).
%Fits with this constraint give similar results as without the $z^{N+1}$ terms.
%
%Next, we perform fits where data at the largest heavy quark mass ($am_h = 0.8$) and the smallest heavy quark mass ($am_h = 0.5$) are excluded from each set.
%Next, we consider fits where data on sets fine, superfine and ultrafine are excluded in turn.
%Then, we show fits where $M_\mathrm{res}$ in the pole factor $P(q^2)$ in the fit function in Eq.~\eqref{hhisqffff} is varied.
%In Section~\ref{sec:Mres}, we described how we estimate the masses of the heavy-strange(light) vector and scalar mesons used in the pole factor $P(q^2)$ (see Eq.~\eqref{hhisqffff}).
%With the pseudoscalar meson mass fixed, the splitting between the pseudoscalar and vector mesons are changed by $\pm 50\%$, similarly for the splitting of the pseudoscalar and scalar mesons.
%Finally, we show a fit that uses correlation functions in which the priors for $V_{\mathrm{nn}, 00}$ of each insertion are $25\%$ wider.
%Fig.~\ref{fig:ff_stability} shows the form factor values at $q^2=0$ and at zero-recoil achieved by these alternative fits alongside the fit including all the data (labelled \lq final\rq).
Good agreement is observed between the fits shown in Fig.~\ref{fig:ff_stability}.
Hence, we conclude that our fit of the form factors is robust.
%In particularly, very little effect is observed by increasing the prior widths varying the pole masses.
%We conclude that the approximations made in Section~\ref{sec:Mres} are appropriate here.
%Excluding data at specific heavy quark masses and data on the finest lattices increases the error on the form factors.
%The two fits where set \emph{uf5} and large $am_h$ excluded show the largest errors, indicative that it is this is data that is most influential on the final form factors at the physical-continuum limit.
%This is unsurprising since parameters associated with set \emph{uf5} and $am_h = 0.8$ are those closest to the physical-continuum point at which we ultimately wish to obtain the form factors $f_{0,+}$.
%Comparing the \lq final\rq result with those without non-zero-recoil data on \emph{uf5} on Fig.~\ref{fig:ff_stability}, it can be seen that including the full dataset on \emph{uf5} does appear to reduce the errors of the form factors.
%Thus, the HISQ propagators at the mass of the strange quark with non-zero momentum are valuable to this study.

%%
%\begin{figure*}
%	\centering
%	\includegraphics[width=0.85\textwidth]{stability_ff_BcDs}
%	\caption{For each of the $81$ different fits indexed by $J$ (see Equation (\ref{eqn:defn_J})), we show the parameters $(c_{0,+})^{(0,1)}_{0000}$ of the physical-continuum form factors for $B_c \to D$. These plots show results from all possible combinations of the correlation function fittings described in Table \ref{tab:ff_stab}.}
%	\label{fig:ff_stability_with corrf_BcDs}
%\end{figure*}
%%
%update1this
%

\section{Reconstructing the form factors} \label{app:reconstruct_ff}

We now provide instructions for reconstructing our form factors in the continuum limit with physical quark masses.
For the convenience of the reader, we have provided the script \texttt{construct\_ffs.py} which constructs our form factors.

The form factors in the continuum limit ($a \to 0$) and the limit of physical masses ($\delta m = 0$ in Eq.~\eqref{eqn:defn_delta_m}) are shown in Fig.~\ref{fig:final_ffs}.
In these limits, the fit form collapses to the physical-continuum parametrisation
\begin{align}
	f(q^2) &= P(q^2)^{-1} \sum_{n=0}^{N_n} c^{(n)} \hat{z}^{(n, N_n)}. \label{eqn:physcont_ffs}
\end{align}
The values for the pole factors $P(q^2)=1-q^2/M^2_{\text{res}}$ in the case $m_h < m_b$ are discussed in Section~\ref{sec:Mres}.
For $m_h = m_b$, we use the $M_{\mathrm{res}}$ values given in Table~\ref{tab:res_pole_masses}.
Recall that we define $ \hat{z}^{(n, N_n)}$ in Eq.~\eqref{eq:zhat} and we take $t_0 = 0$ in Eq.~\eqref{eqn:BcD_littlez}.
In the limit of vanishing lattice spacing and physical quark masses, the coefficients $c^{(n)}$ of the $\hat{z}^{(n, N_n)}$-polynomial $P(q^2) f(q^2)$ are given by
\begin{align}
	c^{(n)}&= \mathcal{L} \sum_{r=0}^{N_r} A^{(nr00)} \Omega^{(n)} \left( \frac{\Lambda}{M_{H_{l(s)}}}\right)^r  \label{eqn:def_cn}
\end{align}
Here, the factor $\mathcal{L}$ is given in Eq.~\eqref{eq:curlyL} and we use the physical ratio $m_l / m_s$ given in  Eq.~\eqref{eq:mlms_ratio} to evaluate $x_{\pi} = m_l / 5.63 m_s^{\mathrm{tuned}}$.
The coefficients $\zeta$ are determined by the fit.
Also, the factors $\Omega^{(n)}$ given in Eq.~\eqref{eq:Omegadefn} are evaluated for $M_{H_{l(s)}} = M_{B_{l(s)}}$.

We now give values for the parameters needed to reconstruct the form factors using the form in Eq.~\eqref{eqn:physcont_ffs}.
Firstly, we take $N_n = 3$ and $N_r = 2$.
For $B_c \to D_l$ and $B_c \to D_s$, coefficients $c^{(n)}$ are given in 
%Tables~\ref{tab:ff_coeffs_BcDu} and~\ref{tab:ff_coeffs_BcDs} respectively, and we provide these coefficients 
in the files \texttt{cn\_BcDl.py} and \texttt{cn\_BcDs.py}.
Table~\ref{tab:meson_masses} gives all meson masses required to construct the form factors.
For  $B_c \to D_l$, we use $t_- = (M_{B_c (0^-)} - M_{D^(0-)})^2$ and $t_+ = (M_{B(0^-)} + M_{\pi (0^-)})^2$.
For  $B_c \to D_s$, we use $t_- = (M_{B_c (0^-)} - M_{D_s (0^-)})^2$ and $t_+ = (M_{B (0^-)} + M_{K (0^-)})^2$.
\begin{table}
	\centering
	\caption{Values we take for various different meson masses. To reconstruct out form factors, these masses should be used in conjunction with the coefficients $c_{0,+,T}^{(n)}$ in the files \texttt{cn\_BcDl.py} and \texttt{cn\_BcDs.py} via Eq.~\eqref{eqn:physcont_ffs}. The bottom four rows gives the masses used in the pole factor $P(q^2)$, and the middle three rows are used to construct $t_+$ needed to transform $q^2$ into $z$ via Eq.~\eqref{eqn:BcD_littlez}. The top three rows allow $q^2_{\mathrm{max}} = t_-$ to be found. These masses feature in the script \texttt{construct\_ffs.py} which we provide.}
	\begin{tabular}{c | c}
		\hline\hline
		meson ($J^P$) & mass GeV\\ [0.1ex]
		\hline
		$B_c (0^-)$ & $6.2749$~\cite{PDG}  \\
		$D (0^-)$ & $1.8648$~\cite{PDG}  \\
		$D_s (0^-)$ & $1.9690$~\cite{PDG}  \\
		\hline
		$B (0^-)$ & $5.27964$~\cite{PDG} \\
		$\pi (0^-)$ & $0.134977$~\cite{PDG}  \\
		$K (0^-)$ & $0.497611$~\cite{PDG}  \\
		\hline
		$B (1^-)$ & $5.324$~\cite{PDG}  \\
		$B (0^+)$ & $5.627$~\cite{Bardeen:2003kt}  \\
		$B_s (1^-)$ & $5.4158$~\cite{PDG}  \\
		$B_s (0^+)$ & $5.711$~\cite{Lang:2015hza}  \\
		\hline\hline
	\end{tabular}
	\label{tab:meson_masses}
\end{table}
Recall that the pole factor is given by $P (q^2) = 1-q^2/M^2_{\mathrm{res}}$ where
For $M_{\mathrm{res}}$, we take the masses of the mesons $B(0^+)$, $B(1^-)$, $B_s (0^+)$ and $B_s(1^-)$ for $f_{0}^l$, $f_{+}^l$, $f_{0}^s$ and $f_{+,T}^s$ respectively.
%For these masses, we use $5.627(35)$, $5.324(70)$, $5.711(13)(19)$ and $5.4158(15)$ respectively, again in GeV.
%For these masses, we use $5.627$, $5.324$, $5.711$ and $5.4158$ respectively, again in GeV.
The masses of the pseudoscalar and vector mesons are obtained from PDG~\cite{PDG}.
Estimates for the masses of the scalar mesons are obtained from~\cite{Bardeen:2003kt} and~\cite{Lang:2015hza}, though precise values are not necessary for our calculation.
We do not include an error on these values.
The reader should use these masses to exactly replicate the form factors shown in Fig.~\ref{fig:final_ffs}.
%Note that in Appendix~\ref{sec:ff_stability} we showed that the final form factors are not sensitive to the precise values of these pole masses.
%However, the coefficients given in Tables~\ref{tab:ff_coeffs_BcDu} and~\ref{tab:ff_coeffs_BcDs} correspond to these particular values.
%Hence, the reader should use these masses to exactly replicate the form factors shown in Fig.~\ref{fig:final_ffs}.

%===========================================================================
\bibliography{BcD_BcDs_form_factors}{}

%merlin.mbs apsrev4-1.bst 2010-07-25 4.21a (PWD, AO, DPC) hacked
%Control: key (0)
%Control: author (72) initials jnrlst
%Control: editor formatted (1) identically to author
%Control: production of article title (-1) disabled
%Control: page (0) single
%Control: year (1) truncated
%Control: production of eprint (0) enabled
\begin{thebibliography}{68}%
\makeatletter
\providecommand \@ifxundefined [1]{%
 \@ifx{#1\undefined}
}%
\providecommand \@ifnum [1]{%
 \ifnum #1\expandafter \@firstoftwo
 \else \expandafter \@secondoftwo
 \fi
}%
\providecommand \@ifx [1]{%
 \ifx #1\expandafter \@firstoftwo
 \else \expandafter \@secondoftwo
 \fi
}%
\providecommand \natexlab [1]{#1}%
\providecommand \enquote  [1]{``#1''}%
\providecommand \bibnamefont  [1]{#1}%
\providecommand \bibfnamefont [1]{#1}%
\providecommand \citenamefont [1]{#1}%
\providecommand \href@noop [0]{\@secondoftwo}%
\providecommand \href [0]{\begingroup \@sanitize@url \@href}%
\providecommand \@href[1]{\@@startlink{#1}\@@href}%
\providecommand \@@href[1]{\endgroup#1\@@endlink}%
\providecommand \@sanitize@url [0]{\catcode `\\12\catcode `\$12\catcode
  `\&12\catcode `\#12\catcode `\^12\catcode `\_12\catcode `\%12\relax}%
\providecommand \@@startlink[1]{}%
\providecommand \@@endlink[0]{}%
\providecommand \url  [0]{\begingroup\@sanitize@url \@url }%
\providecommand \@url [1]{\endgroup\@href {#1}{\urlprefix }}%
\providecommand \urlprefix  [0]{URL }%
\providecommand \Eprint [0]{\href }%
\providecommand \doibase [0]{http://dx.doi.org/}%
\providecommand \selectlanguage [0]{\@gobble}%
\providecommand \bibinfo  [0]{\@secondoftwo}%
\providecommand \bibfield  [0]{\@secondoftwo}%
\providecommand \translation [1]{[#1]}%
\providecommand \BibitemOpen [0]{}%
\providecommand \bibitemStop [0]{}%
\providecommand \bibitemNoStop [0]{.\EOS\space}%
\providecommand \EOS [0]{\spacefactor3000\relax}%
\providecommand \BibitemShut  [1]{\csname bibitem#1\endcsname}%
\let\auto@bib@innerbib\@empty
%</preamble>
\bibitem [{LHC()}]{LHCbImplications}%
  \BibitemOpen
  \href@noop {} {\bibinfo  {journal} {B. Khanji \textit{et al.} (LHCb) (2020),
  \textit{Proceedings of the Implications of LHCb measurements and future
  prospects}}\ }\BibitemShut {NoStop}%
\bibitem [{\citenamefont {Amhis}\ \emph {et~al.}(2021)\citenamefont {Amhis}
  \emph {et~al.}}]{HFLAV:2019otj}%
  \BibitemOpen
\bibfield  {journal} {  }\bibfield  {author} {\bibinfo {author} {\bibfnamefont
  {Y.~S.}\ \bibnamefont {Amhis}} \emph {et~al.} (\bibinfo {collaboration}
  {HFLAV}),\ }\href {\doibase 10.1140/epjc/s10052-020-8156-7} {\bibfield
  {journal} {\bibinfo  {journal} {Eur. Phys. J. C}\ }\textbf {\bibinfo {volume}
  {81}},\ \bibinfo {pages} {226} (\bibinfo {year} {2021})},\ \Eprint
  {http://arxiv.org/abs/1909.12524} {arXiv:1909.12524 [hep-ex]} \BibitemShut
  {NoStop}%
\bibitem [{\citenamefont {Harrison}\ \emph {et~al.}(2020)\citenamefont
  {Harrison}, \citenamefont {Davies},\ and\ \citenamefont
  {Lytle}}]{Harrison:2020gvo}%
  \BibitemOpen
  \bibfield  {author} {\bibinfo {author} {\bibfnamefont {J.}~\bibnamefont
  {Harrison}}, \bibinfo {author} {\bibfnamefont {C.~T.~H.}\ \bibnamefont
  {Davies}}, \ and\ \bibinfo {author} {\bibfnamefont {A.}~\bibnamefont {Lytle}}
  (\bibinfo {collaboration} {HPQCD}),\ }\href {\doibase
  10.1103/PhysRevD.102.094518} {\bibfield  {journal} {\bibinfo  {journal}
  {Phys. Rev. D}\ }\textbf {\bibinfo {volume} {102}},\ \bibinfo {pages}
  {094518} (\bibinfo {year} {2020})},\ \Eprint
  {http://arxiv.org/abs/2007.06957} {arXiv:2007.06957 [hep-lat]} \BibitemShut
  {NoStop}%
\bibitem [{\citenamefont {Follana}\ \emph {et~al.}(2007)\citenamefont
  {Follana}, \citenamefont {Mason}, \citenamefont {Davies}, \citenamefont
  {Hornbostel}, \citenamefont {Lepage}, \citenamefont {Shigemitsu},
  \citenamefont {Trottier},\ and\ \citenamefont {Wong}}]{Follana:2006rc}%
  \BibitemOpen
  \bibfield  {author} {\bibinfo {author} {\bibfnamefont {E.}~\bibnamefont
  {Follana}}, \bibinfo {author} {\bibfnamefont {Q.}~\bibnamefont {Mason}},
  \bibinfo {author} {\bibfnamefont {C.}~\bibnamefont {Davies}}, \bibinfo
  {author} {\bibfnamefont {K.}~\bibnamefont {Hornbostel}}, \bibinfo {author}
  {\bibfnamefont {G.~P.}\ \bibnamefont {Lepage}}, \bibinfo {author}
  {\bibfnamefont {J.}~\bibnamefont {Shigemitsu}}, \bibinfo {author}
  {\bibfnamefont {H.}~\bibnamefont {Trottier}}, \ and\ \bibinfo {author}
  {\bibfnamefont {K.}~\bibnamefont {Wong}} (\bibinfo {collaboration} {HPQCD,
  UKQCD}),\ }\href {\doibase 10.1103/PhysRevD.75.054502} {\bibfield  {journal}
  {\bibinfo  {journal} {Phys. Rev.}\ }\textbf {\bibinfo {volume} {D75}},\
  \bibinfo {pages} {054502} (\bibinfo {year} {2007})},\ \Eprint
  {http://arxiv.org/abs/hep-lat/0610092} {arXiv:hep-lat/0610092 [hep-lat]}
  \BibitemShut {NoStop}%
%%CITATION = HEP-LAT/0610092;%%
\bibitem [{\citenamefont {McLean}\ \emph {et~al.}(2019)\citenamefont {McLean},
  \citenamefont {Davies}, \citenamefont {Lytle},\ and\ \citenamefont
  {Koponen}}]{McLean:2019sds}%
  \BibitemOpen
  \bibfield  {author} {\bibinfo {author} {\bibfnamefont {E.}~\bibnamefont
  {McLean}}, \bibinfo {author} {\bibfnamefont {C.~T.~H.}\ \bibnamefont
  {Davies}}, \bibinfo {author} {\bibfnamefont {A.~T.}\ \bibnamefont {Lytle}}, \
  and\ \bibinfo {author} {\bibfnamefont {J.}~\bibnamefont {Koponen}},\ }\href
  {\doibase 10.1103/PhysRevD.99.114512} {\bibfield  {journal} {\bibinfo
  {journal} {Phys. Rev.}\ }\textbf {\bibinfo {volume} {D99}},\ \bibinfo {pages}
  {114512} (\bibinfo {year} {2019})},\ \Eprint
  {http://arxiv.org/abs/1904.02046} {arXiv:1904.02046 [hep-lat]} \BibitemShut
  {NoStop}%
%%CITATION = ARXIV:1904.02046;%%
\bibitem [{\citenamefont {McLean}\ \emph {et~al.}(2020)\citenamefont {McLean},
  \citenamefont {Davies}, \citenamefont {Koponen},\ and\ \citenamefont
  {Lytle}}]{McLean:2019qcx}%
  \BibitemOpen
  \bibfield  {author} {\bibinfo {author} {\bibfnamefont {E.}~\bibnamefont
  {McLean}}, \bibinfo {author} {\bibfnamefont {C.~T.~H.}\ \bibnamefont
  {Davies}}, \bibinfo {author} {\bibfnamefont {J.}~\bibnamefont {Koponen}}, \
  and\ \bibinfo {author} {\bibfnamefont {A.~T.}\ \bibnamefont {Lytle}},\ }\href
  {\doibase 10.1103/PhysRevD.101.074513} {\bibfield  {journal} {\bibinfo
  {journal} {Phys. Rev. D}\ }\textbf {\bibinfo {volume} {101}},\ \bibinfo
  {pages} {074513} (\bibinfo {year} {2020})},\ \Eprint
  {http://arxiv.org/abs/1906.00701} {arXiv:1906.00701 [hep-lat]} \BibitemShut
  {NoStop}%
\bibitem [{\citenamefont {Cooper}\ \emph {et~al.}(2020)\citenamefont {Cooper},
  \citenamefont {Davies}, \citenamefont {Harrison}, \citenamefont {Komijani},\
  and\ \citenamefont {Wingate}}]{Cooper:2020wnj}%
  \BibitemOpen
  \bibfield  {author} {\bibinfo {author} {\bibfnamefont {L.~J.}\ \bibnamefont
  {Cooper}}, \bibinfo {author} {\bibfnamefont {C.~T.~H.}\ \bibnamefont
  {Davies}}, \bibinfo {author} {\bibfnamefont {J.}~\bibnamefont {Harrison}},
  \bibinfo {author} {\bibfnamefont {J.}~\bibnamefont {Komijani}}, \ and\
  \bibinfo {author} {\bibfnamefont {M.}~\bibnamefont {Wingate}} (\bibinfo
  {collaboration} {HPQCD}),\ }\href {\doibase 10.1103/PhysRevD.102.014513}
  {\bibfield  {journal} {\bibinfo  {journal} {Phys. Rev. D}\ }\textbf {\bibinfo
  {volume} {102}},\ \bibinfo {pages} {014513} (\bibinfo {year} {2020})},\
  \bibinfo {note} {[Erratum: Phys.Rev.D 103, 099901 (2021)]},\ \Eprint
  {http://arxiv.org/abs/2003.00914} {arXiv:2003.00914 [hep-lat]} \BibitemShut
  {NoStop}%
\bibitem [{\citenamefont {Parrott}\ \emph {et~al.}(2021)\citenamefont
  {Parrott}, \citenamefont {Bouchard}, \citenamefont {Davies},\ and\
  \citenamefont {Hatton}}]{Parrott:2020vbe}%
  \BibitemOpen
  \bibfield  {author} {\bibinfo {author} {\bibfnamefont {W.~G.}\ \bibnamefont
  {Parrott}}, \bibinfo {author} {\bibfnamefont {C.}~\bibnamefont {Bouchard}},
  \bibinfo {author} {\bibfnamefont {C.~T.~H.}\ \bibnamefont {Davies}}, \ and\
  \bibinfo {author} {\bibfnamefont {D.}~\bibnamefont {Hatton}},\ }\href
  {\doibase 10.1103/PhysRevD.103.094506} {\bibfield  {journal} {\bibinfo
  {journal} {Phys. Rev. D}\ }\textbf {\bibinfo {volume} {103}},\ \bibinfo
  {pages} {094506} (\bibinfo {year} {2021})},\ \Eprint
  {http://arxiv.org/abs/2010.07980} {arXiv:2010.07980 [hep-lat]} \BibitemShut
  {NoStop}%
\bibitem [{\citenamefont {Harrison}\ and\ \citenamefont
  {Davies}(2021)}]{Harrison:2021tol}%
  \BibitemOpen
  \bibfield  {author} {\bibinfo {author} {\bibfnamefont {J.}~\bibnamefont
  {Harrison}}\ and\ \bibinfo {author} {\bibfnamefont {C.~T.~H.}\ \bibnamefont
  {Davies}} (\bibinfo {collaboration} {LATTICE-HPQCD}),\ }\href@noop {} {\
  (\bibinfo {year} {2021})},\ \Eprint {http://arxiv.org/abs/2105.11433}
  {arXiv:2105.11433 [hep-lat]} \BibitemShut {NoStop}%
\bibitem [{\citenamefont {Sirlin}(1982)}]{Sirlin:1981ie}%
  \BibitemOpen
  \bibfield  {author} {\bibinfo {author} {\bibfnamefont {A.}~\bibnamefont
  {Sirlin}},\ }\href {\doibase 10.1016/0550-3213(82)90303-0} {\bibfield
  {journal} {\bibinfo  {journal} {Nucl. Phys. B}\ }\textbf {\bibinfo {volume}
  {196}},\ \bibinfo {pages} {83} (\bibinfo {year} {1982})}\BibitemShut
  {NoStop}%
\bibitem [{\citenamefont {Bouchard}\ \emph {et~al.}(2013)\citenamefont
  {Bouchard}, \citenamefont {Lepage}, \citenamefont {Monahan}, \citenamefont
  {Na},\ and\ \citenamefont {Shigemitsu}}]{Bouchard:2013eph}%
  \BibitemOpen
  \bibfield  {author} {\bibinfo {author} {\bibfnamefont {C.}~\bibnamefont
  {Bouchard}}, \bibinfo {author} {\bibfnamefont {G.}~\bibnamefont {Lepage}},
  \bibinfo {author} {\bibfnamefont {C.}~\bibnamefont {Monahan}}, \bibinfo
  {author} {\bibfnamefont {H.}~\bibnamefont {Na}}, \ and\ \bibinfo {author}
  {\bibfnamefont {J.}~\bibnamefont {Shigemitsu}} (\bibinfo {collaboration}
  {HPQCD}),\ }\href {\doibase 10.1103/PhysRevD.88.054509} {\bibfield  {journal}
  {\bibinfo  {journal} {Phys. Rev. D}\ }\textbf {\bibinfo {volume} {88}},\
  \bibinfo {pages} {054509} (\bibinfo {year} {2013})},\ \bibinfo {note}
  {[Erratum: Phys.Rev.D 88, 079901 (2013)]},\ \Eprint
  {http://arxiv.org/abs/1306.2384} {arXiv:1306.2384 [hep-lat]} \BibitemShut
  {NoStop}%
\bibitem [{\citenamefont {Buras}\ \emph {et~al.}(2015)\citenamefont {Buras},
  \citenamefont {Girrbach-Noe}, \citenamefont {Niehoff},\ and\ \citenamefont
  {Straub}}]{Buras:2014fpa}%
  \BibitemOpen
  \bibfield  {author} {\bibinfo {author} {\bibfnamefont {A.~J.}\ \bibnamefont
  {Buras}}, \bibinfo {author} {\bibfnamefont {J.}~\bibnamefont {Girrbach-Noe}},
  \bibinfo {author} {\bibfnamefont {C.}~\bibnamefont {Niehoff}}, \ and\
  \bibinfo {author} {\bibfnamefont {D.~M.}\ \bibnamefont {Straub}},\ }\href
  {\doibase 10.1007/JHEP02(2015)184} {\bibfield  {journal} {\bibinfo  {journal}
  {JHEP}\ }\textbf {\bibinfo {volume} {02}},\ \bibinfo {pages} {184} (\bibinfo
  {year} {2015})},\ \Eprint {http://arxiv.org/abs/1409.4557} {arXiv:1409.4557
  [hep-ph]} \BibitemShut {NoStop}%
\bibitem [{\citenamefont {Altmannshofer}\ \emph {et~al.}(2009)\citenamefont
  {Altmannshofer}, \citenamefont {Buras}, \citenamefont {Straub},\ and\
  \citenamefont {Wick}}]{Altmannshofer:2009ma}%
  \BibitemOpen
  \bibfield  {author} {\bibinfo {author} {\bibfnamefont {W.}~\bibnamefont
  {Altmannshofer}}, \bibinfo {author} {\bibfnamefont {A.~J.}\ \bibnamefont
  {Buras}}, \bibinfo {author} {\bibfnamefont {D.~M.}\ \bibnamefont {Straub}}, \
  and\ \bibinfo {author} {\bibfnamefont {M.}~\bibnamefont {Wick}},\ }\href
  {\doibase 10.1088/1126-6708/2009/04/022} {\bibfield  {journal} {\bibinfo
  {journal} {JHEP}\ }\textbf {\bibinfo {volume} {04}},\ \bibinfo {pages} {022}
  (\bibinfo {year} {2009})},\ \Eprint {http://arxiv.org/abs/0902.0160}
  {arXiv:0902.0160 [hep-ph]} \BibitemShut {NoStop}%
\bibitem [{\citenamefont {Bazavov}\ \emph {et~al.}(2010)\citenamefont {Bazavov}
  \emph {et~al.}}]{Bazavov:2010ru}%
  \BibitemOpen
  \bibfield  {author} {\bibinfo {author} {\bibfnamefont {A.}~\bibnamefont
  {Bazavov}} \emph {et~al.} (\bibinfo {collaboration} {MILC}),\ }\href
  {\doibase 10.1103/PhysRevD.82.074501} {\bibfield  {journal} {\bibinfo
  {journal} {Phys. Rev.}\ }\textbf {\bibinfo {volume} {D82}},\ \bibinfo {pages}
  {074501} (\bibinfo {year} {2010})},\ \Eprint {http://arxiv.org/abs/1004.0342}
  {arXiv:1004.0342 [hep-lat]} \BibitemShut {NoStop}%
%%CITATION = ARXIV:1004.0342;%%
\bibitem [{\citenamefont {Bazavov}\ \emph {et~al.}(2013)\citenamefont {Bazavov}
  \emph {et~al.}}]{Bazavov:2012xda}%
  \BibitemOpen
  \bibfield  {author} {\bibinfo {author} {\bibfnamefont {A.}~\bibnamefont
  {Bazavov}} \emph {et~al.} (\bibinfo {collaboration} {MILC}),\ }\href
  {\doibase 10.1103/PhysRevD.87.054505} {\bibfield  {journal} {\bibinfo
  {journal} {Phys. Rev.}\ }\textbf {\bibinfo {volume} {D87}},\ \bibinfo {pages}
  {054505} (\bibinfo {year} {2013})},\ \Eprint {http://arxiv.org/abs/1212.4768}
  {arXiv:1212.4768 [hep-lat]} \BibitemShut {NoStop}%
%%CITATION = ARXIV:1212.4768;%%
\bibitem [{\citenamefont {Bazavov}\ \emph {et~al.}(2016)\citenamefont {Bazavov}
  \emph {et~al.}}]{Bazavov:2015yea}%
  \BibitemOpen
  \bibfield  {author} {\bibinfo {author} {\bibfnamefont {A.}~\bibnamefont
  {Bazavov}} \emph {et~al.} (\bibinfo {collaboration} {MILC}),\ }\href
  {\doibase 10.1103/PhysRevD.93.094510} {\bibfield  {journal} {\bibinfo
  {journal} {Phys. Rev.}\ }\textbf {\bibinfo {volume} {D93}},\ \bibinfo {pages}
  {094510} (\bibinfo {year} {2016})},\ \Eprint
  {http://arxiv.org/abs/1503.02769} {arXiv:1503.02769 [hep-lat]} \BibitemShut
  {NoStop}%
%%CITATION = ARXIV:1503.02769;%%
\bibitem [{\citenamefont {Hart}\ \emph {et~al.}(2009)\citenamefont {Hart},
  \citenamefont {von Hippel},\ and\ \citenamefont {Horgan}}]{Hart:2008sq}%
  \BibitemOpen
  \bibfield  {author} {\bibinfo {author} {\bibfnamefont {A.}~\bibnamefont
  {Hart}}, \bibinfo {author} {\bibfnamefont {G.~M.}\ \bibnamefont {von
  Hippel}}, \ and\ \bibinfo {author} {\bibfnamefont {R.~R.}\ \bibnamefont
  {Horgan}} (\bibinfo {collaboration} {HPQCD}),\ }\href {\doibase
  10.1103/PhysRevD.79.074008} {\bibfield  {journal} {\bibinfo  {journal} {Phys.
  Rev.}\ }\textbf {\bibinfo {volume} {D79}},\ \bibinfo {pages} {074008}
  (\bibinfo {year} {2009})},\ \Eprint {http://arxiv.org/abs/0812.0503}
  {arXiv:0812.0503 [hep-lat]} \BibitemShut {NoStop}%
%%CITATION = ARXIV:0812.0503;%%
\bibitem [{\citenamefont {Borsanyi}\ \emph {et~al.}(2012)\citenamefont
  {Borsanyi} \emph {et~al.}}]{Borsanyi:2012zs}%
  \BibitemOpen
  \bibfield  {author} {\bibinfo {author} {\bibfnamefont {S.}~\bibnamefont
  {Borsanyi}} \emph {et~al.},\ }\href {\doibase 10.1007/JHEP09(2012)010}
  {\bibfield  {journal} {\bibinfo  {journal} {JHEP}\ }\textbf {\bibinfo
  {volume} {09}},\ \bibinfo {pages} {010} (\bibinfo {year} {2012})},\ \Eprint
  {http://arxiv.org/abs/1203.4469} {arXiv:1203.4469 [hep-lat]} \BibitemShut
  {NoStop}%
%%CITATION = ARXIV:1203.4469;%%
\bibitem [{\citenamefont {Dowdall}\ \emph {et~al.}(2013)\citenamefont
  {Dowdall}, \citenamefont {Davies}, \citenamefont {Lepage},\ and\
  \citenamefont {McNeile}}]{Dowdall:2013rya}%
  \BibitemOpen
  \bibfield  {author} {\bibinfo {author} {\bibfnamefont {R.~J.}\ \bibnamefont
  {Dowdall}}, \bibinfo {author} {\bibfnamefont {C.~T.~H.}\ \bibnamefont
  {Davies}}, \bibinfo {author} {\bibfnamefont {G.~P.}\ \bibnamefont {Lepage}},
  \ and\ \bibinfo {author} {\bibfnamefont {C.}~\bibnamefont {McNeile}},\ }\href
  {\doibase 10.1103/PhysRevD.88.074504} {\bibfield  {journal} {\bibinfo
  {journal} {Phys. Rev. D}\ }\textbf {\bibinfo {volume} {88}},\ \bibinfo
  {pages} {074504} (\bibinfo {year} {2013})},\ \Eprint
  {http://arxiv.org/abs/1303.1670} {arXiv:1303.1670 [hep-lat]} \BibitemShut
  {NoStop}%
\bibitem [{\citenamefont {Chakraborty}\ \emph {et~al.}(2017)\citenamefont
  {Chakraborty}, \citenamefont {Davies}, \citenamefont {de~Oliviera},
  \citenamefont {Koponen}, \citenamefont {Lepage},\ and\ \citenamefont {Van~de
  Water}}]{Chakraborty:2016mwy}%
  \BibitemOpen
  \bibfield  {author} {\bibinfo {author} {\bibfnamefont {B.}~\bibnamefont
  {Chakraborty}}, \bibinfo {author} {\bibfnamefont {C.~T.~H.}\ \bibnamefont
  {Davies}}, \bibinfo {author} {\bibfnamefont {P.~G.}\ \bibnamefont
  {de~Oliviera}}, \bibinfo {author} {\bibfnamefont {J.}~\bibnamefont
  {Koponen}}, \bibinfo {author} {\bibfnamefont {G.~P.}\ \bibnamefont {Lepage}},
  \ and\ \bibinfo {author} {\bibfnamefont {R.~S.}\ \bibnamefont {Van~de
  Water}},\ }\href {\doibase 10.1103/PhysRevD.96.034516} {\bibfield  {journal}
  {\bibinfo  {journal} {Phys. Rev. D}\ }\textbf {\bibinfo {volume} {96}},\
  \bibinfo {pages} {034516} (\bibinfo {year} {2017})},\ \Eprint
  {http://arxiv.org/abs/1601.03071} {arXiv:1601.03071 [hep-lat]} \BibitemShut
  {NoStop}%
\bibitem [{\citenamefont {Chakraborty}\ \emph {et~al.}(2015)\citenamefont
  {Chakraborty}, \citenamefont {Davies}, \citenamefont {Galloway},
  \citenamefont {Knecht}, \citenamefont {Koponen}, \citenamefont {Donald},
  \citenamefont {Dowdall}, \citenamefont {Lepage},\ and\ \citenamefont
  {McNeile}}]{Chakraborty:2014aca}%
  \BibitemOpen
  \bibfield  {author} {\bibinfo {author} {\bibfnamefont {B.}~\bibnamefont
  {Chakraborty}}, \bibinfo {author} {\bibfnamefont {C.~T.~H.}\ \bibnamefont
  {Davies}}, \bibinfo {author} {\bibfnamefont {B.}~\bibnamefont {Galloway}},
  \bibinfo {author} {\bibfnamefont {P.}~\bibnamefont {Knecht}}, \bibinfo
  {author} {\bibfnamefont {J.}~\bibnamefont {Koponen}}, \bibinfo {author}
  {\bibfnamefont {G.~C.}\ \bibnamefont {Donald}}, \bibinfo {author}
  {\bibfnamefont {R.~J.}\ \bibnamefont {Dowdall}}, \bibinfo {author}
  {\bibfnamefont {G.~P.}\ \bibnamefont {Lepage}}, \ and\ \bibinfo {author}
  {\bibfnamefont {C.}~\bibnamefont {McNeile}},\ }\href {\doibase
  10.1103/PhysRevD.91.054508} {\bibfield  {journal} {\bibinfo  {journal} {Phys.
  Rev.}\ }\textbf {\bibinfo {volume} {D91}},\ \bibinfo {pages} {054508}
  (\bibinfo {year} {2015})},\ \Eprint {http://arxiv.org/abs/1408.4169}
  {arXiv:1408.4169 [hep-lat]} \BibitemShut {NoStop}%
%%CITATION = ARXIV:1408.4169;%%
\bibitem [{\citenamefont {Bazavov}\ \emph {et~al.}(2014)\citenamefont {Bazavov}
  \emph {et~al.}}]{FermilabLattice:2014tsy}%
  \BibitemOpen
  \bibfield  {author} {\bibinfo {author} {\bibfnamefont {A.}~\bibnamefont
  {Bazavov}} \emph {et~al.} (\bibinfo {collaboration} {Fermilab Lattice,
  MILC}),\ }\href {\doibase 10.1103/PhysRevD.90.074509} {\bibfield  {journal}
  {\bibinfo  {journal} {Phys. Rev. D}\ }\textbf {\bibinfo {volume} {90}},\
  \bibinfo {pages} {074509} (\bibinfo {year} {2014})},\ \Eprint
  {http://arxiv.org/abs/1407.3772} {arXiv:1407.3772 [hep-lat]} \BibitemShut
  {NoStop}%
\bibitem [{\citenamefont {Bazavov}\ \emph {et~al.}(2018)\citenamefont {Bazavov}
  \emph {et~al.}}]{Bazavov:2017lyh}%
  \BibitemOpen
  \bibfield  {author} {\bibinfo {author} {\bibfnamefont {A.}~\bibnamefont
  {Bazavov}} \emph {et~al.},\ }\href {\doibase 10.1103/PhysRevD.98.074512}
  {\bibfield  {journal} {\bibinfo  {journal} {Phys. Rev. D}\ }\textbf {\bibinfo
  {volume} {98}},\ \bibinfo {pages} {074512} (\bibinfo {year} {2018})},\
  \Eprint {http://arxiv.org/abs/1712.09262} {arXiv:1712.09262 [hep-lat]}
  \BibitemShut {NoStop}%
\bibitem [{\citenamefont {Koponen}\ \emph {et~al.}(2017)\citenamefont
  {Koponen}, \citenamefont {Zimermmane-Santos}, \citenamefont {Davies},
  \citenamefont {Lepage},\ and\ \citenamefont {Lytle}}]{Koponen:2017fvm}%
  \BibitemOpen
  \bibfield  {author} {\bibinfo {author} {\bibfnamefont {J.}~\bibnamefont
  {Koponen}}, \bibinfo {author} {\bibfnamefont {A.~C.}\ \bibnamefont
  {Zimermmane-Santos}}, \bibinfo {author} {\bibfnamefont {C.~T.~H.}\
  \bibnamefont {Davies}}, \bibinfo {author} {\bibfnamefont {G.~P.}\
  \bibnamefont {Lepage}}, \ and\ \bibinfo {author} {\bibfnamefont {A.~T.}\
  \bibnamefont {Lytle}},\ }\href {\doibase 10.1103/PhysRevD.96.054501}
  {\bibfield  {journal} {\bibinfo  {journal} {Phys. Rev.}\ }\textbf {\bibinfo
  {volume} {D96}},\ \bibinfo {pages} {054501} (\bibinfo {year} {2017})},\
  \Eprint {http://arxiv.org/abs/1701.04250} {arXiv:1701.04250 [hep-lat]}
  \BibitemShut {NoStop}%
%%CITATION = ARXIV:1701.04250;%%
\bibitem [{MIL()}]{MILCgithub}%
  \BibitemOpen
  \href@noop {} {}\bibinfo {note} {MILC Code Repository,
  https://github.com/milc-qcd}\BibitemShut {NoStop}%
\bibitem [{\citenamefont {Bernard}\ and\ \citenamefont
  {Toussaint}(2018)}]{Bernard:2017npd}%
  \BibitemOpen
  \bibfield  {author} {\bibinfo {author} {\bibfnamefont {C.}~\bibnamefont
  {Bernard}}\ and\ \bibinfo {author} {\bibfnamefont {D.}~\bibnamefont
  {Toussaint}} (\bibinfo {collaboration} {MILC}),\ }\href {\doibase
  10.1103/PhysRevD.97.074502} {\bibfield  {journal} {\bibinfo  {journal} {Phys.
  Rev. D}\ }\textbf {\bibinfo {volume} {97}},\ \bibinfo {pages} {074502}
  (\bibinfo {year} {2018})},\ \Eprint {http://arxiv.org/abs/1707.05430}
  {arXiv:1707.05430 [hep-lat]} \BibitemShut {NoStop}%
\bibitem [{\citenamefont {Tanabashi}\ \emph
  {et~al.}(2018{\natexlab{a}})\citenamefont {Tanabashi} \emph {et~al.}}]{PDG}%
  \BibitemOpen
  \bibfield  {author} {\bibinfo {author} {\bibfnamefont {M.}~\bibnamefont
  {Tanabashi}} \emph {et~al.} (\bibinfo {collaboration} {Particle Data
  Group}),\ }\href {\doibase 10.1103/PhysRevD.98.030001} {\bibfield  {journal}
  {\bibinfo  {journal} {Phys. Rev. D}\ }\textbf {\bibinfo {volume} {98}},\
  \bibinfo {pages} {030001} (\bibinfo {year} {2018}{\natexlab{a}})}\BibitemShut
  {NoStop}%
\bibitem [{\citenamefont {Sachrajda}\ and\ \citenamefont
  {Villadoro}(2005)}]{Sachrajda:2004mi}%
  \BibitemOpen
  \bibfield  {author} {\bibinfo {author} {\bibfnamefont {C.~T.}\ \bibnamefont
  {Sachrajda}}\ and\ \bibinfo {author} {\bibfnamefont {G.}~\bibnamefont
  {Villadoro}},\ }\href {\doibase 10.1016/j.physletb.2005.01.033} {\bibfield
  {journal} {\bibinfo  {journal} {Phys. Lett. B}\ }\textbf {\bibinfo {volume}
  {609}},\ \bibinfo {pages} {73} (\bibinfo {year} {2005})},\ \Eprint
  {http://arxiv.org/abs/hep-lat/0411033} {arXiv:hep-lat/0411033} \BibitemShut
  {NoStop}%
\bibitem [{\citenamefont {Guadagnoli}\ \emph {et~al.}(2006)\citenamefont
  {Guadagnoli}, \citenamefont {Mescia},\ and\ \citenamefont
  {Simula}}]{Guadagnoli:2005be}%
  \BibitemOpen
  \bibfield  {author} {\bibinfo {author} {\bibfnamefont {D.}~\bibnamefont
  {Guadagnoli}}, \bibinfo {author} {\bibfnamefont {F.}~\bibnamefont {Mescia}},
  \ and\ \bibinfo {author} {\bibfnamefont {S.}~\bibnamefont {Simula}},\ }\href
  {\doibase 10.1103/PhysRevD.73.114504} {\bibfield  {journal} {\bibinfo
  {journal} {Phys. Rev.}\ }\textbf {\bibinfo {volume} {D73}},\ \bibinfo {pages}
  {114504} (\bibinfo {year} {2006})},\ \Eprint
  {http://arxiv.org/abs/hep-lat/0512020} {arXiv:hep-lat/0512020 [hep-lat]}
  \BibitemShut {NoStop}%
%%CITATION = HEP-LAT/0512020;%%
\bibitem [{\citenamefont {Hatton}\ \emph {et~al.}(2019)\citenamefont {Hatton},
  \citenamefont {Davies}, \citenamefont {Lepage},\ and\ \citenamefont
  {Lytle}}]{Hatton:2019gha}%
  \BibitemOpen
  \bibfield  {author} {\bibinfo {author} {\bibfnamefont {D.}~\bibnamefont
  {Hatton}}, \bibinfo {author} {\bibfnamefont {C.~T.~H.}\ \bibnamefont
  {Davies}}, \bibinfo {author} {\bibfnamefont {G.~P.}\ \bibnamefont {Lepage}},
  \ and\ \bibinfo {author} {\bibfnamefont {A.~T.}\ \bibnamefont {Lytle}}
  (\bibinfo {collaboration} {HPQCD}),\ }\href {\doibase
  10.1103/PhysRevD.100.114513} {\bibfield  {journal} {\bibinfo  {journal}
  {Phys. Rev. D}\ }\textbf {\bibinfo {volume} {100}},\ \bibinfo {pages}
  {114513} (\bibinfo {year} {2019})},\ \Eprint
  {http://arxiv.org/abs/1909.00756} {arXiv:1909.00756 [hep-lat]} \BibitemShut
  {NoStop}%
\bibitem [{\citenamefont {Hatton}\ \emph
  {et~al.}(2020{\natexlab{a}})\citenamefont {Hatton}, \citenamefont {Davies},
  \citenamefont {Lepage},\ and\ \citenamefont {Lytle}}]{Hatton:2020vzp}%
  \BibitemOpen
  \bibfield  {author} {\bibinfo {author} {\bibfnamefont {D.}~\bibnamefont
  {Hatton}}, \bibinfo {author} {\bibfnamefont {C.~T.~H.}\ \bibnamefont
  {Davies}}, \bibinfo {author} {\bibfnamefont {G.~P.}\ \bibnamefont {Lepage}},
  \ and\ \bibinfo {author} {\bibfnamefont {A.~T.}\ \bibnamefont {Lytle}}
  (\bibinfo {collaboration} {HPQCD}),\ }\href {\doibase
  10.1103/PhysRevD.102.094509} {\bibfield  {journal} {\bibinfo  {journal}
  {Phys. Rev. D}\ }\textbf {\bibinfo {volume} {102}},\ \bibinfo {pages}
  {094509} (\bibinfo {year} {2020}{\natexlab{a}})},\ \Eprint
  {http://arxiv.org/abs/2008.02024} {arXiv:2008.02024 [hep-lat]} \BibitemShut
  {NoStop}%
\bibitem [{\citenamefont {Lepage}(tter)}]{corrfitter}%
  \BibitemOpen
  \bibfield  {author} {\bibinfo {author} {\bibfnamefont {G.~P.}\ \bibnamefont
  {Lepage}},\ }\href@noop {} {\bibfield  {journal} {\bibinfo  {journal}
  {Corrfitter Version 6.0.7}\ } (\bibinfo {year}
  {github.com/gplepage/corrfitter})}\BibitemShut {NoStop}%
\bibitem [{\citenamefont {Lepage}\ \emph {et~al.}(2002)\citenamefont {Lepage},
  \citenamefont {Clark}, \citenamefont {Davies}, \citenamefont {Hornbostel},
  \citenamefont {Mackenzie}, \citenamefont {Morningstar},\ and\ \citenamefont
  {Trottier}}]{Lepage:2001ym}%
  \BibitemOpen
  \bibfield  {author} {\bibinfo {author} {\bibfnamefont {G.~P.}\ \bibnamefont
  {Lepage}}, \bibinfo {author} {\bibfnamefont {B.}~\bibnamefont {Clark}},
  \bibinfo {author} {\bibfnamefont {C.~T.~H.}\ \bibnamefont {Davies}}, \bibinfo
  {author} {\bibfnamefont {K.}~\bibnamefont {Hornbostel}}, \bibinfo {author}
  {\bibfnamefont {P.~B.}\ \bibnamefont {Mackenzie}}, \bibinfo {author}
  {\bibfnamefont {C.}~\bibnamefont {Morningstar}}, \ and\ \bibinfo {author}
  {\bibfnamefont {H.}~\bibnamefont {Trottier}},\ }\bibfield  {booktitle} {\emph
  {\bibinfo {booktitle} {{Lattice field theory. Proceedings, 19th International
  Symposium, Lattice 2001, Berlin, Germany, August 19-24, 2001}}},\ }\href
  {\doibase 10.1016/S0920-5632(01)01638-3} {\bibfield  {journal} {\bibinfo
  {journal} {Nucl. Phys. Proc. Suppl.}\ }\textbf {\bibinfo {volume} {106}},\
  \bibinfo {pages} {12} (\bibinfo {year} {2002})},\ \Eprint
  {http://arxiv.org/abs/hep-lat/0110175} {arXiv:hep-lat/0110175 [hep-lat]}
  \BibitemShut {NoStop}%
%%CITATION = HEP-LAT/0110175;%%
\bibitem [{\citenamefont {Hornbostel}\ \emph {et~al.}(2012)\citenamefont
  {Hornbostel}, \citenamefont {Lepage}, \citenamefont {Davies}, \citenamefont
  {Dowdall}, \citenamefont {Na},\ and\ \citenamefont
  {Shigemitsu}}]{Hornbostel:2011hu}%
  \BibitemOpen
  \bibfield  {author} {\bibinfo {author} {\bibfnamefont {K.}~\bibnamefont
  {Hornbostel}}, \bibinfo {author} {\bibfnamefont {G.~P.}\ \bibnamefont
  {Lepage}}, \bibinfo {author} {\bibfnamefont {C.~T.~H.}\ \bibnamefont
  {Davies}}, \bibinfo {author} {\bibfnamefont {R.~J.}\ \bibnamefont {Dowdall}},
  \bibinfo {author} {\bibfnamefont {H.}~\bibnamefont {Na}}, \ and\ \bibinfo
  {author} {\bibfnamefont {J.}~\bibnamefont {Shigemitsu}},\ }\href {\doibase
  10.1103/PhysRevD.85.031504} {\bibfield  {journal} {\bibinfo  {journal} {Phys.
  Rev. D}\ }\textbf {\bibinfo {volume} {85}},\ \bibinfo {pages} {031504}
  (\bibinfo {year} {2012})},\ \Eprint {http://arxiv.org/abs/1111.1363}
  {arXiv:1111.1363 [hep-lat]} \BibitemShut {NoStop}%
\bibitem [{\citenamefont {Bouchard}\ \emph {et~al.}(2014)\citenamefont
  {Bouchard}, \citenamefont {Lepage}, \citenamefont {Monahan}, \citenamefont
  {Na},\ and\ \citenamefont {Shigemitsu}}]{Bouchard:2014ypa}%
  \BibitemOpen
  \bibfield  {author} {\bibinfo {author} {\bibfnamefont {C.~M.}\ \bibnamefont
  {Bouchard}}, \bibinfo {author} {\bibfnamefont {G.~P.}\ \bibnamefont
  {Lepage}}, \bibinfo {author} {\bibfnamefont {C.}~\bibnamefont {Monahan}},
  \bibinfo {author} {\bibfnamefont {H.}~\bibnamefont {Na}}, \ and\ \bibinfo
  {author} {\bibfnamefont {J.}~\bibnamefont {Shigemitsu}},\ }\href {\doibase
  10.1103/PhysRevD.90.054506} {\bibfield  {journal} {\bibinfo  {journal} {Phys.
  Rev.}\ }\textbf {\bibinfo {volume} {D90}},\ \bibinfo {pages} {054506}
  (\bibinfo {year} {2014})},\ \Eprint {http://arxiv.org/abs/1406.2279}
  {arXiv:1406.2279 [hep-lat]} \BibitemShut {NoStop}%
%%CITATION = ARXIV:1406.2279;%%
\bibitem [{\citenamefont {Lepage}(qfit)}]{lsqfit}%
  \BibitemOpen
  \bibfield  {author} {\bibinfo {author} {\bibfnamefont {G.~P.}\ \bibnamefont
  {Lepage}},\ }\href@noop {} {\bibfield  {journal} {\bibinfo  {journal} {lsqfit
  Version 11.1}\ } (\bibinfo {year} {github.com/gplepage/lsqfit})}\BibitemShut
  {NoStop}%
\bibitem [{\citenamefont {Boyd}\ and\ \citenamefont
  {Savage}(1997)}]{Boyd:1997qw}%
  \BibitemOpen
  \bibfield  {author} {\bibinfo {author} {\bibfnamefont {C.~G.}\ \bibnamefont
  {Boyd}}\ and\ \bibinfo {author} {\bibfnamefont {M.~J.}\ \bibnamefont
  {Savage}},\ }\href {\doibase 10.1103/PhysRevD.56.303} {\bibfield  {journal}
  {\bibinfo  {journal} {Phys. Rev. D}\ }\textbf {\bibinfo {volume} {56}},\
  \bibinfo {pages} {303} (\bibinfo {year} {1997})},\ \Eprint
  {http://arxiv.org/abs/hep-ph/9702300} {arXiv:hep-ph/9702300} \BibitemShut
  {NoStop}%
\bibitem [{\citenamefont {Bourrely}\ \emph {et~al.}(2009)\citenamefont
  {Bourrely}, \citenamefont {Caprini},\ and\ \citenamefont
  {Lellouch}}]{Bourrely:2008za}%
  \BibitemOpen
  \bibfield  {author} {\bibinfo {author} {\bibfnamefont {C.}~\bibnamefont
  {Bourrely}}, \bibinfo {author} {\bibfnamefont {I.}~\bibnamefont {Caprini}}, \
  and\ \bibinfo {author} {\bibfnamefont {L.}~\bibnamefont {Lellouch}},\ }\href
  {\doibase 10.1103/PhysRevD.82.099902} {\bibfield  {journal} {\bibinfo
  {journal} {Phys. Rev. D}\ }\textbf {\bibinfo {volume} {79}},\ \bibinfo
  {pages} {013008} (\bibinfo {year} {2009})},\ \bibinfo {note} {[Erratum:
  Phys.Rev.D 82, 099902 (2010)]},\ \Eprint {http://arxiv.org/abs/0807.2722}
  {arXiv:0807.2722 [hep-ph]} \BibitemShut {NoStop}%
\bibitem [{\citenamefont {Chakraborty}\ \emph {et~al.}(2021)\citenamefont
  {Chakraborty}, \citenamefont {Parrott}, \citenamefont {Bouchard},
  \citenamefont {Davies}, \citenamefont {Koponen},\ and\ \citenamefont
  {Lepage}}]{Chakraborty:2021qav}%
  \BibitemOpen
  \bibfield  {author} {\bibinfo {author} {\bibfnamefont {B.}~\bibnamefont
  {Chakraborty}}, \bibinfo {author} {\bibfnamefont {W.~G.}\ \bibnamefont
  {Parrott}}, \bibinfo {author} {\bibfnamefont {C.}~\bibnamefont {Bouchard}},
  \bibinfo {author} {\bibfnamefont {C.~T.~H.}\ \bibnamefont {Davies}}, \bibinfo
  {author} {\bibfnamefont {J.}~\bibnamefont {Koponen}}, \ and\ \bibinfo
  {author} {\bibfnamefont {G.~P.}\ \bibnamefont {Lepage}} (\bibinfo
  {collaboration} {(HPQCD Collaboration)\textsection{}, HPQCD}),\ }\href
  {\doibase 10.1103/PhysRevD.104.034505} {\bibfield  {journal} {\bibinfo
  {journal} {Phys. Rev. D}\ }\textbf {\bibinfo {volume} {104}},\ \bibinfo
  {pages} {034505} (\bibinfo {year} {2021})},\ \Eprint
  {http://arxiv.org/abs/2104.09883} {arXiv:2104.09883 [hep-lat]} \BibitemShut
  {NoStop}%
\bibitem [{\citenamefont {Charles}\ \emph {et~al.}(1999)\citenamefont
  {Charles}, \citenamefont {Le~Yaouanc}, \citenamefont {Oliver}, \citenamefont
  {Pene},\ and\ \citenamefont {Raynal}}]{Charles:1998dr}%
  \BibitemOpen
  \bibfield  {author} {\bibinfo {author} {\bibfnamefont {J.}~\bibnamefont
  {Charles}}, \bibinfo {author} {\bibfnamefont {A.}~\bibnamefont {Le~Yaouanc}},
  \bibinfo {author} {\bibfnamefont {L.}~\bibnamefont {Oliver}}, \bibinfo
  {author} {\bibfnamefont {O.}~\bibnamefont {Pene}}, \ and\ \bibinfo {author}
  {\bibfnamefont {J.~C.}\ \bibnamefont {Raynal}},\ }\href {\doibase
  10.1103/PhysRevD.60.014001} {\bibfield  {journal} {\bibinfo  {journal} {Phys.
  Rev. D}\ }\textbf {\bibinfo {volume} {60}},\ \bibinfo {pages} {014001}
  (\bibinfo {year} {1999})},\ \Eprint {http://arxiv.org/abs/hep-ph/9812358}
  {arXiv:hep-ph/9812358} \BibitemShut {NoStop}%
\bibitem [{\citenamefont {Tanabashi}\ \emph
  {et~al.}(2018{\natexlab{b}})\citenamefont {Tanabashi} \emph
  {et~al.}}]{Tanabashi:2018oca}%
  \BibitemOpen
  \bibfield  {author} {\bibinfo {author} {\bibfnamefont {M.}~\bibnamefont
  {Tanabashi}} \emph {et~al.} (\bibinfo {collaboration} {Particle Data
  Group}),\ }\href {\doibase 10.1103/PhysRevD.98.030001} {\bibfield  {journal}
  {\bibinfo  {journal} {Phys. Rev. D}\ }\textbf {\bibinfo {volume} {98}},\
  \bibinfo {pages} {030001} (\bibinfo {year} {2018}{\natexlab{b}})}\BibitemShut
  {NoStop}%
\bibitem [{\citenamefont {Hatton}\ \emph
  {et~al.}(2020{\natexlab{b}})\citenamefont {Hatton}, \citenamefont {Davies},
  \citenamefont {Galloway}, \citenamefont {Koponen}, \citenamefont {Lepage},\
  and\ \citenamefont {Lytle}}]{Hatton:2020qhk}%
  \BibitemOpen
  \bibfield  {author} {\bibinfo {author} {\bibfnamefont {D.}~\bibnamefont
  {Hatton}}, \bibinfo {author} {\bibfnamefont {C.~T.~H.}\ \bibnamefont
  {Davies}}, \bibinfo {author} {\bibfnamefont {B.}~\bibnamefont {Galloway}},
  \bibinfo {author} {\bibfnamefont {J.}~\bibnamefont {Koponen}}, \bibinfo
  {author} {\bibfnamefont {G.~P.}\ \bibnamefont {Lepage}}, \ and\ \bibinfo
  {author} {\bibfnamefont {A.~T.}\ \bibnamefont {Lytle}} (\bibinfo
  {collaboration} {HPQCD}),\ }\href {\doibase 10.1103/PhysRevD.102.054511}
  {\bibfield  {journal} {\bibinfo  {journal} {Phys. Rev. D}\ }\textbf {\bibinfo
  {volume} {102}},\ \bibinfo {pages} {054511} (\bibinfo {year}
  {2020}{\natexlab{b}})},\ \Eprint {http://arxiv.org/abs/2005.01845}
  {arXiv:2005.01845 [hep-lat]} \BibitemShut {NoStop}%
\bibitem [{\citenamefont {Bardeen}\ \emph {et~al.}(2003)\citenamefont
  {Bardeen}, \citenamefont {Eichten},\ and\ \citenamefont
  {Hill}}]{Bardeen:2003kt}%
  \BibitemOpen
  \bibfield  {author} {\bibinfo {author} {\bibfnamefont {W.~A.}\ \bibnamefont
  {Bardeen}}, \bibinfo {author} {\bibfnamefont {E.~J.}\ \bibnamefont
  {Eichten}}, \ and\ \bibinfo {author} {\bibfnamefont {C.~T.}\ \bibnamefont
  {Hill}},\ }\href {\doibase 10.1103/PhysRevD.68.054024} {\bibfield  {journal}
  {\bibinfo  {journal} {Phys. Rev. D}\ }\textbf {\bibinfo {volume} {68}},\
  \bibinfo {pages} {054024} (\bibinfo {year} {2003})},\ \Eprint
  {http://arxiv.org/abs/hep-ph/0305049} {arXiv:hep-ph/0305049} \BibitemShut
  {NoStop}%
\bibitem [{\citenamefont {Lang}\ \emph {et~al.}(2015)\citenamefont {Lang},
  \citenamefont {Mohler}, \citenamefont {Prelovsek},\ and\ \citenamefont
  {Woloshyn}}]{Lang:2015hza}%
  \BibitemOpen
  \bibfield  {author} {\bibinfo {author} {\bibfnamefont {C.~B.}\ \bibnamefont
  {Lang}}, \bibinfo {author} {\bibfnamefont {D.}~\bibnamefont {Mohler}},
  \bibinfo {author} {\bibfnamefont {S.}~\bibnamefont {Prelovsek}}, \ and\
  \bibinfo {author} {\bibfnamefont {R.~M.}\ \bibnamefont {Woloshyn}},\ }\href
  {\doibase 10.1016/j.physletb.2015.08.038} {\bibfield  {journal} {\bibinfo
  {journal} {Phys. Lett.}\ }\textbf {\bibinfo {volume} {B750}},\ \bibinfo
  {pages} {17} (\bibinfo {year} {2015})},\ \Eprint
  {http://arxiv.org/abs/1501.01646} {arXiv:1501.01646 [hep-lat]} \BibitemShut
  {NoStop}%
%%CITATION = ARXIV:1501.01646;%%
\bibitem [{\citenamefont {Falk}\ and\ \citenamefont
  {Neubert}(1993)}]{Falk:1992ws}%
  \BibitemOpen
  \bibfield  {author} {\bibinfo {author} {\bibfnamefont {A.~F.}\ \bibnamefont
  {Falk}}\ and\ \bibinfo {author} {\bibfnamefont {M.}~\bibnamefont {Neubert}},\
  }\href {\doibase 10.1103/PhysRevD.47.2982} {\bibfield  {journal} {\bibinfo
  {journal} {Phys. Rev.}\ }\textbf {\bibinfo {volume} {D47}},\ \bibinfo {pages}
  {2982} (\bibinfo {year} {1993})},\ \Eprint
  {http://arxiv.org/abs/hep-ph/9209269} {arXiv:hep-ph/9209269 [hep-ph]}
  \BibitemShut {NoStop}%
%%CITATION = HEP-PH/9209269;%%
\bibitem [{\citenamefont {Georgi}(1990)}]{Georgi:1990um}%
  \BibitemOpen
  \bibfield  {author} {\bibinfo {author} {\bibfnamefont {H.}~\bibnamefont
  {Georgi}},\ }\href {\doibase 10.1016/0370-2693(90)91128-X} {\bibfield
  {journal} {\bibinfo  {journal} {Phys. Lett.}\ }\textbf {\bibinfo {volume}
  {B240}},\ \bibinfo {pages} {447} (\bibinfo {year} {1990})}\BibitemShut
  {NoStop}%
%%CITATION = PHLTA,B240,447;%%
\bibitem [{\citenamefont {Dowdall}\ \emph {et~al.}(2019)\citenamefont
  {Dowdall}, \citenamefont {Davies}, \citenamefont {Horgan}, \citenamefont
  {Lepage}, \citenamefont {Monahan}, \citenamefont {Shigemitsu},\ and\
  \citenamefont {Wingate}}]{Dowdall:2019bea}%
  \BibitemOpen
  \bibfield  {author} {\bibinfo {author} {\bibfnamefont {R.~J.}\ \bibnamefont
  {Dowdall}}, \bibinfo {author} {\bibfnamefont {C.~T.~H.}\ \bibnamefont
  {Davies}}, \bibinfo {author} {\bibfnamefont {R.~R.}\ \bibnamefont {Horgan}},
  \bibinfo {author} {\bibfnamefont {G.~P.}\ \bibnamefont {Lepage}}, \bibinfo
  {author} {\bibfnamefont {C.~J.}\ \bibnamefont {Monahan}}, \bibinfo {author}
  {\bibfnamefont {J.}~\bibnamefont {Shigemitsu}}, \ and\ \bibinfo {author}
  {\bibfnamefont {M.}~\bibnamefont {Wingate}},\ }\href {\doibase
  10.1103/PhysRevD.100.094508} {\bibfield  {journal} {\bibinfo  {journal}
  {Phys. Rev. D}\ }\textbf {\bibinfo {volume} {100}},\ \bibinfo {pages}
  {094508} (\bibinfo {year} {2019})},\ \Eprint
  {http://arxiv.org/abs/1907.01025} {arXiv:1907.01025 [hep-lat]} \BibitemShut
  {NoStop}%
\bibitem [{\citenamefont {Lubicz}\ \emph {et~al.}(2017)\citenamefont {Lubicz},
  \citenamefont {Riggio}, \citenamefont {Salerno}, \citenamefont {Simula},\
  and\ \citenamefont {Tarantino}}]{Lubicz:2017syv}%
  \BibitemOpen
  \bibfield  {author} {\bibinfo {author} {\bibfnamefont {V.}~\bibnamefont
  {Lubicz}}, \bibinfo {author} {\bibfnamefont {L.}~\bibnamefont {Riggio}},
  \bibinfo {author} {\bibfnamefont {G.}~\bibnamefont {Salerno}}, \bibinfo
  {author} {\bibfnamefont {S.}~\bibnamefont {Simula}}, \ and\ \bibinfo {author}
  {\bibfnamefont {C.}~\bibnamefont {Tarantino}} (\bibinfo {collaboration}
  {ETM}),\ }\href {\doibase 10.1103/PhysRevD.96.054514} {\bibfield  {journal}
  {\bibinfo  {journal} {Phys. Rev. D}\ }\textbf {\bibinfo {volume} {96}},\
  \bibinfo {pages} {054514} (\bibinfo {year} {2017})},\ \bibinfo {note}
  {[Erratum: Phys.Rev.D 99, 099902 (2019), Erratum: Phys.Rev.D 100, 079901
  (2019)]},\ \Eprint {http://arxiv.org/abs/1706.03017} {arXiv:1706.03017
  [hep-lat]} \BibitemShut {NoStop}%
\bibitem [{\citenamefont {Lepage}(gvar)}]{gvar}%
  \BibitemOpen
  \bibfield  {author} {\bibinfo {author} {\bibfnamefont {G.~P.}\ \bibnamefont
  {Lepage}},\ }\href@noop {} {\bibfield  {journal} {\bibinfo  {journal} {gvar
  Version 9.22}\ } (\bibinfo {year} {github.com/gplepage/gvar})}\BibitemShut
  {NoStop}%
\bibitem [{\citenamefont {Zyla}\ \emph
  {et~al.}(2020{\natexlab{a}})\citenamefont {Zyla} \emph
  {et~al.}}]{Zyla:2020zbs}%
  \BibitemOpen
  \bibfield  {author} {\bibinfo {author} {\bibfnamefont {P.~A.}\ \bibnamefont
  {Zyla}} \emph {et~al.} (\bibinfo {collaboration} {Particle Data Group}),\
  }\href {\doibase 10.1093/ptep/ptaa104} {\bibfield  {journal} {\bibinfo
  {journal} {PTEP}\ }\textbf {\bibinfo {volume} {2020}},\ \bibinfo {pages}
  {083C01} (\bibinfo {year} {2020}{\natexlab{a}})}\BibitemShut {NoStop}%
\bibitem [{\citenamefont {Aaij}\ \emph {et~al.}(2015)\citenamefont {Aaij} \emph
  {et~al.}}]{Aaij:2014gka}%
  \BibitemOpen
  \bibfield  {author} {\bibinfo {author} {\bibfnamefont {R.}~\bibnamefont
  {Aaij}} \emph {et~al.} (\bibinfo {collaboration} {LHCb}),\ }\href {\doibase
  10.1016/j.physletb.2015.01.010} {\bibfield  {journal} {\bibinfo  {journal}
  {Phys. Lett. B}\ }\textbf {\bibinfo {volume} {742}},\ \bibinfo {pages} {29}
  (\bibinfo {year} {2015})},\ \Eprint {http://arxiv.org/abs/1411.6899}
  {arXiv:1411.6899 [hep-ex]} \BibitemShut {NoStop}%
\bibitem [{\citenamefont {Jenkins}\ \emph {et~al.}(1993)\citenamefont
  {Jenkins}, \citenamefont {Luke}, \citenamefont {Manohar},\ and\ \citenamefont
  {Savage}}]{Jenkins:1992nb}%
  \BibitemOpen
  \bibfield  {author} {\bibinfo {author} {\bibfnamefont {E.~E.}\ \bibnamefont
  {Jenkins}}, \bibinfo {author} {\bibfnamefont {M.~E.}\ \bibnamefont {Luke}},
  \bibinfo {author} {\bibfnamefont {A.~V.}\ \bibnamefont {Manohar}}, \ and\
  \bibinfo {author} {\bibfnamefont {M.~J.}\ \bibnamefont {Savage}},\ }\href
  {\doibase 10.1016/0550-3213(93)90464-Z} {\bibfield  {journal} {\bibinfo
  {journal} {Nucl. Phys. B}\ }\textbf {\bibinfo {volume} {390}},\ \bibinfo
  {pages} {463} (\bibinfo {year} {1993})},\ \Eprint
  {http://arxiv.org/abs/hep-ph/9204238} {arXiv:hep-ph/9204238} \BibitemShut
  {NoStop}%
\bibitem [{\citenamefont {Leljak}\ and\ \citenamefont
  {Melic}(2020)}]{Leljak:2019fqa}%
  \BibitemOpen
  \bibfield  {author} {\bibinfo {author} {\bibfnamefont {D.}~\bibnamefont
  {Leljak}}\ and\ \bibinfo {author} {\bibfnamefont {B.}~\bibnamefont {Melic}},\
  }\href {\doibase 10.1007/JHEP02(2020)171} {\bibfield  {journal} {\bibinfo
  {journal} {JHEP}\ }\textbf {\bibinfo {volume} {02}},\ \bibinfo {pages} {171}
  (\bibinfo {year} {2020})},\ \Eprint {http://arxiv.org/abs/1909.01213}
  {arXiv:1909.01213 [hep-ph]} \BibitemShut {NoStop}%
\bibitem [{\citenamefont {Becirevic}\ \emph {et~al.}(2012)\citenamefont
  {Becirevic}, \citenamefont {Kosnik}, \citenamefont {Mescia},\ and\
  \citenamefont {Schneider}}]{Becirevic:2012fy}%
  \BibitemOpen
  \bibfield  {author} {\bibinfo {author} {\bibfnamefont {D.}~\bibnamefont
  {Becirevic}}, \bibinfo {author} {\bibfnamefont {N.}~\bibnamefont {Kosnik}},
  \bibinfo {author} {\bibfnamefont {F.}~\bibnamefont {Mescia}}, \ and\ \bibinfo
  {author} {\bibfnamefont {E.}~\bibnamefont {Schneider}},\ }\href {\doibase
  10.1103/PhysRevD.86.034034} {\bibfield  {journal} {\bibinfo  {journal} {Phys.
  Rev. D}\ }\textbf {\bibinfo {volume} {86}},\ \bibinfo {pages} {034034}
  (\bibinfo {year} {2012})},\ \Eprint {http://arxiv.org/abs/1205.5811}
  {arXiv:1205.5811 [hep-ph]} \BibitemShut {NoStop}%
\bibitem [{\citenamefont {Khodjamirian}\ \emph {et~al.}(2013)\citenamefont
  {Khodjamirian}, \citenamefont {Mannel},\ and\ \citenamefont
  {Wang}}]{Khodjamirian:2012rm}%
  \BibitemOpen
  \bibfield  {author} {\bibinfo {author} {\bibfnamefont {A.}~\bibnamefont
  {Khodjamirian}}, \bibinfo {author} {\bibfnamefont {T.}~\bibnamefont
  {Mannel}}, \ and\ \bibinfo {author} {\bibfnamefont {Y.~M.}\ \bibnamefont
  {Wang}},\ }\href {\doibase 10.1007/JHEP02(2013)010} {\bibfield  {journal}
  {\bibinfo  {journal} {JHEP}\ }\textbf {\bibinfo {volume} {02}},\ \bibinfo
  {pages} {010} (\bibinfo {year} {2013})},\ \Eprint
  {http://arxiv.org/abs/1211.0234} {arXiv:1211.0234 [hep-ph]} \BibitemShut
  {NoStop}%
\bibitem [{\citenamefont {Hambrock}\ \emph {et~al.}(2015)\citenamefont
  {Hambrock}, \citenamefont {Khodjamirian},\ and\ \citenamefont
  {Rusov}}]{Hambrock:2015wka}%
  \BibitemOpen
  \bibfield  {author} {\bibinfo {author} {\bibfnamefont {C.}~\bibnamefont
  {Hambrock}}, \bibinfo {author} {\bibfnamefont {A.}~\bibnamefont
  {Khodjamirian}}, \ and\ \bibinfo {author} {\bibfnamefont {A.}~\bibnamefont
  {Rusov}},\ }\href {\doibase 10.1103/PhysRevD.92.074020} {\bibfield  {journal}
  {\bibinfo  {journal} {Phys. Rev. D}\ }\textbf {\bibinfo {volume} {92}},\
  \bibinfo {pages} {074020} (\bibinfo {year} {2015})},\ \Eprint
  {http://arxiv.org/abs/1506.07760} {arXiv:1506.07760 [hep-ph]} \BibitemShut
  {NoStop}%
\bibitem [{\citenamefont {Charles}\ \emph {et~al.}(2005)\citenamefont
  {Charles}, \citenamefont {Hocker}, \citenamefont {Lacker}, \citenamefont
  {Laplace}, \citenamefont {Le~Diberder}, \citenamefont {Malcles},
  \citenamefont {Ocariz}, \citenamefont {Pivk},\ and\ \citenamefont
  {Roos}}]{Charles:2004jd}%
  \BibitemOpen
  \bibfield  {author} {\bibinfo {author} {\bibfnamefont {J.}~\bibnamefont
  {Charles}}, \bibinfo {author} {\bibfnamefont {A.}~\bibnamefont {Hocker}},
  \bibinfo {author} {\bibfnamefont {H.}~\bibnamefont {Lacker}}, \bibinfo
  {author} {\bibfnamefont {S.}~\bibnamefont {Laplace}}, \bibinfo {author}
  {\bibfnamefont {F.~R.}\ \bibnamefont {Le~Diberder}}, \bibinfo {author}
  {\bibfnamefont {J.}~\bibnamefont {Malcles}}, \bibinfo {author} {\bibfnamefont
  {J.}~\bibnamefont {Ocariz}}, \bibinfo {author} {\bibfnamefont
  {M.}~\bibnamefont {Pivk}}, \ and\ \bibinfo {author} {\bibfnamefont
  {L.}~\bibnamefont {Roos}} (\bibinfo {collaboration} {CKMfitter Group}),\
  }\href {\doibase 10.1140/epjc/s2005-02169-1} {\bibfield  {journal} {\bibinfo
  {journal} {Eur. Phys. J. C}\ }\textbf {\bibinfo {volume} {41}},\ \bibinfo
  {pages} {1} (\bibinfo {year} {2005})},\ \Eprint
  {http://arxiv.org/abs/hep-ph/0406184} {arXiv:hep-ph/0406184} \BibitemShut
  {NoStop}%
\bibitem [{\citenamefont {Lytle}\ \emph {et~al.}(2018)\citenamefont {Lytle},
  \citenamefont {Davies}, \citenamefont {Hatton}, \citenamefont {Lepage},\ and\
  \citenamefont {Sturm}}]{Lytle:2018evc}%
  \BibitemOpen
  \bibfield  {author} {\bibinfo {author} {\bibfnamefont {A.~T.}\ \bibnamefont
  {Lytle}}, \bibinfo {author} {\bibfnamefont {C.~T.~H.}\ \bibnamefont
  {Davies}}, \bibinfo {author} {\bibfnamefont {D.}~\bibnamefont {Hatton}},
  \bibinfo {author} {\bibfnamefont {G.~P.}\ \bibnamefont {Lepage}}, \ and\
  \bibinfo {author} {\bibfnamefont {C.}~\bibnamefont {Sturm}} (\bibinfo
  {collaboration} {HPQCD}),\ }\href {\doibase 10.1103/PhysRevD.98.014513}
  {\bibfield  {journal} {\bibinfo  {journal} {Phys. Rev. D}\ }\textbf {\bibinfo
  {volume} {98}},\ \bibinfo {pages} {014513} (\bibinfo {year} {2018})},\
  \Eprint {http://arxiv.org/abs/1805.06225} {arXiv:1805.06225 [hep-lat]}
  \BibitemShut {NoStop}%
\bibitem [{\citenamefont {Hatton}\ \emph {et~al.}(2021)\citenamefont {Hatton},
  \citenamefont {Davies}, \citenamefont {Koponen}, \citenamefont {Lepage},\
  and\ \citenamefont {Lytle}}]{Hatton:2021syc}%
  \BibitemOpen
  \bibfield  {author} {\bibinfo {author} {\bibfnamefont {D.}~\bibnamefont
  {Hatton}}, \bibinfo {author} {\bibfnamefont {C.~T.~H.}\ \bibnamefont
  {Davies}}, \bibinfo {author} {\bibfnamefont {J.}~\bibnamefont {Koponen}},
  \bibinfo {author} {\bibfnamefont {G.~P.}\ \bibnamefont {Lepage}}, \ and\
  \bibinfo {author} {\bibfnamefont {A.~T.}\ \bibnamefont {Lytle}},\ }\href
  {\doibase 10.1103/PhysRevD.103.114508} {\bibfield  {journal} {\bibinfo
  {journal} {Phys. Rev. D}\ }\textbf {\bibinfo {volume} {103}},\ \bibinfo
  {pages} {114508} (\bibinfo {year} {2021})},\ \Eprint
  {http://arxiv.org/abs/2102.09609} {arXiv:2102.09609 [hep-lat]} \BibitemShut
  {NoStop}%
\bibitem [{\citenamefont {Melnikov}\ and\ \citenamefont
  {Ritbergen}(2000)}]{Melnikov:2000qh}%
  \BibitemOpen
  \bibfield  {author} {\bibinfo {author} {\bibfnamefont {K.}~\bibnamefont
  {Melnikov}}\ and\ \bibinfo {author} {\bibfnamefont {T.~v.}\ \bibnamefont
  {Ritbergen}},\ }\href {\doibase 10.1016/S0370-2693(00)00507-4} {\bibfield
  {journal} {\bibinfo  {journal} {Phys. Lett. B}\ }\textbf {\bibinfo {volume}
  {482}},\ \bibinfo {pages} {99} (\bibinfo {year} {2000})},\ \Eprint
  {http://arxiv.org/abs/hep-ph/9912391} {arXiv:hep-ph/9912391} \BibitemShut
  {NoStop}%
\bibitem [{\citenamefont {Beneke}\ and\ \citenamefont
  {Braun}(1994)}]{Beneke:1994sw}%
  \BibitemOpen
  \bibfield  {author} {\bibinfo {author} {\bibfnamefont {M.}~\bibnamefont
  {Beneke}}\ and\ \bibinfo {author} {\bibfnamefont {V.~M.}\ \bibnamefont
  {Braun}},\ }\href {\doibase 10.1016/0550-3213(94)90314-X} {\bibfield
  {journal} {\bibinfo  {journal} {Nucl. Phys. B}\ }\textbf {\bibinfo {volume}
  {426}},\ \bibinfo {pages} {301} (\bibinfo {year} {1994})},\ \Eprint
  {http://arxiv.org/abs/hep-ph/9402364} {arXiv:hep-ph/9402364} \BibitemShut
  {NoStop}%
\bibitem [{\citenamefont {Aaij}\ \emph {et~al.}(2012)\citenamefont {Aaij} \emph
  {et~al.}}]{Aaij:2012cq}%
  \BibitemOpen
  \bibfield  {author} {\bibinfo {author} {\bibfnamefont {R.}~\bibnamefont
  {Aaij}} \emph {et~al.} (\bibinfo {collaboration} {LHCb}),\ }\href {\doibase
  10.1007/JHEP07(2012)133} {\bibfield  {journal} {\bibinfo  {journal} {JHEP}\
  }\textbf {\bibinfo {volume} {07}},\ \bibinfo {pages} {133} (\bibinfo {year}
  {2012})},\ \Eprint {http://arxiv.org/abs/1205.3422} {arXiv:1205.3422
  [hep-ex]} \BibitemShut {NoStop}%
\bibitem [{\citenamefont {Du}\ \emph {et~al.}(2016)\citenamefont {Du},
  \citenamefont {El-Khadra}, \citenamefont {Gottlieb}, \citenamefont
  {Kronfeld}, \citenamefont {Laiho}, \citenamefont {Lunghi}, \citenamefont
  {Van~de Water},\ and\ \citenamefont {Zhou}}]{Du:2015tda}%
  \BibitemOpen
  \bibfield  {author} {\bibinfo {author} {\bibfnamefont {D.}~\bibnamefont
  {Du}}, \bibinfo {author} {\bibfnamefont {A.~X.}\ \bibnamefont {El-Khadra}},
  \bibinfo {author} {\bibfnamefont {S.}~\bibnamefont {Gottlieb}}, \bibinfo
  {author} {\bibfnamefont {A.~S.}\ \bibnamefont {Kronfeld}}, \bibinfo {author}
  {\bibfnamefont {J.}~\bibnamefont {Laiho}}, \bibinfo {author} {\bibfnamefont
  {E.}~\bibnamefont {Lunghi}}, \bibinfo {author} {\bibfnamefont {R.~S.}\
  \bibnamefont {Van~de Water}}, \ and\ \bibinfo {author} {\bibfnamefont
  {R.}~\bibnamefont {Zhou}},\ }\href {\doibase 10.1103/PhysRevD.93.034005}
  {\bibfield  {journal} {\bibinfo  {journal} {Phys. Rev. D}\ }\textbf {\bibinfo
  {volume} {93}},\ \bibinfo {pages} {034005} (\bibinfo {year} {2016})},\
  \Eprint {http://arxiv.org/abs/1510.02349} {arXiv:1510.02349 [hep-ph]}
  \BibitemShut {NoStop}%
\bibitem [{\citenamefont {Zyla}\ \emph
  {et~al.}(2020{\natexlab{b}})\citenamefont {Zyla} \emph {et~al.}}]{pdg20}%
  \BibitemOpen
  \bibfield  {author} {\bibinfo {author} {\bibfnamefont {P.}~\bibnamefont
  {Zyla}} \emph {et~al.} (\bibinfo {collaboration} {Particle Data Group}),\
  }\href@noop {} {\bibfield  {journal} {\bibinfo  {journal} {Prog. Theor. Exp.
  Phys.}\ ,\ \bibinfo {pages} {083C01}} (\bibinfo {year}
  {2020}{\natexlab{b}})}\BibitemShut {NoStop}%
\bibitem [{\citenamefont {Bobeth}\ \emph {et~al.}(2007)\citenamefont {Bobeth},
  \citenamefont {Hiller},\ and\ \citenamefont {Piranishvili}}]{Bobeth:2007dw}%
  \BibitemOpen
  \bibfield  {author} {\bibinfo {author} {\bibfnamefont {C.}~\bibnamefont
  {Bobeth}}, \bibinfo {author} {\bibfnamefont {G.}~\bibnamefont {Hiller}}, \
  and\ \bibinfo {author} {\bibfnamefont {G.}~\bibnamefont {Piranishvili}},\
  }\href {\doibase 10.1088/1126-6708/2007/12/040} {\bibfield  {journal}
  {\bibinfo  {journal} {JHEP}\ }\textbf {\bibinfo {volume} {12}},\ \bibinfo
  {pages} {040} (\bibinfo {year} {2007})},\ \Eprint
  {http://arxiv.org/abs/0709.4174} {arXiv:0709.4174 [hep-ph]} \BibitemShut
  {NoStop}%
\bibitem [{\citenamefont {Brod}\ \emph {et~al.}(2011)\citenamefont {Brod},
  \citenamefont {Gorbahn},\ and\ \citenamefont {Stamou}}]{Brod:2010hi}%
  \BibitemOpen
  \bibfield  {author} {\bibinfo {author} {\bibfnamefont {J.}~\bibnamefont
  {Brod}}, \bibinfo {author} {\bibfnamefont {M.}~\bibnamefont {Gorbahn}}, \
  and\ \bibinfo {author} {\bibfnamefont {E.}~\bibnamefont {Stamou}},\ }\href
  {\doibase 10.1103/PhysRevD.83.034030} {\bibfield  {journal} {\bibinfo
  {journal} {Phys. Rev. D}\ }\textbf {\bibinfo {volume} {83}},\ \bibinfo
  {pages} {034030} (\bibinfo {year} {2011})},\ \Eprint
  {http://arxiv.org/abs/1009.0947} {arXiv:1009.0947 [hep-ph]} \BibitemShut
  {NoStop}%
\bibitem [{\citenamefont {Aaij}\ \emph {et~al.}(2018)\citenamefont {Aaij} \emph
  {et~al.}}]{Bediaga:2018lhg}%
  \BibitemOpen
  \bibfield  {author} {\bibinfo {author} {\bibfnamefont {R.}~\bibnamefont
  {Aaij}} \emph {et~al.} (\bibinfo {collaboration} {LHCb}),\ }\href@noop {} {\
  (\bibinfo {year} {2018})},\ \Eprint {http://arxiv.org/abs/1808.08865}
  {arXiv:1808.08865 [hep-ex]} \BibitemShut {NoStop}%
\bibitem [{\citenamefont {Michael}(1994)}]{Michael:1993yj}%
  \BibitemOpen
  \bibfield  {author} {\bibinfo {author} {\bibfnamefont {C.}~\bibnamefont
  {Michael}},\ }\href {\doibase 10.1103/PhysRevD.49.2616} {\bibfield  {journal}
  {\bibinfo  {journal} {Phys. Rev.}\ }\textbf {\bibinfo {volume} {D49}},\
  \bibinfo {pages} {2616} (\bibinfo {year} {1994})},\ \Eprint
  {http://arxiv.org/abs/hep-lat/9310026} {arXiv:hep-lat/9310026 [hep-lat]}
  \BibitemShut {NoStop}%
%%CITATION = HEP-LAT/9310026;%%
\end{thebibliography}%
\bibliographystyle{apsrev4-1}

\end{document}